\documentclass[fleqn,usenatbib]{mnras}

\usepackage{amsmath}
\usepackage{txfonts} 
\usepackage{graphicx} 
\usepackage{xcolor} 

\newcommand{\revisedfig}[2]{%
  {\setlength{\fboxsep}{2pt}\setlength{\fboxrule}{1.2pt}%
   \fcolorbox{black}{white}{\includegraphics[#1]{#2}}}%
}
\newcommand{\Msun}{\mathrm{M}_{\odot}}
\newcommand{\Mdot}{\dot{M}}
\newcommand{\Sigdot}{\dot{\Sigma}}
\newcommand{\aDW}{\alpha_{\mathrm{DW}}}
\newcommand{\aSS}{\alpha_{\mathrm{SS}}}
\newcommand{\cs}{c_{\mathrm{s}}}
\newcommand{\csw}{c_{\mathrm{s,w}}}
\newcommand{\vK}{v_{\mathrm{K}}}
\newcommand{\Rin}{R_{\mathrm{in}}}
\newcommand{\Rout}{R_{\mathrm{out}}}

\title[Magnetothermal winds]{Bridging magnetothermal winds and
photoevaporation to model discs dispersal}

\author[G. Picogna et al.]{
Giovanni Picogna$^{1}$\thanks{E-mail: picogna@usm.lmu.de}
and Barbara Ercolano$^{1,2,3}$
\\
$^{1}$Universit\"{a}ts-Sternwarte, Ludwig-Maximilians-Universit\"{a}t M\"{u}nchen, Scheinerstr. 1, D-81679 M\"{u}nchen, Germany\\
$^{2}$Excellence Cluster Origins, Boltzmannstrasse 2, D-85748 Garching bei M\"{u}nchen, Germany\\
$^{3}$Max-Planck-Institut f\"{u}r Extraterrestrische Physik, Giessenbachstr. 1, 85748 Garching, Germany
}

\date{Draft -- compiled \today}
\pubyear{2026}

\begin{document}
\label{firstpage}
\maketitle

\begin{abstract}
Protoplanetary disc dispersal is driven by two processes usually modelled
separately: photoevaporative and magnetohydrodynamic (MHD) disc winds.
Global simulations indicate that in the inner disc these are not distinct outflows
but a single magnetothermal wind.
We assemble a closed-form, two-phase model that respects it.
A single-field-line wind, whose base is fixed by the irradiated temperature and penetration
column, supplies the launch and feeds a secular evolution, with photoevaporation convolved
on as a sink.
The flux closure $B_z\propto\Sigma^q$ is self-limiting: for $q\le1/2$ depletion alone
cannot demagnetise the disc, so dispersal requires independent flux loss, parameterised
by the magnetic Reynolds number $\mathcal{R}_m$.
Integrating the coupled system yields two regimes.
Efficient flux loss ($\mathcal{R}_m\lesssim1$) lets the magnetisation front recede by
over an order of magnitude and opens a photoevaporative gap. Flux retention
($\mathcal{R}_m\gg1$) drives the front outward, sustains accretion, and defers dispersal
by $\approx2.7$~Myr.
Deriving the base from stellar irradiation instead of prescribing it, we find that the
cold-launch approximation is valid during the early stages of disc evolution: anchoring
the base at plasma equipartition ($\beta_{\rm base} \sim 1$) confines irradiation's
influence on the magnetic lever arm to the magnetothermal annulus, decoupling the peak
accretion rate from the incident flux.
Both regimes clear the disc inside-out, through either a photoevaporatively amplified
cavity wall or an expanding magnetothermal front.
\end{abstract}

\begin{keywords}
accretion, accretion discs -- magnetohydrodynamics (MHD) -- protoplanetary discs
-- stars: winds, outflows
\end{keywords}

\section{Introduction}

The evolution, and eventual dispersal, of a protoplanetary disc is set by how
efficiently it transports angular momentum and loses mass.
Two pictures lead the recent literature.
In the first, a turbulent (``viscous'') stress drives accretion.
A thermally launched photoevaporative wind, powered by stellar extreme-ultraviolet (EUV),
far-ultraviolet (FUV) and X-ray irradiation, removes mass from the upper layers
\citep{clarke2001,alexander2006,gorti2009,owen2010,owen2012,picogna2019,komaki2021}.
In the second, a large-scale magnetic field threading the disc exerts a torque that
drives a magnetised wind, removing both mass and angular momentum
\citep{blandford1982,ferreira1997,casse2000a,casse2000b,suzuki2016,tabone2022}.
Both reproduce the two-timescale character of dispersal, set by stellar irradiation in
the first case and by the initial disc magnetisation in the second \citep{armitage2013}.
Both are invoked to explain dispersal timescales and demographics \citep{pascucci2023}.
Observationally the two wind types are difficult to separate
\citep[especially once external photoevaporation is included,][]{facchini2016,winterhaworth2022,anania2025}.

At the launch point, however, they are not separable mechanisms at all.
Local and global non-ideal MHD simulations of the inner disc show the wind leaving
several scale heights above the mid-plane, from layers that are simultaneously
magnetically torqued, heated and ionised by FUV and X-ray photons
\citep{bai2016,bai2017,bethune2017,wang2019,gressel2020}.
This is the same non-ideal environment in which MRI turbulence is suppressed over much of
the disc's lifetime, leaving the magnetised wind the only viable driver of the observed
accretion rates \citep{bai2017,lesur2021}.
\citet{bai2016} termed the resulting outflow ``magnetothermal'': a single wind whose
acceleration is shared between magnetic pressure and tension, and the thermal pressure of
the irradiated surface gas.
The cold magnetocentrifugal wind and the thermally dominated wind are the two end members
of a continuum, not independent additive winds.

Models combining an MHD-wind torque with a separate photoevaporative mass sink already exist.
The two terms can act at the same radii simultaneously, but each is computed from decoupled
driving physics \citep{kunitomo2020,weder2023,coleman2024,tabone2025}.
We propose a closed-form treatment that instead
(i) sets the wind-launch thermodynamics from the same irradiation that drives
photoevaporation, so the launch is genuinely magnetothermal as opposed to a superposed cold
wind plus a thermal sink, and
(ii) folds photoevaporation into the wind-driven secular evolution analytically.
Its components are established: the single-field-line magnetothermal wind solver of
\citet{bai2016}, the secular wind-driven operator of \citet{tabone2022}, and the
Green's-function treatment of photoevaporation of \citet{ruden2004}.
The novelty is in how they are coupled.

Our central result is that how a disc disperses depends on whether and when it leaves
the magnetothermal phase.
If the vertical magnetic field is tied to the gas surface density through the standard
closure $B_z\propto\Sigma^q$, the midplane plasma $\beta$ (the ratio of the
midplane thermal pressure to the poloidal magnetic pressure,
$\beta=8\pi P_{\mathrm{mid}}/B_z^2$ with $P_{\mathrm{mid}}=\Sigma\cs\Omega/\sqrt{2\pi}$)
evolves as:
\begin{equation}
    \label{eq:betaclosure_intro}
    \beta\propto \frac{\Sigma \cs^2}{B_z^2} \propto \Sigma^{1-2q},
\end{equation}
at fixed midplane temperature.
For $q\le1/2$, depletion alone cannot raise $\beta$ enough to demagnetise the disc.
The hand-over to photoevaporation-dominated dispersal is therefore not an automatic
consequence of disc evolution.
It requires genuine loss of magnetic flux.

That inequality is this paper's load-bearing assumption, and we do not derive it.
$q$ parametrises a flux transport the secular model does not solve.
We adopt it because global non-ideal MHD calculations find the poloidal flux advected
inward and concentrated instead of carried outward with the draining gas
\citep{baistone2017,suriano2018,suriano2019,cui2021,lesur2021}.
Were $q>1/2$ instead, draining alone would demagnetise the disc.
There would be no distinct flux-loss event, and the two branches below would collapse
into one gas-depletion-driven transition.
Section~\ref{sec:flux} sets out the evidence and its qualifications,
Section~\ref{sec:numsetup} the fiducial value.

We parametrise that loss by a single global timescale $\tau_B$,
\begin{equation}
    \label{eq:taub_intro}
    \beta\propto\Sigma^{1-2q}\exp\!\left(t/\tau_B\right),
\end{equation}
so that $\tau_B\to\infty$ retains the flux and finite $\tau_B$ loses it.
Here $\tau_B\sim R_{\mathrm{disc}}^2/\eta_{\mathrm{eff}}$ is the diffusion time set by
the effective diffusivity of the poloidal induction equation, and is interchangeable
with the magnetic Reynolds number $\mathcal{R}_m$ used throughout.
$\mathcal{R}_m\gg1$ means $\tau_B\to\infty$, and $\mathcal{R}_m\lesssim1$ means $\tau_B$
short compared with the disc lifetime.
It is a proxy for outward flux diffusion, not a solution of the induction equation.

As \citet{armitage2013} first suggested, this splits the late evolution into two branches:
\begin{itemize}
\item if flux is efficiently lost ($\mathcal{R}_m\lesssim1$), the magnetothermal driving weakens and internal photoevaporation
can open a gap;
\item if flux is retained ($\mathcal{R}_m\gg1$), the field concentrates as the disc drains, accretion is
sustained and dispersal deferred --- the regime in which \citet{martel2022} obtain an
accreting cavity.
\end{itemize}

Photoevaporative dispersal is thus the outcome of the flux-losing branch specifically,
not the default fate of every disc, and we make that choice the model's explicit control
parameter.

Section~\ref{sec:launch} sets out the wind launch physics and fixes the magnetothermal
wind base from photoevaporation physics;
Section~\ref{sec:secular} constructs the Green's function of the wind operator and the
photoevaporation convolution;
Section~\ref{sec:flux} ties the magnetothermal front to the magnetic flux transport;
Section~\ref{sec:results} integrates the two flux-transport branches and analyses their
evolution and dispersal.
Throughout, the photoevaporation is the star's own: external (FUV-driven) photoevaporation
is deliberately left out, and Section~\ref{sec:caveats} sets out why, at what cost, and
what would be needed to include it consistently.
Section~\ref{sec:discussion} compares our model with prior work and lists the caveats,
and Section~\ref{sec:conclusions} gives our main conclusions.
\section{The primordial phase: the magnetothermal wind with an irradiation-set base}
\label{sec:launch}

We consider a disc around a star of mass $M_\ast$, with cylindrical radius $R$,
Keplerian frequency $\Omega=\sqrt{GM_\ast/R^3}$, sound speed $\cs$,
and pressure scale height $H=\cs/\Omega$.
Table~\ref{tab:params} shows all the adopted variables grouped by role.

\begin{table}
\centering
\caption{Quantities in the model, grouped by role, with the values adopted in
Section~\ref{sec:results}.}
\label{tab:params}
\small
\setlength{\tabcolsep}{4pt}
\begin{tabular}{llr}
\hline
Symbol & Meaning & Value or definition \\
\hline
\multicolumn{3}{l}{\textit{Evolved field}}\\
$\Sigma(R,t)$ & disc surface density (dependent variable) \\
\hline
\multicolumn{3}{l}{\textit{Prescribed inputs}}\\
$M_\ast$ & stellar mass & $1\,\Msun$ \\
$M_{\rm disc}$ & initial disc mass & $0.02\,\Msun$ \\
$\Sigma_0(R)$ & initial surface density, $\propto(R/R_c)^{-1}e^{-R/R_c}$ & $R_c=30$~AU \\
$T_{\mathrm{mid},0}$ & passive mid-plane temperature at 1 AU & $280$~K \\
$p$ & power law mid-plane temperature index & $1/2$ \\
$\mu$ & mean molecular weight at the wind base & $2.34$ (molecular) \\
$q$ & flux-closure index, $B_z\propto\Sigma^q$ & 0.3 \\
$p_B$ & wind-stress index, $\aDW\propto\beta^{-p_B}$ & $\approx1$ \\
$q_{\mathrm{div}}$ & field-divergence index (Sec.~\ref{sec:bai}) & $\to\infty$ in eq.~(\ref{eq:spruit}) \\
$\beta_0(R)$ & initial mid-plane plasma-$\beta$ & $30\,(R/\mathrm{AU})^{3/2}$ \\
$\beta_{\mathrm{crit}}$ & launch-cutoff plasma-$\beta$ (midplane) & $10^4$ \\
$\beta_{\mathrm{couple}}$ & coupling-height plasma-$\beta$ (local) & $1$ \\
$\aDW^0$ & wind transport coefficient, initial value & $10^{-3}$ \\
$\aDW^{\max}$ & ceiling on $\aDW$ (eq.~\ref{eq:adwfull}) & $1$ \\
$\aSS$ & turbulent (viscous) $\alpha$ & $10^{-5}$ \\
$\tau_B$ & flux-loss timescale (the branch control) & $\infty$ / $0.5$~Myr \\
$\mathcal{R}_m$ & large-scale magnetic Reynolds number & set by $\tau_B$ \\
$L_X$ & stellar X-ray luminosity & $2\times10^{30}$~erg~s$^{-1}$ \\
$N_{\mathrm b}$ & radiation penetration column (eq.~\ref{eq:Nrad}) & $10^{21}$~cm$^{-2}$ \\
$f_{\mathrm{gap},\sigma}$ & depletion threshold defining $R_{\rm cav}$ & $0.05$ \\
\hline
\multicolumn{3}{l}{\textit{Calculated (primordial-phase launch / identities)}}\\
$\lambda$ & magnetic lever arm, $(R_{\mathrm A}/R_0)^2$ & derived (Sec.~\ref{sec:bai}) \\
$\xi$ & ejection index, $=1/[2(\lambda-1)]$ & derived \\
$z_0(R,t)$ & wind-base height & $\min(z_{0,\mathrm{th}},z_{0,\mathrm{mag}})$ \\
$\omega_B$ & flux-decay index, $p_B(1-2q)$ & $0.4$ \\
$\Xi$ & X-ray ionisation parameter & $L_X/(nr^2)$ \\
$T_{\mathrm b}(R)$ & irradiated wind-base temperature & Sec.~\ref{sec:bridge} \\
$\beta_{\mathrm{base}}$ & plasma-$\beta$ at $z_0$; sets $\lambda$ \\
$S(\beta)$ & launch switch & eq.~(\ref{eq:launchswitch}) \\
$\psi$ & $\aDW/\aSS$; analytic limit assumes $\psi\gg1$ \\
\hline
\multicolumn{3}{l}{\textit{Diagnostics (from the evolving $\Sigma(R,t)$)}}\\
$\beta(R,t)$ & evolving mid-plane plasma-$\beta$ & eq.~(\ref{eq:betaclosure}) \\
$R_\beta(t)$ & magnetisation front \\
$R_{\mathrm c}(t)$ & photoevaporation cut-off radius \\
$R_{\mathrm{gap}}(t)$ & photoevaporation gap front (rate-based) \\
$R_{\mathrm{cav}}(t)$ & inner-cavity edge (density-based) \\
$\Mdot_\ast(t)$ & stellar accretion rate, $\equiv\Mdot_{\mathrm{acc}}$ \\
\hline
\end{tabular}
\end{table}

The central quantity is the magnetic lever arm
$\lambda\equiv(R_{\mathrm{A}}/R_0)^2$, the squared ratio of the Alfv\'en radius to
the launch radius on a wind field line \citep{blandford1982}. In a radially
self-similar accretion--ejection flow the wind-driven mass loss and accretion are
related through the ejection index
\begin{equation}
\xi \;\equiv\; \left.\frac{\mathrm{d}\Mdot_{\mathrm{w}}/\mathrm{d}\ln R}{\Mdot_{\mathrm{acc}}}\right|_{R_0}
\;=\; \frac{1}{2(\lambda-1)},
\label{eq:xilambda}
\end{equation}
\citep{ferreira1997,bai2016}.
The launch calculation of Section~\ref{sec:launch} returns $\xi$ (hence
$\lambda$), and the secular evolution of Section~\ref{sec:secular} adopts the
same $\xi$ as its wind mass-loss coefficient.

We construct the launch calculation in three steps.
Section~\ref{sec:bai} sets out the single-field-line solution that returns
$\lambda$ from a prescribed base state. Section~\ref{sec:bridge} fixes the base
state itself from the irradiation and magnetic-coupling conditions at the disc surface.
Section~\ref{sec:launchout} then collects the resulting wind properties and their
limiting behaviour.

\subsection{The single-field-line solution}
\label{sec:bai}

We model the wind launching mechanism using the single-field-line
formulation of \citet{bai2016}, which reduces the steady, axisymmetric,
ideal-MHD equations to one-dimensional conservation laws along a
prescribed poloidal field line \citep{weber1967,spruit1996}.

The field line is straight, of inclination $\theta$ measured from the disc
mid-plane, so that arc length $s$ from the base $(R_0,z_0)$ traces
$R(s)=R_0+s\cos\theta$, $z(s)=z_0+s\sin\theta$. The poloidal field strength along
it is prescribed to steepen from parallel ($B_{\mathrm p}\propto R^{-1}$) to
diverging ($\propto R^{-2}$) geometry, with the transition set by a parameter
$q_{\mathrm{div}}$.
Under this kinematic prescription $\nabla \cdot \mathbf{B} = 0$ ties
$B_{\mathrm p}(s)$ to the flux-tube cross-section. This means $q_{\mathrm{div}}$
governs how neighbouring field lines diverge, and the full trans-field force
balance need not be solved.
The transition sits at roughly $q_{\mathrm div}^{-1/2} R_0$. For their fiducial
$q_{\rm div} = 0.25$ (transition at $\simeq 2R_0$) \citet{bai2016} find that the
Alfv\'en radius varies by a factor of a few across their parameter space.

In Gaussian cgs units, continuity, induction and azimuthal momentum give three
constants along the line: the mass loading $k$, the field-line angular velocity
$\omega$, and the specific angular momentum $l$,
\begin{equation}
k=\frac{4\pi\rho v_{\mathrm{p}}}{B_{\mathrm{p}}},\quad
\omega=\frac{v_\phi}{R}-\frac{kB_\phi}{4\pi\rho R},\quad
l=Rv_\phi-\frac{RB_\phi}{k},
\label{eq:invariants}
\end{equation}
with $\rho$ the gas density, $v_{\mathrm p}$ and $v_\phi$ the poloidal and
azimuthal fluid speeds, and $B_\phi$ the toroidal field. At the base
$B_\phi=0$ and the field co-rotates, so $\omega\simeq\Omega_{\mathrm K}(R_0)$, but
$v_\phi/R$ departs from it downstream as the Maxwell stress spins the gas up. The
energy equation supplies a fourth constant, the Bernoulli function
\begin{equation}
E=\tfrac{1}{2}\left(v_{\mathrm p}^2+v_\phi^2\right)+\mathcal{H}
-\frac{GM_\ast}{r}-\frac{\omega RB_\phi}{k},
\label{eq:bernoulli}
\end{equation}
with $r$ the spherical radius, in which thermodynamics enters only through the
enthalpy $\mathcal{H}=\int\mathrm{d}P/\rho$ and the last term is the work done by
the magnetic torque \citep{blandford1982,spruit1996}. For the isothermal closure
$P=\rho\csw^2$ adopted by \citet{bai2016}, $\mathcal{H}=\csw^2\ln\rho$, so the
entire thermal content of the problem is carried by the single base sound
speed $\csw$.

The standard reduction \citep{weber1967,spruit1996} eliminates $B_\phi$ between
the invariants and casts the flow as one algebraic relation along the line in the
variable $x\equiv v_{\mathrm{Ap}}^2/v_{\mathrm p}^2$, the squared ratio of the
poloidal Alfv\'en speed to the flow speed. That relation is singular where
$x=1$, and finiteness of $v_\phi$ there forces $l=\omega R_{\mathrm A}^2$, with
$R_{\mathrm A}$ the cylindrical radius of the Alfv\'en crossing. This is the
sense in which the wind's angular momentum sets its lever arm
$\lambda=(R_{\mathrm A}/R_0)^2$: the conserved $l$ and $\omega$, both fixed at
the base, determine where the Alfv\'en surface must sit. Requiring the flow to
pass smoothly through the slow and fast magnetosonic points in addition imposes
three conditions on two unknowns at each critical point, so the problem is an
eigenvalue problem. The mass loading $k$ and the Alfv\'en radius
$R_{\mathrm{A}}$ are not free: $\omega$ and $E$ select them, and hence so do the
base conditions.

This is the property the rest of the paper relies on. Given a footpoint
$(R_0,z_0)$ and the base state $(\csw,\rho_0,B_{\mathrm{p}0},\theta)$, the solver
returns the lever arm $\lambda=(R_{\mathrm{A}}/R_0)^2$, and with it the ejection
index $\xi$ through equation~(\ref{eq:xilambda}), the wind mass flux, and the
surface torque.
Rather than solve the full eigenvalue problem at every footpoint, we read the
lever arm per cell off the \citet{bai2016} solution family (his Fig.~5). This
collapses onto a single monotone function $\lambda(\beta_{\mathrm{base}})$ of the
wind-base plasma-$\beta$,
$\beta_{\mathrm{base}}=\beta_{\mathrm{mid}}(\csw^2/\cs^2)\exp[-(z_0/H)^2/2]$
(the mid-plane $\beta$ scaled to the anchor height $z_0$). Because $\csw$ and
$z_0$ are set per cell by the irradiation and magnetic-coupling conditions at
the disc surface, $\lambda$ (and hence $\xi$) responds self-consistently to
the wind-base state as the disc evolves. A hotter or more depleted base raises
$\beta_{\mathrm{base}}$ and lowers $\lambda$; a cooler or less depleted base
raises it. For the fiducial disc at $t=0$ the derived $\lambda$ runs from
$6.8$ ($\xi=0.086$) across the inner launching region to $5.2$ ($\xi=0.12$) at
the magnetisation front, and the two branches then drive it apart.
Section~\ref{sec:leverarm} follows that evolution.

Because the map is a digitisation of a published figure, its calibration range
bounds where $\lambda$ is interpolated instead of extrapolated. Bai's Fig.~5 spans
$\beta_{\mathrm{base}}=2.5\times10^{-4}$ to $160$. The low-$\beta$ end is never
approached: wherever the base is set magnetically, $\beta_{\mathrm{base}}=1$ by the
definition of $z_{0,\rm mag}$, four decades above the tabulated minimum. At the
high-$\beta$ end, measured over the launching region where $\lambda$ has any effect,
every cell and all of the mass lie inside the tabulated range on the flux-loss
branch at every epoch: $\beta_{\mathrm{base}}$ there never exceeds $7.7$, five per
cent of the upper bound. On the flux-retained branch the same holds through
$t=5$~Myr, where the maximum reaches $91$. Only in that branch's final Myr does
it cross: at $5.5$~Myr, five of $192$ launching cells, carrying $8$ per cent of
the launching region's mass, reach $\beta_{\mathrm{base}}\approx230$. Beyond the
tabulated range the map continues smoothly onto the weak-field Blandford--Payne
limit $\lambda\to3/2$ instead of clamping, so those cells are extrapolated,
not pinned. The window in which they appear is the one
Appendix~\ref{app:trust-near-dispersal} already excludes on resolution grounds.

\subsection{Fixing the base from photoevaporation physics}
\label{sec:bridge}

The single-field-line solution of Section~\ref{sec:bai} takes the base sound speed
$\csw$ and the base height $z_0$ as free parameters. Fixing both per streamline, from
the local irradiation and magnetic-coupling conditions, is what makes the launch
magnetothermal, and is where photoevaporation physics enters the wind solution.
Because each field line is solved independently, no global temperature profile is
imposed. The strict self-similarity of the classical accretion--ejection structures
\citep{ferreira1997,casse2000b} is relaxed, and each footpoint carries its own base
temperature.

\subsubsection{The base temperature}

The wind-base sound speed follows from the irradiated base temperature,
\begin{equation}
\csw(R_0)=\sqrt{\frac{k_{\mathrm{B}}\,T_{\mathrm{b}}(R_0)}{\mu m_{\mathrm{H}}}},
\label{eq:csw}
\end{equation}
where $k_{\mathrm B}$ is the Boltzmann constant, $m_{\mathrm H}$ the hydrogen mass,
$T_{\mathrm b}$ the irradiated base temperature, and $\mu$ the mean molecular weight.
The base sits at the sharp vertical gradient where FUV/X-ray heating overcomes line cooling
\citep{wang2019,sarafidou2024}.

We take the molecular $\mu=2.34$ throughout the internally irradiated base. The
\citet{sellek2024} $\Xi$--$T$ relation adopted below returns
$T_{\mathrm{b}}=95$--$450$~K over $0.5$--$200$~AU at the penetration column
$N_{\mathrm b}=10^{21}\,\mathrm{cm^{-2}}$ that defines the base
(Appendix~\ref{app:windbase-robustness}, panel c). Over that range hydrogen is
neither dissociated nor ionised, and the same models report H$_2$ surviving
into the wind as its dominant coolant at tens of AU. The non-ideal MHD
wind-launching literature the Section~\ref{sec:bai} closure is drawn from
likewise takes the launching region to be molecular
\citep[e.g.][]{bai2016,gressel2020}. A fully
ionised $\mu=0.65$ would overstate $\csw^2$ by a factor $3.6$. The one exception is
the escape temperature of equation~(\ref{eq:tesc}), which retains the ionised value
where its $10^4$~K ceiling binds and the gas genuinely is ionised.

We set $T_{\mathrm b}$ from the local X-ray ionisation state instead of
prescribing it. Writing the ionisation parameter as
$\Xi\equiv L_X/(n r^2)$ \citep{tarter1969}, with $n$ the number density at the base and
$r$ the distance from the star, we adopt the empirical
$\Xi$--$T$ relation of \citet{sellek2024}. We take
$L_X=2\times10^{30}\,\mathrm{erg\,s^{-1}}$, consistent with the
$\log L_X$--$\log M_\ast$ relation of \citet{gudel2007} at $M_\ast=1\,\Msun$.
The same radiation field and the same local density then drive both the wind-base
thermodynamics and the photoevaporative mass loss, instead of entering as
independent inputs. Since $\Xi\propto1/\Sigma$, an
instantaneous $\Sigma\to T_{\mathrm b}$ closure would let the base temperature
respond to depletion without limit. We therefore relax $T_{\mathrm b}$ toward its
equilibrium value over the local recombination time
$\tau_{\mathrm{rec}}=1/(\alpha_{\mathrm r}(T)n)$, with $\alpha_{\mathrm r}$ the
hydrogenic radiative recombination coefficient \citep{vernerferland1996}, so the
gas cannot reconfigure faster than it can recombine. For the same reason we do
not extrapolate the $\Xi$--$T$ relation outside the density range it was fitted
over: below $\Sigma/\Sigma_0=10^{-2}$ the base is blended smoothly back to
$T_{\mathrm{mid}}$. That blend bites only late and only where the disc is nearly
empty. By $t\approx3.5$~Myr the flux-loss branch has $\Sigma/\Sigma_0\sim10^{-4}$
inside a few AU, so its base sits on the mid-plane temperature even though the
irradiation front has reached it. It errs in one direction only: $\Xi\propto1/\Sigma$
implies that gas should in reality be hotter than the mid-plane, so the late
inner disc is under-heated, not over-heated.
Appendix~\ref{app:windbase-robustness} gives the numerical treatment and the
sensitivity of the results to it.

Whether that base can drive a wind on its own is set by the escape condition
\begin{equation}
T_{\mathrm{esc}}(R)=\frac{\mu m_{\mathrm{H}}GM_\ast}{k_{\mathrm{B}}R}\propto R^{-1}.
\label{eq:tesc}
\end{equation}
Gas at $T_{\mathrm b}=T_{\mathrm{esc}}$ has $\csw=\vK$, i.e.\
$\varepsilon_{\mathrm b}\equiv\csw/\vK=1$: the condition under which a purely thermal
wind is self-launching. Thermal winds in practice launch from well inside the radius
at which it is met \citep{owen2012}, and we evaluate the base
temperature from the $\Xi$--$T$ relation above, at the launch column, so
\begin{equation}
\varepsilon_{\mathrm b}(R,t)=\left[\frac{T_{\mathrm b}(R,t)}{T_{\mathrm{esc}}(R)}\right]^{1/2}
\label{eq:epsb}
\end{equation}
is a local, evolving diagnostic rather than a fixed radius.

Photoionised gas equilibrates near $10^4$~K. The photoelectrons carry only the
excess energy above the $13.6$~eV hydrogen ionisation threshold, and balancing that
heating against recombination and collisionally excited line cooling fixes the
equilibrium temperature at $\sim 10^4$~K, largely independently of the ionising
spectrum and luminosity \citep{osterbrockferland2006}. The $\Xi$--$T$ relation
recovers this asymptote, so the fiducial base temperature is bounded without an
artificial cap.

\subsubsection{The base height}

The launch surface is not a static geometric boundary. In X-ray-heated non-ideal
MHD simulations the outflow originates from an extended vertical transition zone,
and two thresholds are crossed at different heights. Above $z_{\mathrm{0,th}}$ the
gas is hot enough to be irradiation-heated, and above $z_{\mathrm{0,mag}}$ it is
well enough coupled to the field for the ideal-MHD invariants of
Section~\ref{sec:bai} to apply. Neither is individually necessary for an outflow.
Material launched from a coupled but as-yet-unheated layer leaves as a cold
magnetocentrifugal wind and is heated as it rises. The converse state (heated
but as-yet-uncoupled) is harder to sustain, because the photons that heat the
surface layer also ionise it. FUV photoionisation of trace species maintains
$x_{\mathrm e}\sim10^{-5}$ in the heated layer \citep{perezbecker2011}.
Non-ideal MHD calculations that solve the thermochemistry alongside the diffusivities
find the irradiated surface to be well enough coupled to launch, instead of a
decoupled layer that leaves thermally \citep{gressel2015,gressel2020}. We
therefore do not read the two thresholds as selecting between magnetic and
thermal driving. The base is simply the lower of the two,
\begin{equation}
z_0(R_0) = \min \left( z_{\mathrm{0, th}}, \, z_{\mathrm{0, mag}} \right),
\label{eq:z0}
\end{equation}
and which of the two is binding fixes the thermodynamic character of the
base, not the identity of the driving mechanism. Where
$z_{\mathrm{0,mag}}<z_{\mathrm{0,th}}$ the base lies below the irradiation front.
The gas is launched cold and magnetocentrifugally and is heated only once it has
risen past $z_{\mathrm{0,th}}$, so the irradiation adds neither mass loading nor
angular-momentum extraction at the base.
Where $z_{\mathrm{0,th}}<z_{\mathrm{0,mag}}$ the base sits at the front itself.
The launch is genuinely magnetothermal, and the elevated base temperature raises
$\csw$ and with it the mass loading, lowering the lever arm $\lambda$.

On the flux-retained branch $z_{\mathrm{0,mag}}<z_{\mathrm{0,th}}$ over
$72$~per cent of the launching region at $t=0$, falling to $64$~per cent by $5$~Myr.
Taking the minimum therefore places the base below the irradiation front over most
of the disc, whereas \citet{baistone2013} and \citet{bai2016} identify the base with
the FUV front itself.
Recent global simulations support the former.
Carrying all three non-ideal terms with radiation transport, \citet{mori2025} find
that ``wind launching can take place in regions where gas is not fully coupled to
magnetic field with $\mathrm{Am}\gtrsim1$'', below an FUV front that in their
calculation ``is located high in the atmosphere''.
Launching from marginally coupled gas beneath the front is what equation~(\ref{eq:z0})
does by construction.
The two prescriptions coincide precisely where the remaining third of the disc sits,
in the magnetothermal annulus.

\begin{figure}
\centering
\includegraphics[width=\columnwidth]{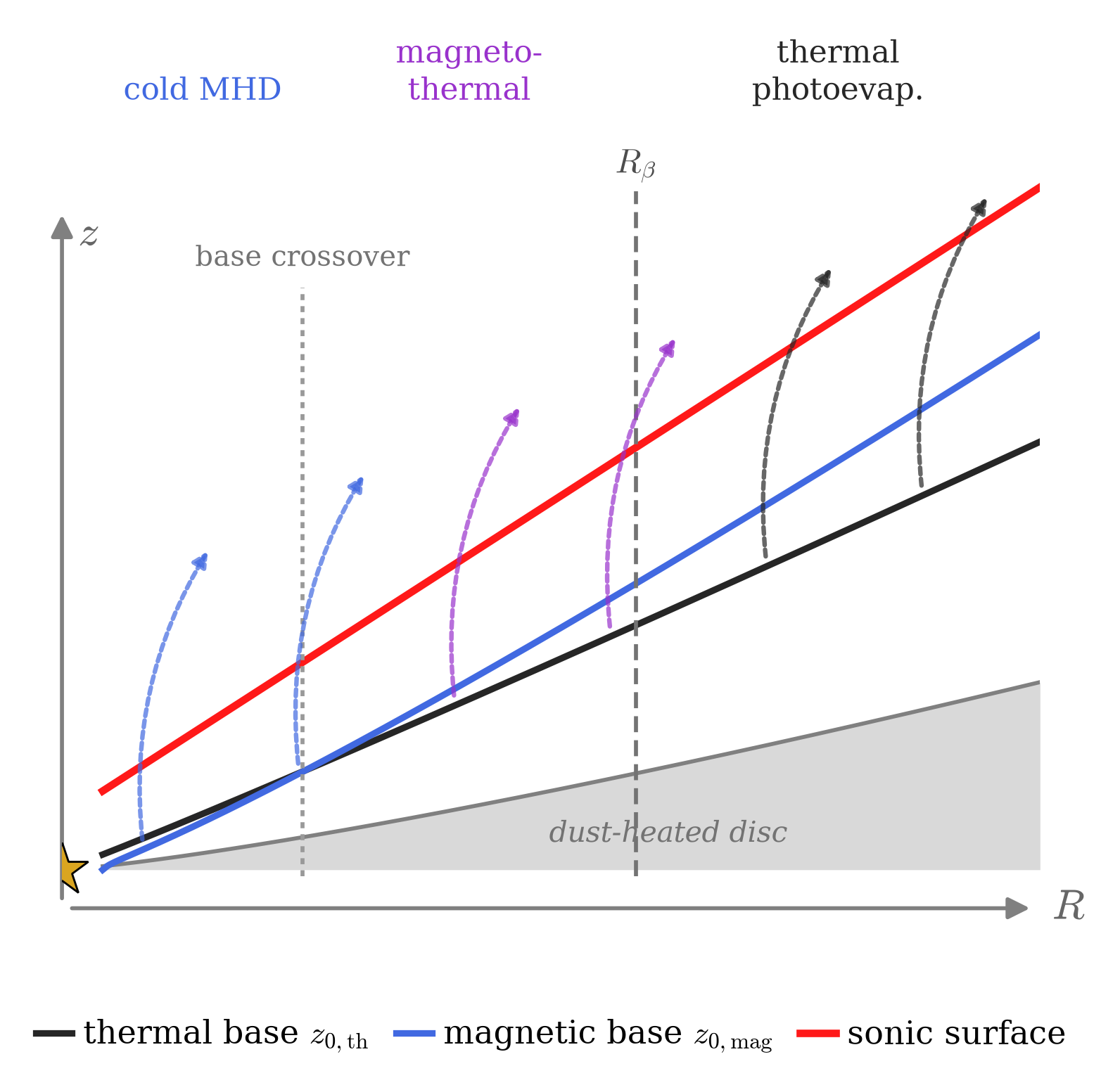}
\caption{Schematic vertical structure of the magnetothermal wind base, divided by
radius into three regimes: an inner cold magnetocentrifugal launch, an
intermediate magnetothermal band bounded by the base crossover
$z_{0,\rm th}=z_{0,\rm mag}$ and the magnetisation front $R_\beta$, and thermal
photoevaporation beyond $R_\beta$ (Section~\ref{sec:bridge}).}
\label{fig:windbase}
\end{figure}

This is not the magnetothermal-to-thermal boundary.
That boundary is set by the field, as opposed to the base.
The outflow is magnetically driven only while $\beta<\beta_{\mathrm{crit}}$, that is
inside the magnetisation front $R_\beta$; beyond it the
wind is purely thermal. Because the radial column falls and $\beta$ rises
outward, $z_{\mathrm{0,th}}$ decreases and $z_{\mathrm{0,mag}}$ increases with
$R$. The disc therefore divides into three regions: an inner cold magnetocentrifugal
launch, an intermediate magnetothermal band between the base crossover
$z_{\mathrm{0,th}}=z_{\mathrm{0,mag}}$ and $R_\beta$, and thermal
photoevaporation beyond $R_\beta$ (Figure~\ref{fig:windbase}). The two radii are
related but distinct, and their ordering (and hence the width of the
magnetothermal band) is a result of the evolution, not an input.

The \emph{thermal base} $z_{\mathrm{0,th}}$ sits where stellar X-ray/EUV irradiation
penetrates to the heating column
$N_{\mathrm{X}}\approx10^{21}$--$10^{22}~\mathrm{cm^{-2}}$ \citep{ercolano2009}.
Because this radiation grazes the disc from the central star, the attenuating
column is the radial, not the vertical, one. For a hydrostatic Gaussian interior of
scale height $H=\cs/\Omega$ and nominal mid-plane H-nucleus density
$n_{\mathrm{mid}}=\Sigma/(\sqrt{2\pi}Hm_{\mathrm{H}})$, the column to the point
$(R,z)$ along a ray of constant opening $u\equiv z/R$ is
\begin{equation}
N_{\mathrm{rad}}(R,u)=\int_0^{R} n_{\mathrm{mid}}(r)\,
\exp\!\left[-\frac{(u\,r)^2}{2H(r)^2}\right]\sqrt{1+u^2}\;\mathrm{d}r ,
\label{eq:Nrad}
\end{equation}
which decreases monotonically with $u$; $z_{\mathrm{0,th}}(R)=u_\star R$ where
$N_{\mathrm{rad}}(R,u_\star)=N_{\mathrm{X}}$.
We adopt the lower bound of that range, $N_{\mathrm X}=10^{21}\,\mathrm{cm^{-2}}$,
and evaluate equation~(\ref{eq:Nrad}) with the column scaled by the proton mass,
$\Sigma/m_{\mathrm H}$.\footnote{The mean mass per hydrogen nucleus is
$\mu_{\mathrm H} m_{\mathrm H}$ with $\mu_{\mathrm H} \simeq 1.4$, distinct from the
mean mass per free particle $\mu m_{\mathrm H}$ ($\mu = 2.34$) that sets the sound
speed.}
Since $N_{\mathrm rad}$ is linear in $n_{\mathrm mid}$, this is exactly equivalent
to using the true H-nucleus density with a threshold
$N_X = 10^{21}/\mu_{\mathrm H}\simeq7\times10^{20}\,\mathrm{{cm}^{-2}}$. Because
the uncertainty on $N_{\mathrm X}$ itself spans an order of magnitude, the factor
$1.4$ is absorbed into that choice rather than left as an uncorrected offset.
The lower integration limit in equation~(\ref{eq:Nrad}) accounts for attenuation by
disc material interior to the computational domain. We extrapolate the initial
surface density profile $\Sigma_0$ inwards to a magnetospheric truncation radius of
$\sim 0.05$~AU, scaled to the time-dependent surface density at the inner boundary
so that the inner shadow recedes as the disc drains. Neglecting it would expose the
innermost grid cells to midplane ionising radiation and collapse
$z_{\mathrm{0,th}}$ unphysically. Beyond $R_{\mathrm c}(t)$ the unshadowed line of
sight assumed by $\Xi$ breaks down, and there $T_{\mathrm b}$ reverts to the midplane
temperature, a strict lower bound on the wind-base kinematics, justified in
Appendix~\ref{app:windbase-robustness}.

The \emph{magnetic base} $z_{\mathrm{0,mag}}$ marks the base of the magnetically
dominated region. Above it the magnetic energy density exceeds the thermal,
gravitational and kinetic energy densities of the gas, so the poloidal velocity
aligns with the poloidal field. This is the approximation on which the
single-field-line solution of Section~\ref{sec:bai} rests.
Assuming a vertically isothermal Gaussian density profile and a constant
$B_z$, the plasma-$\beta$ scales as $\beta(z)=\beta(R,t)\,e^{-z^2/2H^2}$.
Imposing equipartition, $\beta(z_{\mathrm{0,mag}})=\beta_{\mathrm{couple}}\sim 1$, yields
\begin{equation}
z_{\mathrm{0,mag}}=H\sqrt{2\ln\!\left[\beta(R,t)/\beta_{\mathrm{couple}}\right]},
\label{eq:z0mag}
\end{equation}
which is set to zero if the midplane is already magnetically dominated
($\beta(R,t)\le\beta_{\mathrm{couple}}$).
Equation~(\ref{eq:z0mag}) is our own construction, and is not the definition of
the wind base adopted by the works whose launch physics we use. Both
\citet{baistone2013} and \citet{bai2016} place the base at the height where the
azimuthal velocity turns from sub- to super-Keplerian, marking the onset of
centrifugal acceleration. \citet{bai2016} takes that height as an input
boundary condition instead of deriving it. The two criteria are nevertheless
the same transition seen from opposite sides. Gas below the base is
sub-Keplerian because it is pressure supported, and turns super-Keplerian when
the magnetic torque overcomes that support, which occurs where the magnetic
and thermal energy densities cross, $\beta\sim1$. Equation~(\ref{eq:z0mag}) is
thus an analytic proxy for a height we do not otherwise compute, since this
vertically integrated treatment carries no $v_\phi(z)$.
The proxy can be tested two ways. Quantitatively, the fiducial run of
\citet{baistone2013} has a midplane $\beta_0=10^5$ and measures its wind base at
$z_{\rm b}\simeq4.69\,H$. Equation~(\ref{eq:z0mag}) with
$\beta_{\mathrm{couple}}=1$ gives $H\sqrt{2\ln10^5}=4.80\,H$, agreeing to
$2.3$~per cent with no free parameter. Independently, the two-dimensional global
simulations of \citet{mori2025}, which carry all three non-ideal terms (Ohmic,
Hall and ambipolar) together with radiation transport, report that in their
Hall-free run ``wind launching empirically occurs around the region where the
magnetic pressure dominates over the thermal pressure (i.e., plasma
$\beta\sim1$)''. Our $\beta_{\mathrm{couple}}=1$ is therefore not a convenient
choice but the launching surface those calculations find, and we adopt it on that
basis.

Whether the gas is coupled to the field well enough to drive a wind at all
is a separate requirement.
This model does not evaluate it. Semi-analytic wind-driving disc solutions
exist only above an ambipolar Elsasser number $\Lambda_{\mathrm A}\gtrsim1$
\citep{wardlekoenigl1993,koenigl2011}. Published simulations of discs of this type
satisfy that bound over the launching region, and Section~\ref{sec:caveats} gives
the evidence together with what it does not cover. A consistent anchor would be set
by $\Lambda_{\mathrm A}$ in place of $\beta$, which a vertically integrated
model carrying no ionisation network cannot compute. Varying how the two thresholds
of equation~(\ref{eq:z0}) are combined moves the integrated results by only a few
per cent. That bounds the sensitivity to where the anchor sits, but not to the
criterion that defines it.

The isothermal-Gaussian assumption sets a strict lower limit on the base height.
In irradiated discs the heated surface layers decay far more gradually than a
Gaussian, so the true $\beta(z)$ falls more slowly and the physical coupling
transition lies higher in the atmosphere.
The model's topology limits how far that error propagates. The Gaussian fails in the
heated layer above the thermal base, $z > z_{\mathrm{0,th}}$. Where the base is
magnetically anchored ($z_{\mathrm{0,mag}} < z_{\mathrm{0,th}}$) the wind anchors in
the shielded interior, where the isothermal assumption holds. Because
$\beta_{\mathrm{base}}=\beta_{\mathrm{couple}}$ by definition in those cells, the
lever arm $\lambda$ and the mass loading $\xi$ follow from the coupling criterion
alone, independently of the assumed vertical density profile. Where the isothermal
estimate instead places the magnetic base inside the heated layer
($z_{\mathrm{0,mag}} > z_{\mathrm{0,th}}$), the minimum in
equation~(\ref{eq:z0}) defaults to the thermal threshold and bounds the launch
height.
The validity of this assumption breaks down within the thermally anchored cells.
There $\beta_{\mathrm{base}}$ is evaluated on the Gaussian at
$z_{\mathrm{0,th}}/H \simeq 3.5$--$5.5$, and the density underestimate propagates
directly into $\lambda$ and $\xi$. A flared atmosphere would raise
$\beta_{\mathrm{base}}$, lower $\lambda$, and lift $z_{\mathrm{0,mag}}$, moving
further cells into the thermally anchored regime and compounding the error.
Correcting it requires the vertical temperature structure, which a vertically
integrated model does not carry.
This is the principal structural limitation of the base prescription.

Equations~(\ref{eq:z0})--(\ref{eq:z0mag}) make the base height a function of the
local surface density and field strength, and so couple the launch directly to the
disc secular evolution.
Figure~\ref{fig:windbase} shows schematically the resulting geometry.

\subsubsection{Secular behaviour and scope}

Because the base thresholds depend on local surface density and field strength, the launch
kinematics couple directly to the disc's secular evolution.
As the disc drains, the radial column drops below the penetration depth $N_{\mathrm{X}}$
and drives $z_{\mathrm{0,th}}$ toward the mid-plane. However, irradiation alone only moves
the launch height. Changing the launch mechanism requires magnetic flux loss.
If flux dissipates (finite $\tau_B$), the mid-plane $\beta$ rises, driving $z_{\mathrm{0,mag}}$
higher and forcing the magnetisation front $R_\beta$ (where $\beta=\beta_{\mathrm{crit}}$) to
recede. As Figure~\ref{fig:windbase} shows, this sweeps the disc from a magnetically driven
outflow, through a mixed state, to a purely thermal wind. If instead flux is retained
($\tau_B\to\infty$), $R_\beta$ expands and the outflow stays magnetically driven for the
disc's entire lifetime.

\subsection{Wind properties and the cold (Blandford--Payne) limit}
\label{sec:launchout}

For each footpoint the converged solution returns the dimensionless mass loading
$\mu_{\mathrm w}=\omega R_0k/B_{\mathrm{p}0}$, the lever arm $\lambda$,
and the surface-torque accretion and wind mass-loss rates formulated by \citet{bai2016}:
\begin{equation}
\Mdot_{\mathrm{acc}}=\frac{2R_0}{\Omega_{\mathrm K}(R_0)}\,
B_{\mathrm{p}0}^2\left|\frac{B_\phi}{B_{\mathrm{p}}}\right|_{z_0}\!\sin\theta ,
\qquad
\frac{\mathrm{d}\Mdot_{\mathrm{w}}}{\mathrm{d}\ln R}=2\pi R_0^2\rho_0 v_{\mathrm{p}0},
\label{eq:launchoutputs}
\end{equation}
with $B_z=B_{\mathrm p}\sin\theta$ for the equatorial
convention on $\theta$.
The ratio of these rates reproduces the ejection index $\xi$ only to within $\sim 10\%$,
because the torque formulation assumes a strictly co-rotating base \citep{wardlekoenigl1993}
that the solved $v_\phi$ violates slightly \citep[][footnote 9]{bai2016}.
We therefore extract $\xi$ directly from the exact angular-momentum balance in
equation~(\ref{eq:xilambda}), which needs neither the launch-surface area element nor the
field inclination.

The mass loading $\mu_{\mathrm w}$ is dimensionless under a unit base density normalisation,
and single-field-line models fix kinematic ratios ($\lambda$, $\xi$, and velocity profiles)
but not the absolute mass flux. $\Sigdot=2\rho_0v_{\mathrm{p}0}$ scales with the base
density, which here is set to enforce the $\beta_{\mathrm{couple}}$ height condition
(equation~\ref{eq:z0mag}), not to calibrate a physical mass-loss rate.
We therefore take only the robustly determined $\lambda$ from the local solution, prescribe
$\aDW$, and compute the wind mass loss through the transport sink
(equation~\ref{eq:tabone}).

In the cold limit ($c_{\mathrm s,\mathrm{w}}\to0$, $q_{\mathrm{div}}\to\infty$, $\theta=0$,
$z_0\to0$), the solution collapses to the \citet{spruit1996} relations:
\begin{equation}
\lambda=\tfrac32\left(1+\mu_{\mathrm w}^{-2/3}\right),
\qquad
\frac{v_{\mathrm{p}}^\infty}{\omega R_0}=\mu_{\mathrm w}^{-1/3}.
\label{eq:spruit}
\end{equation}
This limit is algebraically identical to the self-similar cold family of \citet{blandford1982},
yielding their invariant
$\beta_{\mathrm{BP}} = \mu_{\mathrm w}(2\lambda-3)^{3/2} = 3^{3/2} \simeq 5.20$.
Our $\mu_{\mathrm w}$ matches \citeauthor{spruit1996}'s dimensionless mass flux exactly, and
aligns with the $\kappa$ of \citet{martel2022} by excluding the base-inclination
factor $1/\sin\theta_0$ present in \citeauthor{blandford1982}'s original definition.

This cold relation is also what carries the lever arm outside the magnetised
disc. Beyond the magnetisation front the launch switch turns off, and $\lambda$
follows the $\lambda(\mu_{\mathrm w})$ of equation~(\ref{eq:spruit}) continuously,
as opposed to the $\beta_{\mathrm{base}}$ map of Section~\ref{sec:bai}. We apply it there
because it is continuous and has the right weak-field limit, not because it is
calibrated there: outside the magnetically dominated regime it is a functional form,
not a calibrated relation.

Two constraints bound its use.
First, it assumes $q_{\mathrm{div}}\to\infty$; finite values (e.g.\ $q_{\mathrm{div}}=0.25$)
preserve the functional form but introduce a zero-point offset \citep{bai2016}.
Second, it enforces $\lambda>3/2$, so $\xi<1$ always.
A weakening magnetic field therefore drives the system toward higher mass loss per unit
accretion, the inverse of what an outer-disc conveyor feeding an inner cavity requires.
It remains a purely cold relation, omitting the thermal driving that raises
$c_{\mathrm s,\mathrm{w}}$ at the base (Discussion Caveat~1).

\section{The primordial phase: secular evolution and the photoevaporation sink}
\label{sec:secular}

\subsection{The wind-driven operator}

The secular surface density evolves according to the \citet{tabone2022} master equation:
\begin{align}
\frac{\partial\Sigma}{\partial t}
=\;& \frac{3}{R}\frac{\partial}{\partial R}\!\left[\frac{1}{R\Omega}
\frac{\partial}{\partial R}\!\left(R^2\aSS\Sigma\cs^2\right)\right]
+ \frac{3}{2R}\frac{\partial}{\partial R}\!\left[\frac{\aDW\Sigma\cs^2}{\Omega}\right]
\nonumber\\
&-\;\frac{3\aDW\Sigma\cs^2}{4(\lambda-1)R^2\Omega}
\;-\;\left[1-S(\beta)\right]\Sigdot_{\mathrm{PE}}(R,t).
\label{eq:tabone}
\end{align}
The first term is the \citet{lyndenbell1974} viscous (diffusive) operator with
$\nu=\aSS\cs^2/\Omega$. The second is the wind torque, a first-order
(advective) term equivalent to an inward velocity
\begin{equation}
v_{\mathrm{DW}}=\frac{3}{2}\frac{\aDW\cs^2}{R\Omega}
=\frac{3}{2}\aDW\frac{\cs^2}{\vK}.
\label{eq:vdw}
\end{equation}
The third is the wind mass loss, which is linear in $\Sigma$ and equal to
$\Sigdot_{\mathrm{DW}}=\xi\,v_{\mathrm{DW}}\Sigma/R$ with the same $\xi$ as
equation~(\ref{eq:xilambda}). The fourth is the internal photoevaporation
sink.

The transport coefficient $\aDW$ is not a static normalisation but evolves with the disc:
\begin{equation}
\aDW(R,t)=\min\!\left[\aDW^{0}
\left(\frac{\Sigma}{\Sigma_0}\right)^{-(1-2q)},\,\aDW^{\max}\right]
S(\beta)\,S_{\rm th},
\label{eq:adwfull}
\end{equation}
in which the depletion factor $(\Sigma/\Sigma_0)^{-(1-2q)}$ follows from the
field--surface-density closure of eq.~\ref{eq:betaclosure_intro} and strengthens the
torque as the disc drains. $\aDW^{\max}=1$ imposes a physical Maxwell-stress ceiling.
$S_{\mathrm{th}}$ cuts off the wind if the base collapses to the mid-plane,
\textbf{
\begin{equation}
S_{\rm th}(R,t)=1-\exp\!\left[-\left(\frac{z_0}{0.01\,H}\right)^{2}\right],
\label{eq:sth}
\end{equation}
a smooth step that is unity once the base has risen even slightly above the
mid-plane and falls to zero only as $z_0\to0$.}
The magnetic switch,
\begin{equation}
S(\beta)=\left[1+\beta/\beta_{\mathrm{crit}}\right]^{-1},
\label{eq:launchswitch}
\end{equation}
terminates the outflow for weak fields ($\beta\gg\beta_{\mathrm{crit}}$).

Because $\beta\propto\Sigma^{1-2q}$ (equation~\ref{eq:betaclosure_intro}), the depletion
factor scales exactly as $\beta^{-1}$.
In the active launching region ($\beta<\beta_{\mathrm{crit}}$ where $S\to1$), this recovers
the $\alpha_{\mathrm{DW}}\propto\beta^{-1}$ wind-stress scaling measured in non-ideal MHD
simulations \citep{baistone2013,bai2016,lesur2021}.
In the weak-field limit, $S(\beta)$ imposes a second power, steepening the fall-off to
$\alpha_{\mathrm{DW}}\propto\beta^{-2}$.
We choose the linear switch because it gives the shallowest available rolloff. A steeper
cutoff would artificially suppress the residual outer-disc torque.

Weighting the photoevaporative sink by $1-S(\beta)$ enforces a unified outflow.
Above the physical base $z_0$, mass loading is conserved along the field line.
Irradiation heats gas the magnetic base has already lofted, setting $\beta_{\mathrm{base}}$
and with it the kinematic parameters ($\lambda$, $\xi$).
Adding $\dot{\Sigma}_{\mathrm{PE}}$ directly to $\dot{\Sigma}_{\mathrm{DW}}$ would count
that same gas twice.
The magnetisation therefore paces the total mass loss: the outflow acts as a
magnetothermal wind ($S\to1$) at low $\beta$, and reverts smoothly to pure photoevaporation
($S\to0$) at high $\beta$.
This matches the radiation-magnetohydrodynamic results of \citet{sarafidou2024}, who
find the total mass-loss rate transitions between the two regimes instead of summing them.

\subsection{Closing the launch--evolution coupling}
\label{sec:match}

Equating the \citet{tabone2022} advective flux,
$\Mdot_{\mathrm{acc}}=2\pi R\Sigma v_{\mathrm{DW}}=3\pi\aDW\Sigma\cs^2 R/\vK$, with the
\citet{bai2016} surface Maxwell torque,
$\Mdot_{\mathrm{acc}}=(2R/\Omega)\vert{}B_zB_\phi\vert{}_{z_0}$, yields the transport coefficient:
\begin{equation}
\aDW=\frac{2}{3\pi}\,\frac{R\,B_{\mathrm{p}0}^2|B_\phi/B_{\mathrm{p}}|_{z_0}\sin\theta}{\Sigma\cs^2}.
\label{eq:adwmatch}
\end{equation}
This identifies $\aDW$ as the surface Maxwell stress normalised by the column thermal
energy density.
Instead of recomputing $\aDW$ from the local stress at every radius and timestep, we
anchor it to a global initial normalisation $\aDW^0=10^{-3}$ (Table~\ref{tab:params})
and evolve it through equation~(\ref{eq:adwfull}).
Because the closure forces $\aDW \propto \Sigma^{-(1-2q)}$, the torque strengthens as the disc
drains. This is the explicit mechanism driving rapid terminal collapse on the
flux-retained branch.
The ejection index $\xi$, by contrast, is derived continuously from the local
angular-momentum balance (equation~\ref{eq:xilambda}) and fed straight into the secular
mass-loss sink.

Parameterising one quantity globally while deriving the other locally is necessary because
the two models account for mass inconsistently. The local launch solution's absolute mass
flux, $\Sigdot=2\rho_0v_{\mathrm{p}0}$, differs from the secular transport sink,
$\xi v_{\mathrm{DW}}\Sigma/R$, by over three orders of magnitude across the disc. This is
because the 1D vertical model's base density is a construct adopted to enforce the
$\beta_{\mathrm{couple}}$ height condition (equation~\ref{eq:z0mag}) and carries no
connection to the physical $\Sigma$ the secular model tracks. We therefore keep only the
kinematic ratios the launch solution determines reliably ($\lambda$, $\xi$), and discard its
unnormalised absolute mass flux.

\subsection{Characteristics and the Green's function}
\label{sec:char}

In the wind-dominated regime ($\psi\equiv\alpha_{\mathrm{DW}}/\alpha_{\mathrm{SS}}\gg1$), the
diffusive term becomes negligible and equation~(\ref{eq:tabone}) reduces to a first-order
hyperbolic equation with linear sinks.
We adopt the irradiated passive-disc profile
$c_{\mathrm{s}}^2=c_{\mathrm{s}}^2(R_{\mathrm{n}})(R/R_{\mathrm{n}})^{-1/2}$, departing from the
classical self-similar $c_{\mathrm{s}}^2\propto R^{-1}$ of magnetised accretion-ejection structures.
Under this profile the ratio $c_{\mathrm{s}}^2/v_{\mathrm{K}}$ is constant, so
$v_{\mathrm{DW}}=\tfrac{3}{2}\alpha_{\mathrm{DW}}c_{\mathrm{s}}^2/v_{\mathrm{K}}$ becomes
radius-independent if $\alpha_{\mathrm{DW}}$ is constant.
(Because equation~\ref{eq:adwmatch} gives
$\alpha_{\mathrm{DW}}\propto R B_z^2/(\Sigma c_{\mathrm{s}}^2)$, a strictly uniform
$\alpha_{\mathrm{DW}}$ with $\Sigma\propto R^{\xi-1}$ demands $B_z \propto R^{(2\xi-5)/4}$.
We do not enforce this $B_z$ dynamically, but treat $\alpha_{\mathrm{DW}}$ as a uniform parameter.
The derivations below hold locally along characteristics even where $B_z(R)$ departs from this
profile.)

The accretion characteristics are then straight lines:
\begin{equation}
\frac{\mathrm{d}R}{\mathrm{d}t}=-A,\quad R(t)=R_i-A\,t,
\qquad A\equiv\frac32\aDW\frac{\cs^2(R_{\mathrm n})\sqrt{R_{\mathrm n}}}{\sqrt{GM_\ast}}.
\label{eq:char}
\end{equation}
This describes finite-time, non-spreading accretion: an annulus reaches the inner boundary in
$t \sim R_i/A$.
Along a characteristic, $\mathrm{d}\Sigma/\mathrm{d}R=(\xi-1)\Sigma/R$, yielding the homogeneous
solution:
\begin{equation}
\Sigma(R)\propto R^{\,\xi-1}.
\label{eq:sigprofile}
\end{equation}
For a general base scaling $c_{\mathrm{s}}^2\propto R^{-p}$, the slope is $\xi+p-\frac{3}{2}$,
recovering the classical self-similar $\xi-\frac{1}{2}$ at $p=1$, and $\xi-1$ for our irradiated
$p=\frac{1}{2}$.
The secular transport therefore recovers the steady-wind surface-density scaling implied by the
local launch solution, which is an internal consistency check on the coupling.

Including the photoevaporation sink, the full transport along a characteristic is
$\mathrm{d}\Sigma/\mathrm{d}t = \mathcal{D}(R)\Sigma - [1-S(\beta)]\dot{\Sigma}_{\mathrm{PE}}$,
with $\mathcal{D}(R) = A(1-\xi)R^{-1}$.
Its integrating-factor solution provides the Green's function:
\begin{equation}
\Sigma\big(R(t)\big)=\frac{1}{\mathcal{I}(t)}
\left[\Sigma_i-\int_0^t \mathcal{I}(t')\,\left[1-S\right]\Sigdot_{\mathrm{PE}}\big(R(t'),t'\big)\,\mathrm{d}t'\right],
\label{eq:greens}
\end{equation}
where $\mathcal{I}(t)=\exp[-\int_0^t\mathcal{D}\,\mathrm{d}t']$.
The first term governs the wind-driven evolution. The second is the photoevaporation correction
convolved along the characteristic.
Because the launch switch $S$ modifies only the source term, it leaves the advective kernel $\mathcal{D}$
unbroken, allowing continuous superposition across the entire disc.

Equation~(\ref{eq:greens}) assumes the strict $\psi\gg1$ limit.
For general $\psi$, writing $\nu \propto R^\gamma$ with $\gamma=\frac{3}{2}-p$,
self-similar mass balance demands an effective inner-disc slope
$\mathrm{d}\ln\Sigma/\mathrm{d}\ln R = \xi(\psi,\lambda)-\gamma$.
Here $\xi(\psi,\lambda)$ is the accreting root of:
\begin{equation}
2b^2+(1+\psi)\,b-\xi_\infty\,\psi=0 ,
\qquad
\xi_\infty\equiv\frac{1}{2(\lambda-1)} ,
\label{eq:psiquad}
\end{equation}
where $b \equiv a+\gamma$ \citep{tabone2022}.
This slope interpolates smoothly between the wind-dominated ($\xi \to \xi_\infty$)
and purely viscous ($\xi \to 0$, giving $\Sigma \propto R^{-\gamma}$) limits.

The error in the closed-form exponent (equation~\ref{eq:sigprofile}) is exactly
$\xi_\infty-\xi(\psi,\lambda)$: negligible at our fiducial $\psi\simeq100$ ($0.004$), but
$0.14$ at $\psi=1$.
Sweeping our numerical integrator across $\psi\in[0.1,10^4]$ at fixed $\alpha_{\mathrm{DW}}^0$
recovers this analytic $\xi(\psi,\lambda)-\gamma$ slope to within $0.09$~dex.

One systematic bias persists even at $\psi\gg1$. Dropping the viscous term turns the PDE from
parabolic to hyperbolic. The full solver imposes boundary conditions at both
$R_{\mathrm{in}}$ and $R_{\mathrm{out}}$, whereas the advective solution transports only from
the outer boundary and leaves no freedom at the inner edge. Against the full two-term
numerical integration, it therefore overestimates the surface density, and so
$\dot{M}_{\mathrm{acc}}$, near the inner boundary by $\sim10\%$.

\subsection{Validity window, gap opening, and the dispersal cut-off}
\label{sec:validity}

The analytic superposition (equation~\ref{eq:greens}) is strictly valid only while
photoevaporation does not reshape the advective characteristics. This requires the
photoevaporative sink to stay negligible against net wind-driven advection,
$[1-S(\beta)]\dot{\Sigma}_{\mathrm{PE}} \ll (1-\xi)\Sigma v_{\mathrm{DW}}/R$.
This bounds the closed form, not the model. The numerical integration carries the same
equation through the reshaping and beyond, and the two coincide inside the bound. The window
below is therefore where the analytic scalings may be read directly off the evolution.

This phase ends when a gap opens at $R_{\mathrm{gap}}$, defined by the balance:
\begin{equation}
\left[1-S(\beta)\right]\Sigdot_{\mathrm{PE}}(R_{\mathrm{gap}})
=(1-\xi)\,\frac{\Sigma(R_{\mathrm{gap}})\,v_{\mathrm{DW}}(R_{\mathrm{gap}})}{R_{\mathrm{gap}}}.
\label{eq:gap}
\end{equation}
The $(1-\xi)$ factor limits the resupply to the net advection remaining after local wind mass
loss. Omitting it overstates the resupply and biases $R_{\mathrm{gap}}$ outward.
The suppression factor $[1-S(\beta)]$ further ensures that photoevaporation can carve a gap
only where the magnetic wind has already shut down, so $R_{\mathrm{gap}}$ appears later than an
unsuppressed comparison would predict. Its location when it does open, however, is largely
unchanged.
An analogous radius-dependent balance (local photoevaporative removal against local
accretion-driven resupply) was derived independently for a purely viscous disc by
\citet{malanga2025}, who minimise the ratio of the viscously supplied surface density to the
photoevaporation rate to obtain both the gap radius and a criterion for inside-out versus
outside-in dispersal. The wind-driven resupply term here plays the role their viscous solution
does.
Globally, the primordial phase follows the closed-form mass history of \citet{tabone2022}.
For a magnetic field decaying with index $\omega_B$, the disc mass and wind-ejection fraction are:
\begin{equation}
M_{\mathrm D}(t)=M_0\Big(1-\frac{\omega_B t}{2 t_{\mathrm{acc},0}}\Big)^{1/\omega_B},
\quad
f_M\equiv\frac{\Mdot_{\mathrm w}}{\Mdot_\ast}=\Big(\frac{R_{0,\mathrm T}}{\Rin}\Big)^{\xi}-1.
\label{eq:tabhist}
\end{equation}
The transition to the dispersal phase occurs at $t_{\mathrm{max}}$, defined when the wind-driven
stellar accretion rate drops to match the integrated photoevaporation rate
($\dot{M}_\ast(t_{\mathrm{max}})=\dot{M}_{\mathrm{PE,int}}$). This is the wind-driven
analogue of the polynomial accretion cut-off of \citet{ruden2004} and
\citet{alexander2007}.

\begin{figure}
\centering
\includegraphics[width=\columnwidth]{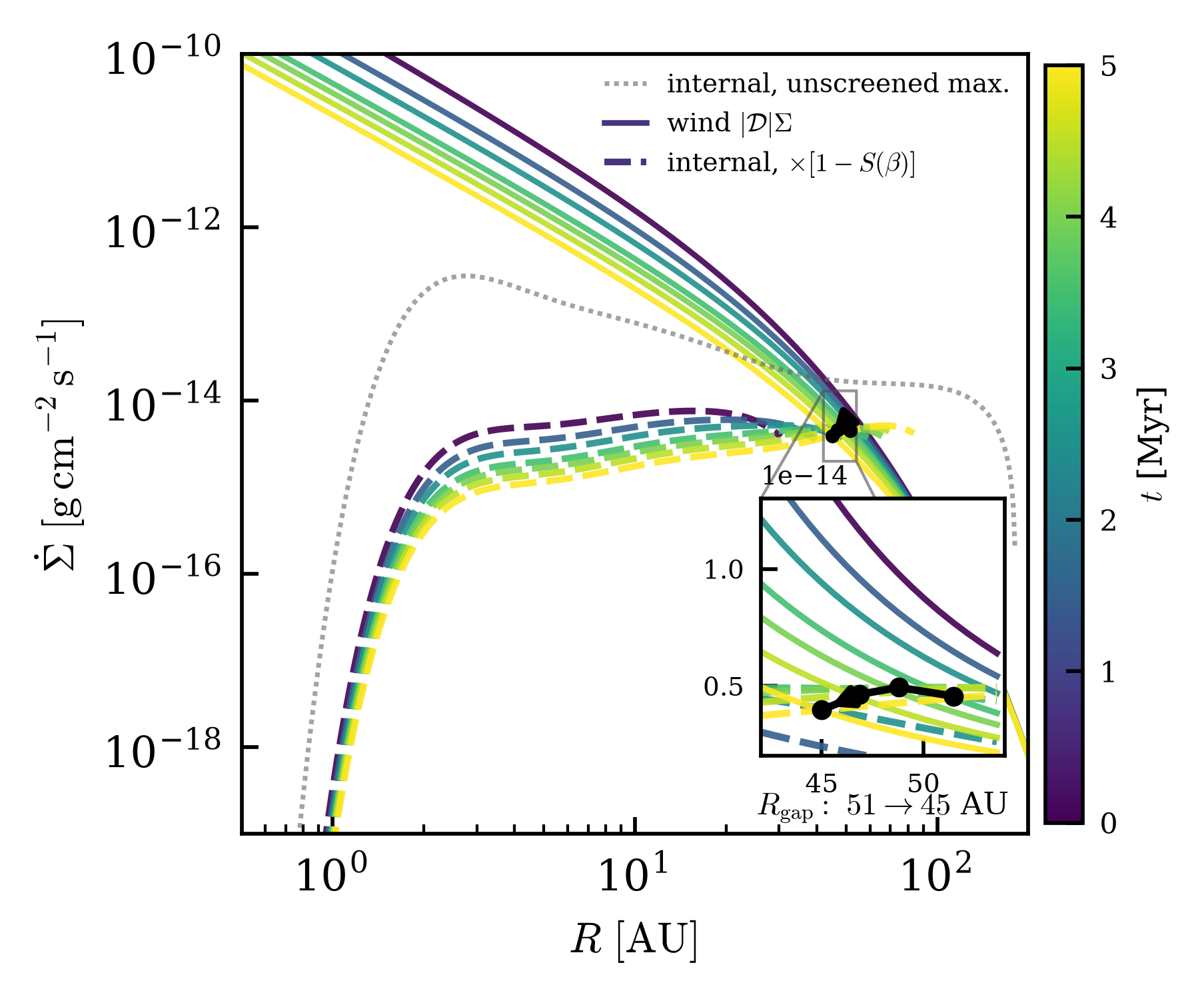}
\caption{Mass-loss terms of equation~(\ref{eq:gap}), evaluated on real
snapshots of the flux-retained run of Figure~\ref{fig:integrator}
(colour, $t=0$--$5$~Myr). Solid: the wind-driven advective resupply. Dashed:
the internal photoevaporation rate after the launch switch $[1-S(\beta)]$,
each curve cut off at that epoch's own screening radius $R_{\rm c}(t)$,
beyond which the profile is already suppressed. The fully unscreened
profile is shown once (grey dotted) as the ceiling every other
curve, at every epoch, lies below. Each intersection (dot) gives the
predicted gap-opening radius $R_{\rm gap}$; the dots, joined in time order,
trace how $R_{\rm gap}$ migrates inward, from $51$ to $45$~AU over
$3.5$--$5.0$~Myr, as $\dot M_{\rm acc}(t)$ declines.}
\label{fig:sinkcompetition}
\end{figure}

The gap does not necessarily open at the peak of the photoevaporation profile
($R_{\mathrm{peak}} \approx 2.8$~AU). Its location is a diagnostic of the mass-loss profile
against the instantaneous accretion rate, not an imposed boundary condition. As
Figure~\ref{fig:sinkcompetition} shows, wind resupply can overwhelm mass loss well inside the
disc, and the crossing point migrates inward over time as $\dot{M}_{\mathrm{acc}}(t)$ drops.
On the flux-retained branch this is not a transition from the wind to photoevaporation through
the launch switch.
Since flux is retained while gas drains, $\beta$ falls and $S(\beta)$ rises as the disc depletes,
so $[1-S(\beta)]$ suppresses the internal sink more over time.
What drives the crossing instead is the retreat of the internal-PE screening
radius $R_{\rm c}(t)$ (Section~\ref{sec:numsetup}): as the inner disc drains, $R_{\rm c}(t)$
expands outward, exposing previously shadowed gas to direct irradiation faster than either
the wind's decline or the strengthening launch switch can compensate.

To see how that crossing scales with stellar mass, we write the mass-loss profile in
scaled form as
$\dot{\Sigma}_{\mathrm{PE}}(R)=[\dot{M}_{\mathrm{PE}}/(2\pi R_{\mathrm{s}}^2)]F(R/R_{\mathrm{s}})$.
Where the wind is fully launched ($S\to1$), equation~(\ref{eq:gap}) reduces to:
\begin{equation}
x^2F(x)=(1-\xi)\,\frac{\Mdot_{\mathrm{acc}}}{\Mdot_{\mathrm{PE}}},
\qquad x\equiv\frac{R_{\mathrm{gap}}}{R_{\mathrm s}}.
\label{eq:gapscaled}
\end{equation}
The radial scale $R_{\mathrm{s}}$ cancels, leaving the dimensionless crossing point dependent
on $M_\ast$ only through the ratio $\dot{M}_{\mathrm{acc}}/\dot{M}_{\mathrm{PE}}$.
Comparing empirical accretion rates ($\dot{M}_{\mathrm{acc}} \propto M_\ast^{1.8}$;
\citealt{manara2023}) with standard photoevaporation scalings yields a weak dependence.
The X-ray scaling of \citet{owen2012} (using $L_X$ from \citealt{gudel2007}) scales as
$M_\ast^{1.69}$, reducing the gap dependence to $\propto M_\ast^{0.11}$. The shallower
hydrodynamic scaling of \citet{picogna2021}
($\dot{M}_{\mathrm{PE}}\propto M_\ast^{1.0}$) yields $\propto M_\ast^{0.8}$.
In both cases the gap opens well outside the mass-loss peak, which is a feature of the mass
balance instead of an artefact of our $1~M_\odot$ reference profile.

\section{Magnetic flux transport and the magnetothermal front}
\label{sec:flux}
The magnetothermal-to-thermal transition (Section~\ref{sec:bridge}) is not imposed
at a fixed radius. It emerges from the recession of the magnetisation front as the
inner disc loses the ability to sustain an MHD wind. We define that front from
quantities the model already evolves instead of solving the flux-transport problem
dynamically.

\subsection{The magnetisation front and the launching threshold}

The local launching condition is a midplane plasma beta below a threshold,
$\beta(R,t)<\beta_{\mathrm{crit}}$, with $\beta_{\mathrm{crit}}\sim10^4$--$10^5$. This is
the net-vertical-field magnetisation at which non-ideal MHD simulations launch a
dynamically significant magnetothermal wind and drive accretion at observed rates.
A midplane $\beta_z\sim10^5$ suppresses the MRI and sustains a laminar wind
accreting at $\sim10^{-8}\,\Msun\,\mathrm{yr}^{-1}$, with the wind-driven rate
scaling smoothly as $\Mdot_{\rm acc}\propto\beta_z^{-1}$
\citep{baistone2013,bai2016,bai2017,lesur2021}.
The condition therefore has no sharp cutoff (it weakens as a power law as $\beta$
rises), and it is an intrinsic property of the field, set by magnetisation and
non-ideal microphysics, not a comparison of mass-loss rates.
Adopting a local field--column closure $B_z\propto\Sigma^{q}$, we obtain:
\begin{equation}
\beta(R,t)=\beta_0(R)\left[\frac{\Sigma(R,t)}{\Sigma_0(R)}\right]^{1-2q},
\label{eq:betaclosure}
\end{equation}
where $\beta_0(R)$ is the initial midplane profile. We write $\beta\equiv\beta_P$
throughout, interchangeably, matching the notation of \citet{sarafidou2024}. The
magnetothermal region is $\{R:\beta<\beta_{\mathrm{crit}}\}$, and its outer boundary
$R_\beta(t)$ (the magnetothermal front) satisfies
$\beta(R_\beta,t)=\beta_{\mathrm{crit}}$ and migrates as $\Sigma(R,t)$ evolves.

The threshold $\beta_{\mathrm{crit}}$ sets where a magnetised wind can be driven
at all. That is distinct from whether the resulting wind is
magnetically or thermally dominated, which depends jointly on the field
strength and the X-ray luminosity. \citet{sarafidou2024} mapped this parameter space
with radiation--non-ideal-MHD models, finding magnetically dominated winds for
$\log\beta_P\lesssim3$ and predominantly thermal winds for $\log\beta_P\gtrsim6$.
These correspond to X-ray thresholds $\log(L_X/\mathrm{erg\,s^{-1}})\approx29.3$ and
$30.8$. For $\beta_P\gtrsim10^5$ the total mass-loss rate converges to the purely
photoevaporative value, whereas a strong field ($\log\beta_P\lesssim4$) enhances it by
roughly an order of magnitude. Early Ohmic-only simulations \citep{rodenkirch2020}
placed that convergence at $\beta_P\gtrsim10^7$. Ambipolar and Hall terms shift it to
weaker fields, as the additional diffusion weakens the magnetic driving
\citep{sarafidou2024,sarafidou2025}.

These dominance boundaries lie at higher $\beta$ than $\beta_{\mathrm{crit}}$ because
they balance the wind against the photoevaporative floor
($\dot M_{\rm MHD}(\beta)\sim\dot M_{\rm PE}$) instead of establishing the
fundamental launching capability.
They therefore depend on the floor assumed: \citet{sarafidou2024} used the
\citet{picogna2019} X-ray prescription, which sits an order of magnitude above the
\citet{sellek2024} rates adopted here.
Because $\beta_{\rm cross}\propto\dot M_{\rm PE}^{-1/p_B}$, a lower floor shifts the
magnetic/thermal boundary to higher $\beta_P$. Under the \citet{sellek2024} rates,
the magnetothermal wind dominates mass loss beyond the $\sim10^5$--$10^6$ range
\citet{sarafidou2024} identify. The launching threshold itself is invariant (the
field's capability to drive a wind does not depend on what a thermal wind would
otherwise have removed), but the switch built on it is not equally immune. Through
$[1-S(\beta)]$, equation~(\ref{eq:launchswitch}) partitions mass loss between the
magnetothermal and thermal channels in equation~(\ref{eq:tabone}), which is a
dominance statement of the same kind as $\beta_{\rm cross}$. At
$\beta=\beta_{\mathrm{crit}}$ half the photoevaporative rate is applied, rising to
$91$ per cent at $\beta=10^5$. For the floor \citet{sarafidou2024} adopt, this tracks
their transition closely. With the lower rates used here it concedes mass loss to the
thermal channel sooner. We retain $\beta_{\mathrm{crit}}=10^4$ because it is the
launching threshold the simulations constrain, but flag the partition as an
independent assumption sensitive to the photoevaporative floor.

\subsection{The poloidal induction equation and the limits of a 1D solution}

The secular evolution of this magnetisation is governed by the poloidal induction
equation:
\begin{equation}
\frac{\partial B_z}{\partial t}=-\frac1R\frac{\partial}{\partial R}\left[R\left(v_\psi B_z-\eta_{\mathrm{eff}}\frac{\partial B_z}{\partial R}\right)\right],
\label{eq:fluxtransport}
\end{equation}
where the behaviour is controlled by the large-scale magnetic Reynolds number
$\mathcal{R}_m=|v_\psi|R/\eta_{\mathrm{eff}}$, with $v_\psi$ the field-line drift speed and
$\eta_{\mathrm{eff}}=\eta_{\mathrm{turb}}+\eta_{\mathrm{AD}}$.
Writing the non-ideal physics as a single scalar $\eta_{\mathrm{eff}}$ omits the
Hall term, which drifts the field with the electrons instead of dissipating it, with a sign
set by the observationally unconstrained polarity of $\mathbf{B}_z\cdot\mathbf{\Omega}$.
We do not integrate equation~(\ref{eq:fluxtransport}) dynamically, because a vertically
averaged 1D model cannot capture the 3D transport mechanisms it would need.
Appendix~\ref{app:flux-transport-test} shows what happens if one tries: integrated with
a uniform-coefficient, wind-free, Hall-free reference model \citep{guilet2014}, it
redistributes flux outward.
Lacking the meridional flows and Hall drift that drive inward advection in higher dimensions,
the 1D diffusion operator smooths the field radially.

\subsection{The self-limiting flux closure}

Global 3D non-ideal calculations, however, report the net effect of poloidal flux
transport to be inward advection and concentration (into rings and an inner zone),
as opposed to outward diffusion with the draining gas, though the sign is not uniform at
every radius or field polarity \citep{baistone2017,suriano2018,suriano2019,cui2021,lesur2021}.
Because a 1D solve is biased toward outward diffusion and misses these 3D concentration mechanisms, we
instead bracket the parameter space of equation~(\ref{eq:fluxtransport}) with two
phenomenological extremes: a flux-retained branch ($\mathcal{R}_m \gg 1$) and a
flux-loss branch ($\mathcal{R}_m \lesssim 1$).

The local closure index $q$ carries this net trend, as a
radius-resolved counterpart to the global field-decay index $\omega_B$ of
Section~\ref{sec:secular}. With the canonical
wind-stress scaling $\aDW\propto\beta^{-p_B}$ \citep[where $p_B \approx 1$ fits the
ambipolar stratified simulations of][]{simon2013b, armitage2013}, the advective flux
yields $\aDW\propto\Sigma^{-p_B(1-2q)}$, so the global decay index is
\begin{equation}
\omega_B=p_B\,(1-2q),\qquad p_B\approx1.
\label{eq:qomega}
\end{equation}
The sign of $1-2q$ governs the secular evolution. For any $q \le \tfrac12$ the ratio
$\beta(R,t)/\beta_0(R) = [\Sigma/\Sigma_0]^{1-2q}$ cannot rise as the disc drains. Gas
depletion alone cannot demagnetise the plasma, and the closure is self-limiting.
We adopt $q \le \tfrac12$ as our fiducial configuration, with the specific value given
in Table~\ref{tab:params}. In this regime, $\beta$ falls during depletion,
driving the same self-reinforcing accretion runaway described by
\citet{armitage2013} and \citet{martel2022}. This is strongest at the $q=0$ endpoint
($\omega_B=1$, strictly retained flux), where the field is frozen to the gas and
their assumption is matched exactly. For $q > \tfrac12$
(approaching flux freezing as $q \to 1$) the field decays fast enough for $\beta$ to
climb during draining. Because gas depletion is fastest in the inner disc, the wind is
then quenched from the inside out: the transition radius $R_\beta$ migrates to larger
radii as the demagnetised region expands, and the disc is handed over to
photoevaporation continuously on the local depletion timescale.

On the flux-loss branch the only route to an expanding demagnetised region is an
independent reduction of $B_z$ at fixed $\Sigma$ by outward diffusion
($\mathcal{R}_m \lesssim 1$). In that diffusion-dominated limit
equation~(\ref{eq:fluxtransport}) yields an effective global field-decay timescale
$\tau_B\sim R_{\mathrm{disc}}^2/\eta_{\mathrm{eff}}$, making $\mathcal{R}_m\lesssim1$
equivalent to $\tau_B\lesssim t_{\mathrm{acc}}$. The thin-disc value of
$v_\psi/\eta_{\mathrm{eff}}$ is a long-standing open problem
\citep{lubow1994,guilet2012,guilet2014,okuzumi2014,baistone2017,leung2019}, so we treat
$\mathcal{R}_m$ (equivalently $\tau_B$) as the controlling unknown that brackets the two
outcomes, a target for inference rather than a prediction. We refer to the resulting
split in outcomes as a two-branch divide: these integrations probe only the two limits
$\mathcal{R}_m\gg1$ and $\mathcal{R}_m\lesssim1$, not a continuous sweep of
$\mathcal{R}_m$. This matches \citeauthor{armitage2013}'s conclusion that flux
loss slower than mass accretion is the critical assumption behind two-timescale
wind-driven dispersal.

\subsection{The primordial-to-dispersal transition and its numerical closure}

We define the primordial-to-dispersal transition operationally through the gap-opening
condition (equation~\ref{eq:gap}). This is the epoch and radius at which photoevaporation first
overruns the local wind resupply, severing the disc and leaving the inner regions to
drain onto the star without outer resupply. The front reaching the peak of the
mass-loss profile, $R_\beta(t)\to R_{\mathrm{peak}}$, is a separate condition, marking
when the most photoevaporative annuli become unmagnetised and any residual inner disc
is swept on the terminal timescale. Which condition occurs first, and their radial
separation, is a property of the particular integration. Both are moving boundaries
derived from $(\mathcal{R}_m,\,q,\,\Sigma(R,t))$, unlike the fixed radii of classical
thermal-wind models.

To follow this two-branch divide without integrating equation~(\ref{eq:fluxtransport}), we use
a conservative finite-volume integration of the wind-dominated surface-density equation
(Section~\ref{sec:secular}), augmenting the algebraic closure~(\ref{eq:betaclosure})
with the global exponential decay introduced in equation~(\ref{eq:taub_intro}):
\begin{equation}
\beta(R,t)=\beta_0(R)\left[\frac{\Sigma(R,t)}{\Sigma_0(R)}\right]^{1-2q}
\exp\left(\frac{t}{\tau_B}\right),
\label{eq:betaclosuredecay}
\end{equation}
This parameterised decay is a proxy for outward flux diffusion. The flux-retained run
takes $\mathcal{R}_m\gg1$ ($\tau_B\to\infty$). The flux-loss run takes
$\mathcal{R}_m\lesssim1$ with $\tau_B=0.5$~Myr, from the characteristic diffusion time
$\tau_B\sim R_{\mathrm{disc}}^2/\eta_{\mathrm{eff}}$ for
$\eta_{\mathrm{eff}}\sim4\times10^{18}$~cm$^2$~s$^{-1}$ \citep{guilet2014} evaluated at
the outer grid edge $R_{\mathrm{disc}}\sim\Rout=500$~AU. We evaluate it there rather
than at the disc's characteristic mass scale ($R_{\rm c}\approx30$~AU) because flux
leaves the global budget permanently only once it diffuses out of the disc. Internal
diffusion merely redistributes it radially.

What this costs should be stated plainly.
A single global $\tau_B$ imposes a scalar approximation on a transport problem
that is neither global nor scalar. It assumes $\eta_{\mathrm{eff}}$ applies
uniformly across the disc, independent of local radial velocity $v_r(R)$ or
gradients in the diffusivity, and it carries no representation of the Hall drift.
Our own vertically averaged flux-transport calculation
(Appendix~\ref{app:flux-transport-test}) shows why 1D approximations fail to
reproduce the inside-out demagnetisation that the closure assumes.
The two runs are therefore a phenomenological bracket on the limiting outcomes
(establishing what follows if flux is retained or lost on a disc-lifetime
timescale) and leave the transport mechanics to targeted 3D studies.
\section{Secular evolution of the two flux-transport branches}
\label{sec:results}

The closed forms of Sections~\ref{sec:secular}--\ref{sec:flux} establish the model's
scalings: the ejection index $\xi$ shared between the launch and secular solutions
(equation~\ref{eq:sigprofile}) and the stellar-mass dependence of the gap radius
(equation~\ref{eq:gapscaled}), together with the result that depletion alone can
never trigger dispersal under an algebraic $B_z(\Sigma)$ closure.
Those are statements about every disc obeying the model, not evolutionary tracks for
a particular one.
Once the closure is allowed to feedback, the characteristics of Section~\ref{sec:char}
are straight only for spatially constant $\aDW$. The depletion factor
$\Sigma_{\rm r}^{-(1-2q)}$ and the launch switch $S(\beta)$ both make $\aDW$ evolve with
the solution instead.
That feedback is what produces the two-branch divide.

We therefore integrate the governing equation numerically, verifying in
Appendix~\ref{app:scheme} that the solver reproduces the analytic characteristics.
The two branches identified in Section~\ref{sec:flux} differ solely in the imposed
behaviour of the large-scale magnetic flux.
We integrate them from identical initial conditions and with identical transport physics,
so that every difference reported below is attributable to that one choice.

\subsection{Numerical setup}
\label{sec:numsetup}

\subsubsection{Initial conditions and grid}

We initialise a $1\,\Msun$ star and a $0.02\,\Msun$ disc on a one-dimensional grid
spanning $\Rin=0.5$~AU to $\Rout=500$~AU, discretised into $N=512$ logarithmically
spaced cells with centres $R_j$, faces $R_{{\rm f},j}$ and cell area
$A_j=2\pi R_j\,\Delta R_j$. Outflow (zero-gradient) and zero-inflow boundary conditions
are applied at the inner and outer edges respectively.

The initial gas surface density follows the viscous self-similar profile of
\citet{lyndenbell1974},
\begin{equation}\label{eq:sigma0}
    \Sigma_0(R)\propto(R/R_{\rm c})^{-1}\exp(-R/R_{\rm c})\,,
\end{equation}
with an exponential cutoff $R_{\rm c}=30$~AU.
We adopt this $\psi\to0$ profile because it is the form discs are conventionally
characterised with observationally, and a reasonable state after the initial
gravito-turbulent phase.
A wind-consistent alternative is in any case ill-posed for a radially varying
lever arm $\lambda$ (Section~\ref{sec:bai}).
We nonetheless verify in Appendix~\ref{app:highmdot} that a wind-consistent initial
profile at a higher accretion rate does not alter the two-branch divide described below.

\subsubsection{Integration scheme and timestepping}

The evolved variable is the cell mass $A_j\Sigma_j$, so mass is conserved by design.
All transport (the wind flux, the viscous flux and the linear wind sink) is
advanced by a single backward-Euler solve.
Only the internal photoevaporative source is left explicit, being local and not stiff.
The system is tridiagonal with positive diagonal and non-positive off-diagonals,
an M-matrix. Its inverse is therefore non-negative, and the solve returns
$\Sigma^{\rm new}>0$ at any $\Delta t$, with no Courant condition.

The equation is nonetheless non-linear since the transport coefficients in
equation~(\ref{eq:adwfull}) are functions of $\Sigma$.
Each coefficient ($\aDW$, $S(\beta)$, the wind-base state, $k_j$ and $a_j$) is
evaluated at the old time level and held fixed across the solve, making the step
linearised-implicit.
Positivity is unconditional, since the M-matrix property follows from the sign structure
of the assembled matrix alone.
Accuracy is not. Lagging the coefficients renders the non-linearity first order in
$\Delta t$, and it is that error, not stability, which the timestep coefficients bound.

The four dimensionless coefficients are the local sink rate, the advective rate,
the viscous diagonal rate and the explicit photoevaporative removal, all evaluated only on
cells retaining one per cent of their initial surface density and on their immediate neighbours.
Integration stops once $\max_R\Sigma$ falls below $1\,\mathrm{g\,cm^{-2}}$.

Appendix~\ref{app:scheme} sets out the discretisation and the four timestep coefficients.
It checks the scheme against the closed-form characteristics solution of Section~\ref{sec:char},
which it reproduces to under a quarter of a per cent over $1$~Myr in the limit where that
solution is exact.
Appendix~\ref{app:tabone2022-validation} additionally benchmarks the transport solver
against the independent closed-form solutions of \citet{tabone2022}.

\subsubsection{Viscosity, temperature and the initial field}

A small turbulent viscosity $\aSS=10^{-5}$ is included to represent residual
hydrodynamic instabilities (e.g.\ the vertical shear instability). With
$\aDW^0=10^{-3}$ this gives a wind-to-viscous ratio $\psi=\aDW/\aSS\simeq100\gg1$, so
transport remains wind-dominated.
The midplane temperature follows the passive-disc irradiation profile
$T_{\rm mid}=280\,(R/\mathrm{AU})^{-1/2}$~K with mean molecular weight $\mu=2.34$,
the same molecular value carried through the irradiated wind base (Section~\ref{sec:bridge}).

The initial midplane magnetisation is $\beta_0(R)=30\,(R/\mathrm{AU})^{3/2}$, which
crosses $\beta_{\rm crit}=10^4$ at $R\approx50$~AU. The wind is therefore launched
out to $R_\beta(0)\approx50$~AU, while the outer disc starts unmagnetised. The launch
coefficient $\aDW$ includes the smooth cutoff $S(\beta)$ of
equation~(\ref{eq:launchswitch}), which weakens the wind as a power law instead of
switching it off at a sharp boundary. Together with the depletion factor of
equation~(\ref{eq:adwfull}) this gives
$\aDW\propto\beta^{-1}$ inside the magnetothermal front, matching the
$\Mdot_{\rm acc}\propto\beta_z^{-1}$ scaling established by non-ideal-MHD
simulations \citep{baistone2013,bai2016,bai2017,lesur2021}. Outside the front it
gives $\aDW\propto\beta^{-2}$ instead, as discussed in
Section~\ref{sec:secular}.
Beyond $R_\beta$ the lever arm is taken from equation~(\ref{eq:spruit}) rather
than held at its launching value, and the two need not match: at
$\beta_{\mathrm{crit}}$ that closure returns $\lambda=3$ identically, while the
derived value just inside $R_\beta$ is whatever the Bai map gives there
($\lambda\approx5$ at $t=0$, falling below $3$ later in the evolution), so the
resulting step can run either way -- visible as the discontinuity in
Figure~\ref{fig:leverarm}'s dashed continuation beyond $R_\beta$. It is
dynamically inert regardless, since it governs $\lambda$ only where the launch
switch is already suppressing the torque. Across the launching region the
derived values ($\lambda\approx4$--$7$, $\xi\approx0.09$--$0.16$) straddle
\citeauthor{bai2016}'s fiducial ($\lambda\simeq5.2$, $\xi\simeq0.12$), as befits
a thermally assisted base.

\subsubsection{The flux-closure index}

The flux-closure index is fixed at $q=0.3$ throughout.
This is not a measured quantity, as the literature constrains the
branch, not the value. The inward flux advection of Section~\ref{sec:flux}
places the disc on the self-limiting $q\le\tfrac12$ side, and within it $q=0.3$
is a mid-range choice. By equation~(\ref{eq:qomega}) it corresponds to a global
decay index $\omega_B=p_B(1-2q)=0.4$, inside the range $\omega_B=0.25$--$1$ over
which \citet{tabone2022} construct their self-similar family. The integrator is
validated against that same family (Appendix~\ref{app:tabone2022-validation}).

\subsubsection{Photoevaporative sinks and shadowing}

Internal photoevaporation is set by the X-ray-driven mass-loss profile of
\citet{sellek2024} (the column-integrated polynomial fit of their Table~3), scaled
for a $1\,\Msun$ star with the same $L_X=2\times10^{30}$~erg~s$^{-1}$ (0.5--5~keV)
adopted in Section~\ref{sec:bridge}. It peaks at $R_{\mathrm{peak}}\approx2.8$~AU and
carries a total rate $\approx4\times10^{-9}\,\Msun\,\mathrm{yr}^{-1}$.
Appendix~\ref{app:sellek2024-validation} benchmarks it, as implemented here, against
\citeauthor{sellek2024}'s own published calculations.
Following \citet{picogna2026}, internal photoevaporation acts only on the
directly-illuminated bulk disc. Beyond the co-moving cut-off radius $R_{\rm c}(t)$
the outer disc is screened by the inner disc (its scale height drops and the outer
midplane no longer sees the ionising field), so $\dot\Sigma_{\rm PE}$ is smoothly
suppressed there, by the factor $[1+(R/R_{\rm c})^{6}]^{-1}$.
We identify $R_{\rm c}(t)$ with the disc's current characteristic mass
radius, the radius enclosing a fraction $1-e^{-1}\approx0.632$ of the
instantaneous disc mass.
For the initial profile (\ref{eq:sigma0}) this returns the exponential-taper scale itself,
$R_{\rm c}(0)=30$~AU, so the initial condition is unchanged. It is, however,
updated from $\Sigma(R,t)$ at every step, not held fixed, and is free to
move in either direction as the disc evolves.
It remains a proxy: a self-consistent treatment would track $H(R,t)/R$ and locate
the shadowing front directly, which we list among the caveats.

\subsection{The two-branch divide: front trajectories and accretion history}
\label{sec:bifurcation}
\begin{figure*}
\centering
\includegraphics[width=0.82\textwidth]{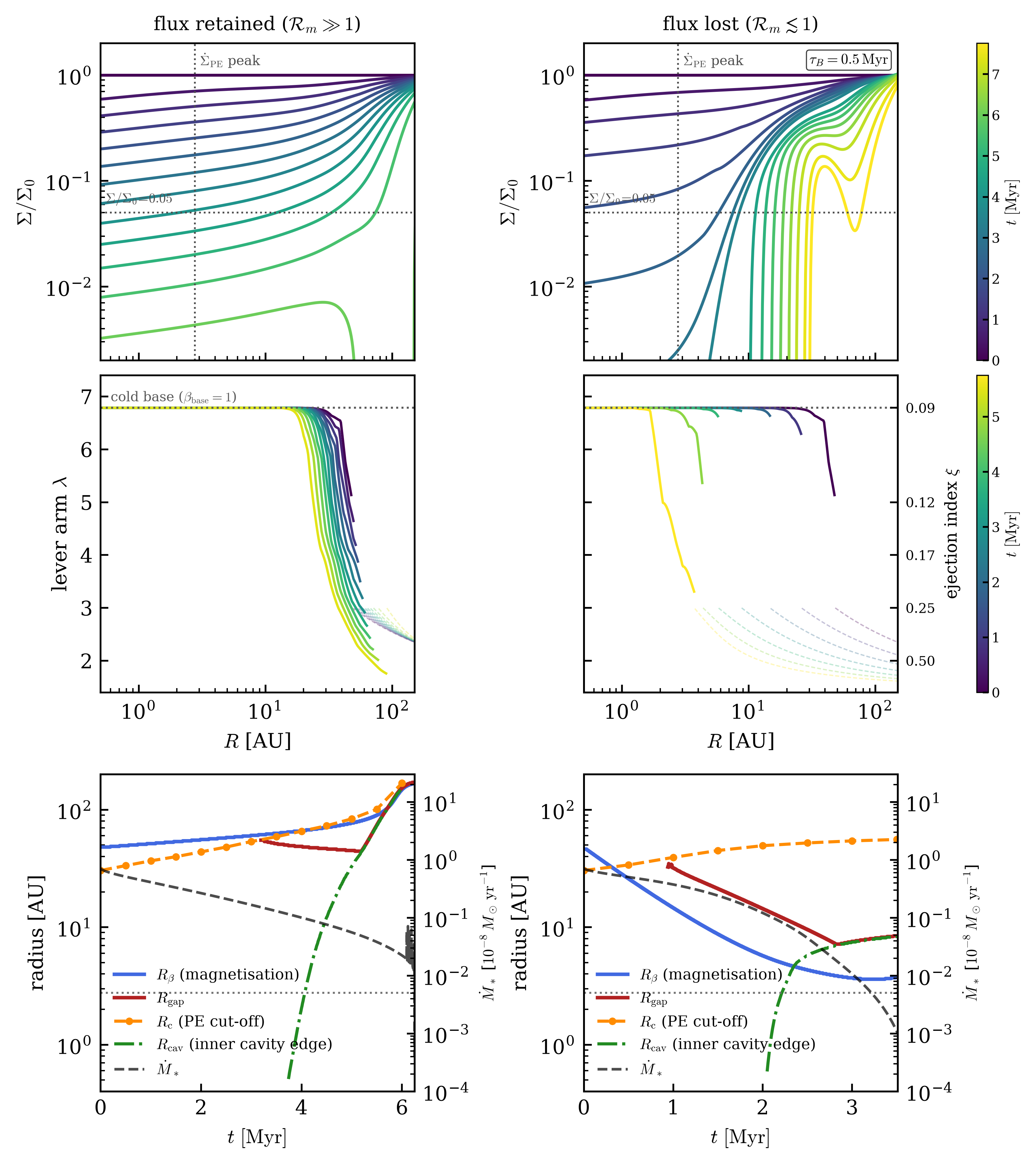}
\caption{Coupled integration of the secular wind equation with the flux closure
(\ref{eq:betaclosure}), under the fiducial $\Xi$--$T$ wind-base prescription
(Section~\ref{sec:launch}), for a flux-retained run ($\mathcal{R}_m\gg1$, left)
and a flux-loss run ($\mathcal{R}_m\lesssim1$, right).
\emph{Top:} surface density $\Sigma(R,t)/\Sigma_0$ evolution.
\emph{Middle:} derived lever arm $\lambda(R)$, solid over the launching region
$S(\beta)>1/2$ and continued faint and dashed beyond $R_\beta$, where
$S(\beta)\le1/2$: the map still returns a value there, but it enters the torque
multiplied by nothing and sources no wind, so it is a prescribed rather than a
physically active quantity. The corresponding ejection index is
$\xi=1/[2(\lambda-1)]$.
\emph{Bottom:} front trajectories and $\Mdot_\ast$, drawn up to and not beyond
each branch's own accretion collapse. The diagnostics degenerate on
floor-level gas immediately past it, and their resolution is only marginal in
the final $\sim0.3$~Myr before it (Appendix~\ref{app:trust-near-dispersal}), so
neither regime is shown.
The evolution, and the distinction between the two gap measures, are described in
Sections~\ref{sec:validity} and~\ref{sec:bifurcation}; convergence in
Appendix~\ref{app:resolution-check}.}
\label{fig:integrator}
\label{fig:leverarm}
\end{figure*}

Figure~\ref{fig:integrator} (bottom panels) tracks three radii:
\begin{enumerate}
\item $R_\beta$, the magnetisation front of Section~\ref{sec:flux}, where 
$\beta=\beta_{\mathrm{crit}}$;
\item $R_{\mathrm{gap}}$, the innermost radius at which the delivered
photoevaporative rate overruns the wind's net advective resupply
(equation~\ref{eq:gap});
\item $R_{\rm cav}$, the density-based inner-cavity edge, set by the
$\Sigma/\Sigma_0=0.05$ threshold.
\end{enumerate}

In the flux retained ($\mathcal{R}_m\gg1$) branch, the magnetisation front
$R_\beta$ expands monotonically as the field concentrates and the wind
strengthens. It grows from $47$~AU to $77$~AU by $t=5$~Myr, and to
$\approx140$~AU by $0.3$~Myr before the branch's terminal collapse at
$t\approx6.25$~Myr.
$R_{\mathrm{gap}}$ is first defined at $t=3.19$~Myr at $55$~AU, $3.1$~Myr
ahead of the collapse. It then migrates inward to $45$~AU by
$t\approx5.2$~Myr as the accretion rate declines (see equation~\ref{eq:gap}).

In the flux lost ($\mathcal{R}_m\lesssim1$) branch, $R_\beta$ instead recedes
monotonically from $47$~AU to a minimum of $3.6$~AU at $t=3.1$~Myr.
$R_{\mathrm{gap}}$ already exceeds it the first instant both are defined
($t=0.90$~Myr, $R_{\mathrm{gap}}=37$ against $R_\beta=16$~AU), $2.6$~Myr ahead of
that branch's collapse.
Over the pre-collapse window the gap spans $R_{\mathrm{gap}}\approx7$--$37$~AU,
bounded above by the cut-off radius $R_{\rm c}(t)$ and lying well outside
$R_{\mathrm{peak}}$. Although $\Sigdot_{\mathrm{PE}}$ peaks there, the inner
disc is still wind-resupplied, so the balance is first lost further out, near
the outer edge of the unscreened bulk disc (see Figure~\ref{fig:sinkcompetition}).

The gap's location relative to the front differs by branch.
On the flux-retained branch it opens inside $R_\beta$, where the wind is still
substantially coupled ($S(\beta)\approx0.6$ at the crossing) and its resupply
is a real competitor that photoevaporation only narrowly overtakes. On the
flux-loss branch it opens outside the front, where the field has
already receded to $S(\beta)\approx0.2$ and the wind has correspondingly
little resupply left to give. Flux retention therefore defers the gap and
moves it inside the front rather than preventing it.

After the cavity edge overtakes the gap location ($t\approx5.2$~Myr on the
flux-retained branch, $t\approx2.8$~Myr on the flux-lost one), they coincide
to a single grid cell. From there only $R_{\rm cav}$ is a sensitive tracer.

The stellar accretion rate (Fig.~\ref{fig:integrator}, black dashed)
distinguishes the branches by when it collapses:
in both branches $\Mdot_\ast$ declines smoothly from its initial value of
$\sim6.9\times10^{-9}\,\Msun\,\mathrm{yr}^{-1}$, and then falls sharply to a
numerically-floor-dominated residual once the innermost active accretion
channel empties.
Adopting $\Mdot_\ast<10^{-11}\,\Msun\,\mathrm{yr}^{-1}$ as the criterion
throughout, this happens at $t\approx6.25$~Myr in the flux-retained branch
and $t\approx3.51$~Myr in the flux-loss one.
The two branches differ by $\approx2.7$~Myr, far beyond the $\lesssim2\%$ grid-resolution
sensitivity of either collapse time (Appendix~\ref{app:resolution-check}), so
the ordering is a robust feature of the two-branch divide.
The collapse follows directly from a model assumption. The
wind in this integration is switched off beyond $R_\beta$, so the screened
outer disc cannot resupply the inner one once that reservoir is exhausted,
and the disc drains to (numerical) completion instead of settling into the fed
cavity steady state of \citet{martel2022} and \citet{sarafidou2025}.
\citet{armitage2013} ran the equivalent experiment (the wind switched off outside
an imposed radius, in their case a fixed $30$~AU as opposed to a moving $R_\beta$)
and did obtain a fed cavity, because in their model the outer disc remains viscous
and supplies it. The difference here is the wind-dominated limit
($\psi\gg1$, Section~\ref{sec:numsetup}). There is no viscous reservoir left to do
the feeding, so the same geometry gives terminal drainage instead of a
quasi-steady cavity. Which of the two a real disc shows therefore depends on the
residual turbulent stress in the screened outer disc, a quantity neither
calculation constrains. Over a disc's lifetime, then, the branches are
distinguished by the timing of this collapse and by the shape of
$R_\beta(t)$ before either branch's own collapse (monotonically expanding
versus monotonically receding), not by a persistently different
asymptotic $\Mdot_\ast$.

This two-branch divide makes a prediction that can be tested against resolved wind
observations, because the two branches disagree about the state of the
magnetothermal region at a given accretion rate rather than at a given
time. \citet{bajaj2026} analyse JWST/MIRI spectra of $72$ inclined, mostly
Class~II discs and find that atomic jets and atomic--molecular winds of
MHD origin dominate while a disc is accreting rapidly, whereas at lower
accretion rates the jets weaken and the winds become predominantly atomic.
Extended H$_2$ and the [O\,\textsc{i}] low-velocity component appear
preferentially above $\dot M_\ast\sim10^{-8.5}\,\Msun\,\mathrm{yr^{-1}}$, and
low-velocity [Ne\,\textsc{ii}] emission preferentially below it. That ordering
is the magnetothermal-to-thermal handover of this model, indexed by the same
variable.
Our two branches reach that accretion rate in comparable times but in opposite
states. The flux-retained branch passes it at $t=1.60$~Myr with $R_\beta$ having
expanded to $54$~AU, while the flux-loss branch passes it at $1.22$~Myr
with $R_\beta$ having receded to $11.6$~AU, a factor of four inside its
initial value. A population in which MHD wind tracers fade as accretion declines
is therefore the flux-loss branch, not the flux-retained one.

Internal photoevaporation never reaches its full potential on the
flux-loss branch. The magnetisation front recedes to a resolution-converged
minimum of $3.61$~AU, stopping $30$~per cent short of
$R_{\mathrm{peak}}=2.766$~AU, so the radius where $\Sigdot_{\mathrm{PE}}$ is
strongest stays wind-resupplied throughout the branch's life. The gap opens
correspondingly far out, $R_{\mathrm{gap}}\approx37$~AU
($\approx13\,R_{\mathrm{peak}}$).

\subsection{The wind base in the evolving disc}
\label{sec:windbasemech}
\begin{figure*}
\centering
\includegraphics[width=\textwidth]{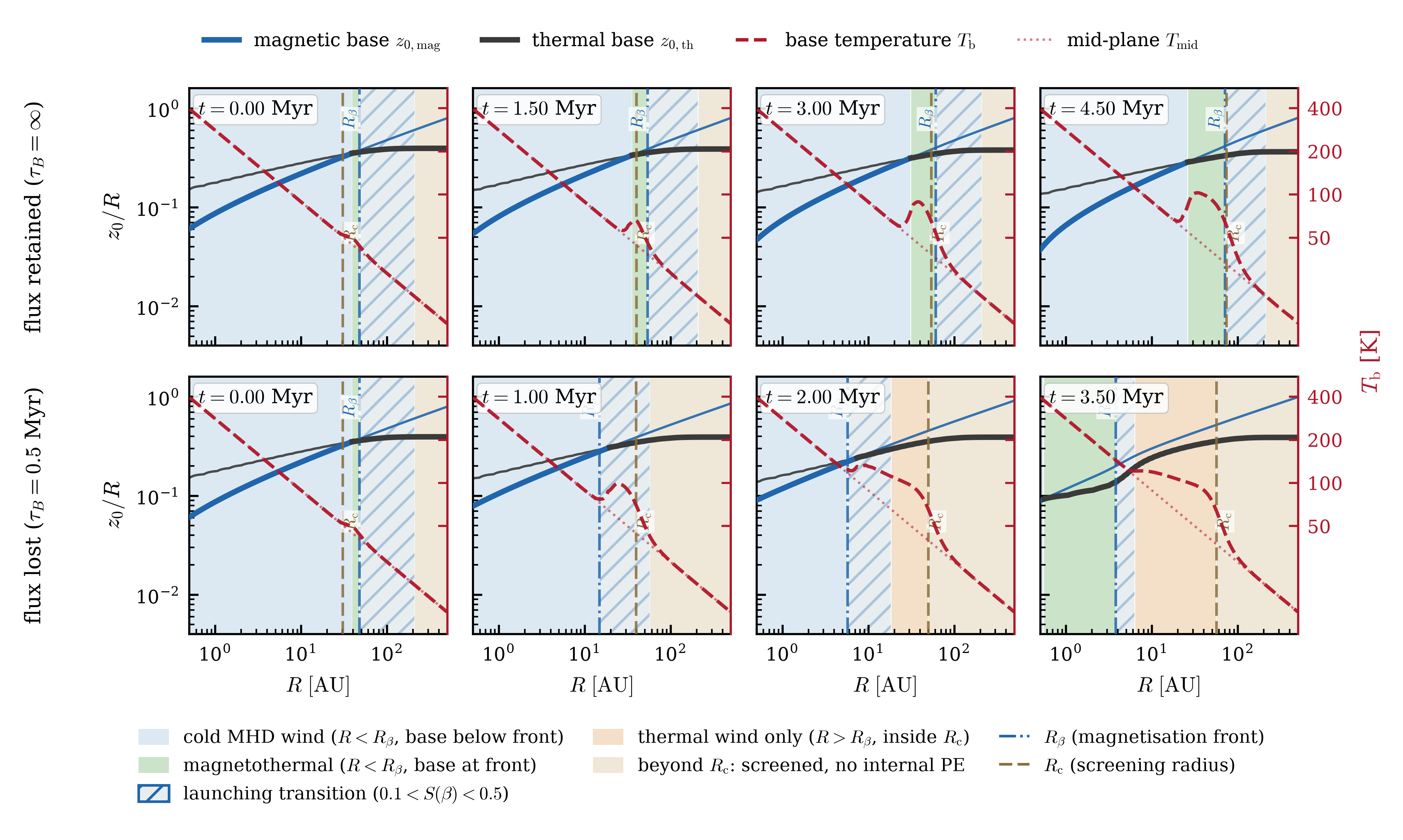}
\caption{The wind-base launch mechanism on both branches. 
\emph{Top row:} flux retained, $R_\beta$ expanding.
\emph{Bottom row:} flux lost, $R_\beta$ receding; epochs are labelled in each
panel. Left axis: the competing base heights $z_{0,\rm mag}$ (blue) and
$z_{0,\rm th}$ (dark), wind base $z_0=\min$ of the two drawn thick.
Right axis: the wind-base temperature $T_{\rm b}$
(red dashed) against the mid-plane $T_{\rm mid}$ (dotted). Shading gives the
launching regime and the vertical lines the two radii bounding it, $R_\beta$ and
$R_{\rm c}$ (defined in Section~\ref{sec:windbasemech}).
The shaded magnetothermal region is drawn in two levels: solid shading marks
$S(\beta)>1/2$, i.e.\ inside $R_\beta$, where the magnetothermal channel takes
most of the mass loss. The hatched band marks $0.1<S(\beta)\le1/2$, a
weakly-launching shoulder outside $R_\beta$ where the wind is
sub-dominant and the complementary weight $[1-S(\beta)]$ leaves the majority of
the mass loss to the thermal channel.}
\label{fig:windbase-mechanism}
\end{figure*}

Figure~\ref{fig:windbase-mechanism} makes the wind-base decomposition of
Section~\ref{sec:bridge} quantitative.

The figure divides the disc by launching regime.
A magnetocentrifugal wind is launched only inside $R_\beta$, where $S(\beta)>1/2$ and
the wind therefore carries more than half the mass loss (equation~\ref{eq:launchswitch}
with $\beta<\beta_{\rm crit}$).
Inside it the base lies either below the irradiation front (a cold MHD wind, blue region)
or at it (magnetothermal, green region).
The switch closes smoothly, as opposed to abruptly, so a wind persists over $0.1<S(\beta)<1/2$
(the shaded region) while photoevaporation dominates it.
Beyond that the outflow is purely thermal for $R<R_{\rm c}$ (pink region), and is
suppressed beyond $R_{\rm c}$ (gray region), where the inner disc's column screens the star.
The wind-base temperature decides which threshold wins. $T_{\rm b}$ falls from
$\approx400$~K at $0.5$~AU and rejoins the mid-plane curve in the outer disc,
which the isolated runs leave unheated ($T_{\rm b}=T_{\rm mid}=28$~K at
$100$~AU; Section~\ref{sec:bridge}). A heated plateau of
$\approx100$--$130$~K sits between the two fronts and moves inward with them.
Beyond $R_\beta$ that heated base matters only in the narrow transitional annulus
with $0.1<S(\beta)<0.5$, where a sub-dominant residual wind torque survives.
Elsewhere it adds no mass loss: the outflow is purely thermal, set by the
photoevaporative sink term.

On the flux-loss branch the base crossover $z_{0,\rm th}=z_{0,\rm mag}$
separates the inner cold-MHD launch from the magnetothermal one. It sweeps
inward as flux is lost, from $39.6$~AU at $t=0$ to $0.7$~AU by $t=3.5$~Myr,
tracking the receding magnetisation front $R_\beta(t)$ (Section~\ref{sec:bifurcation}). The
two are related but distinct radii ($z_{0,\rm th}=z_{0,\rm mag}$ versus
$\beta=\beta_{\rm crit}$), and they do not recede together. A launch is
magnetothermal only where the base has reached the irradiation front and
the field still dominates the mass loss ($S(\beta)>1/2$, i.e. $R<R_\beta$), so
the magnetothermal region is the annulus between the two. At $t=0$ the crossover
lies inside $R_\beta$ and that annulus is open. The front recedes the faster
of the two, however, and overtakes the crossover within $0.5$~Myr. From then
until $\approx2.5$~Myr the crossover sits outside $R_\beta$ and the annulus
is empty: the irradiated gas has fallen outside the launching region altogether
and drives a purely thermal wind.
It reopens over the final $\approx1$~Myr before collapse.
By $t=3.5$~Myr the crossover has receded to $0.55$~AU, essentially the inner
grid edge. The entire surviving launching region ($0.55$--$3.76$~AU) has therefore
turned magnetothermal, the opposite of the branch's first two Myr. The
driving mechanism varies sharply across that region: wind-dominated inside
$\approx1.5$~AU ($\gtrsim95$~per cent of the local mass loss), crossing to
photoevaporation-dominated ($\gtrsim65$~per cent) beyond $\approx3$~AU, with
the $50$/$50$ point falling close to $R_{\mathrm{peak}}\approx2.8$~AU.

On the flux-retained branch the expanding front does not outrun the crossover, so
that magnetothermal band is present from $t=0$ and widens monotonically, from a
factor $1.2$ in radius to $3.1$ by $5$~Myr, with a base reaching $\approx2.4$ times
the mid-plane temperature.

The magnetothermal launch that distinguishes this model from a superposed cold wind
and thermal sink is largely a flux-retained-branch phenomenon by duration
(on the flux-loss branch it is absent for most of the branch's life), but not by
consequence. It reopens decisively in the final $\approx1$~Myr before
collapse, and that is exactly where the branch clears. The results are insensitive
to the choice of anchor in equation~(\ref{eq:z0}).

\subsection{An inside-out disc clearing}
\label{sec:outerclear}
Figure~\ref{fig:integrator}'s top panels track the evolution of $\Sigma(R,t)/\Sigma_0$
on both branches until dispersal.
Within the cut-off radius the dispersal proceeds through two channels acting in concert.
The inner disc, which holds most of the mass, drains by wind-driven accretion, while
internal photoevaporation removes the unscreened gas ($R\lesssim R_{\rm c}$), eventually
opening a gap.
Gas beyond $R_{\rm c}$ receives no internal mass loss at any given instant.
The cut-off expands as the disc evolves, however, and material that was screened is
progressively exposed.

On the flux-loss branch $R_{\rm c}$ grows from $31$~AU at $t=0$ to $62$~AU by $t=5$~Myr.
The outer disc responds: $\Sigma/\Sigma_0$ at $80$--$120$~AU falls to $70$--$92\%$ of its
initial value ($\approx20$ per cent of its mass).
Internal photoevaporation is therefore not merely reshaping the inner disc.
It is dispersing the whole of it, on a timescale set by how fast the screening lifts.
The erosion accelerates accordingly.
By $t=7.75$~Myr, $R_{\rm c}$ has reached $106$~AU, and the same annulus retains only $19$
per cent of its mass.
The region inside $100$~AU (which held $96$ per cent of the initial mass) is down to
$3.1$ per cent of what it started with, and the disc as a whole to $5.0$ per cent.
Internal photoevaporation alone therefore disperses this disc on a $\sim10$~Myr
timescale, with no external irradiation required.
We quote the sequence only to $7.75$~Myr, where the run reaches the surface-density
stopping criterion of Section~\ref{sec:numsetup}, with no sign of stalling.

That clearing proceeds inside-out.
The minimum of $\Sigma/\Sigma_0$ sits at the inner boundary throughout, and every
depletion contour is anchored there and sweeps outward.
The $\Sigma/\Sigma_0=0.5$ contour reaches $8.2$~AU at $t=1$~Myr, $25$~AU at $2$~Myr and
$57$~AU at $5$~Myr.
The direction is set by where the sinks act, not by the direction in which the screening
front travels.
The cut-off does move outward, but it only ever exposes gas to a photoevaporative profile
that is itself centrally peaked, at $R_{\rm peak}\approx2.8$~AU.

The mass-loss profile applied above is the one computed for an unbroken disc.
Once a cavity has opened, the directly illuminated wall is eroded faster, and the profile
is no longer the primordial one.
That enhancement is calibrated and its consequences quantified in
Appendix~\ref{app:cavitype}.

On the flux-retained branch the outer disc is removed too, but almost entirely by the
magnetothermal wind.
There $R_\beta$ expands, from $47$~AU at $t=0$ to $77$~AU at $5$~Myr, so the
$80$--$120$~AU band is progressively overtaken by the wind-active zone.
$\Sigma/\Sigma_0$ there falls to $72\%$ at $t=2.5$~Myr and $31\%$ at $5$~Myr, reaching
the numerical floor by $6$~Myr, shortly before the branch's collapse.
Internal photoevaporation accounts for $7\%$ of the outer-disc removal by $5$~Myr, and
the expanding magnetothermal wind does the rest.
This is so despite $R_{\rm c}$ expanding faster on this branch than on the
flux-loss one, reaching $84$~AU by $5$~Myr against $62$~AU: the cut-off does sweep the
band, but the wind gets there first.

Neither branch clears from the outside in, and neither approaches a disc-wide,
spatially uniform decline (i.e. the homologously depleting disc of
\citet{armitage2013,wood2002}), which needs $\beta_z$ held uniform in radius rather
than tied to the locally evolving $\Sigma$ as it is here.
The two branches are not alike each other either, and the difference is observable.
On both, the depleted region is anchored at the inner boundary and expands outwards.
A cavity at the $\Sigma/\Sigma_0<0.05$ threshold opens at $t=2.05$~Myr on the flux-loss
branch and at $3.74$~Myr on the flux-retained one, in each case at the innermost grid
cell, and no detached gap forms at any epoch on either.
What differs is how fast the cavity grows and what state the disc around it is in.
The flux-loss cavity expands slowly
($5.7$~AU at $2.5$~Myr, $8.3$ at $3.5$ and $13.5$ at $5$~Myr) while its outer
disc is barely touched, $\Sigma/\Sigma_0$ at $80$--$120$~AU still standing at
$0.82$ at $5$~Myr. The flux-retained cavity opens more than a Myr later and then
sweeps outward more than an order of magnitude faster ($2.1$~AU at $4$~Myr,
$33$ at $5$ and $151$ at $6$~Myr) into an outer disc that the expanding wind
has already thinned to $0.30$ of its initial surface density. The flux-loss
branch therefore gives the conventional transition-disc morphology, a modest hole
in an otherwise intact disc. The flux-retained branch instead depletes roughly
evenly across the inner disc while $R_\beta$ still holds the outer one,
then, once dispersal kicks in, clears the remainder so rapidly that
little is left outside by the time the hole is large.

The $\sim10$~Myr dispersal time of the flux-loss branch depends strongly on the
screening radius, which is here diagnosed from the disc's characteristic mass radius,
a function of $\Sigma$ alone, while the shadow it stands for is set by the inner
disc's aspect ratio $H(R,t)/R$, which the proxy never evaluates. What we can conclude
is that internal photoevaporation can clear this disc unaided once the cut-off
is free to move. The timescale on which it does so is not a calibrated prediction.

\subsection{The lever arm of the evolving wind}
\label{sec:leverarm}
The lever arm $\lambda$ is the derived quantity that wind observations constrain directly.
Each cell reads $\lambda=f(\beta_{\mathrm{base}})$ off the \citet{bai2016} map
(Section~\ref{sec:bai}), with $\beta_{\mathrm{base}}$ set by the local field, the
base height and the base temperature.
Its evolution is therefore a prediction of the coupled system.
It is shown in the middle panels of Figure~\ref{fig:leverarm} for the two branches.
Because $\lambda$ is meaningful only where a wind is actually launched, every
statistic quoted below is restricted to the launching region $S(\beta)>1/2$,
i.e.\ $R<R_\beta$; Figure~\ref{fig:leverarm}'s middle panels additionally show
its faint, dashed continuation beyond $R_\beta$, where the launch switch
already suppresses the torque (Section~\ref{sec:bai}).

Two features are important.
The first is the plateau at $\lambda=6.79$ ($\xi=0.086$) that occupies the inner
disc at every epoch on both branches.
Where the base is set magnetically instead of thermally ($z_{0,\rm mag}<z_{0,\rm th}$,
the cold-launch regime of Section~\ref{sec:bridge}), $z_{0,\rm mag}$ is defined as the
height at which the local plasma beta reaches unity.
This makes $\beta_{\mathrm{base}}=1$ identically there, so $\lambda$ takes the single
value the map returns at that argument.
The plateau is thus a direct consequence of the coupling criterion discussed in
Section~\ref{sec:bridge}, and it inherits that criterion's uncertainty.
A coupling height defined by an Elsasser number instead of by equipartition would move
the plateau, though not the contrast between the branches.
The second is that $\lambda$ falls below the plateau only in the magnetothermal annulus,
where the irradiation front lies below the magnetic base and the elevated $\csw$ raises
$\beta_{\mathrm{base}}$.
The lever arm is therefore a direct indicator of which launch regime each radius is in.

The branches separate in the spread of $\lambda$ rather than in its typical value.
On the flux-retained branch the magnetothermal shoulder is present from $t=0$ and both
widens and deepens.
The minimum lever arm over the launching region falls from $5.2$ to $1.5$, the
mass-weighted mean $\langle\lambda\rangle$ from $6.7$ to $4.3$ at $5$~Myr, and the
ejection index implied by that mean rises from $\xi=0.088$ to $0.12$.
On the flux-loss branch $\lambda$ is instead pinned at the cold plateau across
essentially the whole launching region for the first $2$~Myr (the median never leaves
it, and $\langle\lambda\rangle$ stays within $2$ per cent of its initial value), and
then falls steeply to $4.8$ over the final $0.7$~Myr.
That is the same statement as Section~\ref{sec:windbasemech}'s, in the variable an
observation can reach.
The flux-loss branch launches an essentially cold, weakly loaded wind for most of
its life and acquires a magnetothermal character only as it disperses, whereas
the flux-retained branch carries a broad range of lever arms throughout.

Deriving the lever arm instead of prescribing it is not trivial, and it needs a fair baseline.
A cold, weakly loaded launch corresponds to the model's own equipartition-base limit,
$\lambda\approx6.79$ at $\beta_{\mathrm{base}}=1$ (Section~\ref{sec:bai}).
Repeating both named integrations with $\lambda$ held fixed at that value, and altering
nothing else, leaves the sustained accretion rate over $0.5$--$3$~Myr unchanged
on both branches.
For most of each branch's mass-weighted life the derived $\lambda$ sits on or near this
same plateau (Figure~\ref{fig:leverarm}).
It does, however, delay the terminal collapse, by $5.8$ per cent on the flux-retained
branch and $1.6$ per cent on the flux-lost one.
The derived run alone departs from the plateau in its final approach to collapse, where
the mass-weighted mean falls toward $\lambda\approx2$, ejecting more mass than the fixed
cold limit right when it matters most for the total budget.

Both branches straddle \citet{bai2016}'s own fiducial $\lambda\simeq5.2$, which
the flux-retained mean crosses at $t\approx3.6$~Myr and the flux-loss mean only
in its final $0.3$~Myr. An observationally inferred lever arm is weighted by the
material that is actually flowing and emitting, whereas $\lambda$ here is a local
quantity at the wind base. The mapping between the two is not one-to-one, and the
comparison should be read at the level of range and trend, not of a single
number.

\citet{martel2022} report $(\kappa,\lambda)=(2.2\times10^{-2},23)$ near the cavity and $(2.5,3.2)$
in the outer disc, corresponding to $\beta_{\mathrm{BP}}\simeq6.2$ and $15.7$, respectively.
As expected, the cold analytical relation holds tightly in the weakly loaded, magnetically
dominated inner wind, but diverges by a factor of 3 for heavily loaded outer outflows.

The lever arms derived here can also be set against a direct measurement.
\citet{kim2026} resolve a rotating, nested outflow around the Class~0 protostar
HOPS~358 and infer a dimensionless lever arm $\lambda\simeq2.3$, with the
launching region constrained to $10$--$18$~AU. Over those same radii our derived
$\lambda$ is $\approx6.8$, a factor of three larger. Three differences bear on
the comparison. HOPS~358 is a deeply embedded Class~0 source while these
calculations describe the Class~II dispersal phase, at a different surface
density and magnetisation. The observational value is inferred from outflow
kinematics on scales well above the base, whereas ours is the local footpoint
value. Finally, the two need not coincide even for one wind if the mass loading
varies along the flow. The tension is nonetheless real, and it points in a definite
direction: $\lambda\simeq2.3$ implies a substantially heavier mass loading
than the fiducial field strength selects. Equation~(\ref{eq:spruit}) admits such
values, since $\lambda\to3/2$ as $\mu_{\mathrm w}\to\infty$, so the formalism
spans the measurement. What it does not do is select it at the adopted $\beta$.
Since $\xi=1/[2(\lambda-1)]$, a lever arm of $2.3$, compared with $6.8$, would raise
the wind mass loss per unit accretion by a factor of about four, so this is the
sharpest observational test of the launch prescription currently available.

\section{Discussion}
\label{sec:discussion}

\subsection{Relation to existing combined models}

Numerical models that combine an MHD wind with photoevaporation already exist.
\citet{kunitomo2020} first evolved a one-dimensional disc under both a magnetically
driven wind and a photoevaporative wind, finding that the former clears the inner
disc early and the latter the outer disc late. \citet{weder2023} built a population
synthesis with magnetic braking, an MHD wind, and internal/external
photoevaporation, including a semi-analytic prescription for the shielding of EUV
photoevaporation by the MHD wind. \citet{coleman2024} combine the \citet{tabone2022} operator with
\citet{picogna2021}-type internal and \textsc{fried}~v2 external photoevaporation,
and state explicitly that they treat the MHD and photoevaporative winds
``entirely separably'' given the uncertainty in how the two interact. The AGE-PRO
synthesis \citep{tabone2025} does not combine them at all: it contrasts two
mutually exclusive populations, turbulence-driven accretion with
\citet{picogna2021} internal photoevaporation, against MHD disc wind evolution alone.

Where these models do include both, they add them as spatially separate
processes. 
\citet{weder2023} argue that the addition is a simplification, and that
a cold MHD wind and photoevaporation are physically distinct in kind,
so a single wind of an intermediate character should be expected instead.
The present construction is an attempt at that wind. The launch is genuinely
magnetothermal. The same irradiation sets
the wind-base thermodynamics and drives the late-phase thermal wind, and the two
are rationed by magnetisation instead of summed.

\subsection{Relation to existing Green's-function solutions}

Green's functions for the homogeneous magnetically-driven-wind operator have been
derived independently. \citet{shadmehri2024} obtained analytic time-dependent
solutions, and a Green's-function solution in the same spirit has been presented in
the tidal-disruption context \citep{tamilan2026}. The homogeneous kernel is
therefore not new. The novel elements here are (i) the photoevaporation-sink
convolution along the wind characteristics (equation~\ref{eq:greens}); (ii) the
coupling to a magnetothermal launch, so that $(\xi,\lambda,\aDW)$ follow from a
physically set base temperature instead of being free, together with the launch
switch $S(\beta)$, which distributes one wind between its magnetothermal and
thermal channels instead of summing two: the operator's own coefficient
carries the magnetisation, and the thermal sink enters the convolution weighted
by $[1-S(\beta)]$; and (iii) the matched
primordial-to-dispersal transition of Section~\ref{sec:validity}, in which the
gap-opening radius follows from the resupply balance of equation~(\ref{eq:gap})
instead of being identified with the classical gravitational radius.

\subsection{Caveats}
\label{sec:caveats}

\subsubsection{The isothermal launch proxy}

The launch physics is compressed into a single isothermal base sound speed (equation~\ref{eq:csw}).
A rigorous treatment would enforce radiative equilibrium along the streamline, making the flow
non-barotropic (e.g., the entropy-deposition model of \citealt{casse2000b}).
We treat the \citet{bai2016} ideal-MHD slow magnetosonic point not as a solver limit, but as an
internal diagnostic of the magnetothermal-to-thermal boundary.

Applying the self-consistent $\Xi$--$T$ prescription (Section~\ref{sec:launch}) only where $z_0$
reaches the true irradiation front keeps the sustained-accretion lifetime within $4\%$ of a
strictly cold-base launch.
Assuming full ionisation ($\mu=0.65$) at that base would produce an artificial hot-base
acceleration. We adopt molecular $\mu=2.34$ across the internally irradiated base.
While the blend parameter $f_{\mathrm{hot}}$ can explore hotter bases
(Appendix~\ref{app:windbase-robustness}), the wind-base aspect ratio crosses $h\sim0.3$ once
$f_{\rm hot}\gtrsim0.5$. Since the \citet{tabone2022} operator is derived
for a thin, subsonic disc, that sweep is meaningful only to $f_{\rm hot}\lesssim0.25$.

\subsubsection{Prescribed flux evolution and the missing cavity}

The secular flux evolution remains unsolved, parametrised entirely by $\tau_B$ and the
unconstrained field-divergence index $q_{\mathrm{div}}$.
The assumed closure dictates the qualitative outcome: it selects the dispersal branch
instead of deriving it.
While \citet{martel2022} showed that a flux-retaining cavity sustains accretion,
our vertically integrated model does not reproduce it.

Our flux-retained branch does clear from the inside out, and does open a cavity
(Section~\ref{sec:outerclear}), but not an accreting one. The cavity appears only
at $t=3.74$~Myr, by which point $\Mdot_\ast$ has already fallen to $1.2\times10^{-9}$ from
a plateau of $3.0\times10^{-9}\,\Msun\,\mathrm{yr}^{-1}$, and the branch collapses at
$6.25$~Myr. The cavity is a late feature of a disc on its way out, not a long-lived
accreting structure.
The \citet{tabone2022} operator imposes a subsonic radial velocity ceiling
($v_{\mathrm{DW}}\sim\alpha_{\mathrm{DW}}h c_{\mathrm{s}}\ll c_{\mathrm{s}}$) that
cannot capture the transsonic radial flows expected in evacuated cavities.
Reaching such flows would in any case require the inner disc to become strongly
magnetised. With our adopted depletion scaling ($\beta \propto \Sigma^{1-2q}$ for
$q=0.3$), however, evacuating the cavity by three orders of magnitude lowers $\beta$
by only one, so it never does before the whole disc collapses.
Sustaining a steady transition-disc cavity requires a physical ``conveyor'' of poloidal
flux inward, which our model lacks. Nor does \citet{martel2022} establish that such a
conveyor persists on secular timescales.

Whether the absent cavity is a defect of the model or of the parameters we adopt
therefore has two answers.
The subsonic ceiling follows from the form of the \citet{tabone2022} operator, and no
choice of $(\psi,q,\beta_0,\tau_B)$ removes it, so on that count the model cannot produce
a transsonic cavity flow at any point in its parameter space.
The magnetisation threshold, by contrast, is parametric.
Because $\beta\propto\Sigma^{1-2q}$, the decade of $\beta$ bought by three decades of
evacuation at $q=0.3$ becomes three decades at $q=0$, and that endpoint is precisely
where this model inherits the \citet{armitage2013} and \citet{martel2022} accretion runaway.
A cavity may well form elsewhere in $(q,\beta_0,\tau_B)$.

\subsubsection{Internal photoevaporation and the screening proxy}

Internal photoevaporation operates exclusively inside the screening radius
$R_{\mathrm{c}}(t)$ \citep{picogna2026}.
We diagnose this cut-off from the mass radius, a proxy for the aspect ratio $H/R$.
Its expansion as the disc drains is robust, but the timescale is not. The proxy
depends only on $\Sigma$ and ignores temperature, so lowering $T_0$ from $280$ to
$150$~K flattens $H/R$ by $\sim 25\%$ and leaves $R_{\mathrm{c}}$ unchanged.
A physical shadow would shrink instead, accelerating outer-disc clearing.
Tracking $H(R,t)/R$ dynamically is the required next step.

\subsubsection{Vertical symmetry and multi-dimensional feedback}
These calculations are vertically integrated and symmetric about the mid-plane by
construction. Two results from multi-dimensional non-ideal MHD bear on that.

First, the sign of $\mathbf{B}_z\cdot\mathbf{\Omega}$ breaks the symmetry.
\citet{mori2025} find fast one-sided surface accretion, at $\sim10$ per cent of
the Keplerian speed, and correspondingly asymmetric winds when the field is
aligned with the disc angular momentum (the wind failing on the side that
carries the accretion layer), while anti-aligned fields give more nearly
symmetric outflows. \citet{sarafidou2024} likewise find polarity-dependent
structure in Hall-MHD. A vertically integrated treatment returns the average over
the two surfaces and cannot represent that asymmetry. Since the polarity of any
individual disc is observationally unconstrained (Section~\ref{sec:flux}), the
results here should be read as that average, as opposed to either surface.

Second, the wind reacts back on the irradiation front that defines its own base.
\citet{mori2025} find that the outflow enhances radiative heating by elevating
that front. Our base height $z_{\mathrm{0,th}}$ is set by exactly this front,
through the penetration column of equation~(\ref{eq:Nrad}), so the feedback acts
directly on the quantity the base prescription computes, and is absent here.
Relatedly, the thermal assistance in this model is evaluated locally at each
footpoint, whereas in multi-dimensional calculations radiative transfer couples
neighbouring streamlines \citep{gressel2020}. The base-temperature sweep of
Appendix~\ref{app:windbase-robustness} bounds the sensitivity to the base
temperature itself, but not to a front whose height the wind has moved.

\subsubsection{Scope of the closures}

The result that $q \le 1/2$ prevents demagnetisation assumes a fixed mid-plane
temperature.
Because $\beta \propto \Sigma^{1-2q}T^{1/2}$, an optically thinning cavity that heats
up sharply could offset the $\Sigma$ dependence and force demagnetisation even for
$q \le 1/2$.
Separately, the closed-form characteristics (Section~\ref{sec:char}) assume
$\psi \gg 1$. The numerical integrations retain the full operator, so this restricts
only the analytic approximation, not the integrated results.

\subsubsection{The unverified coupling criterion}

The wind base is placed where the field reaches equipartition,
$\beta_{\mathrm{couple}}=1$, and the model never evaluates whether the gas there
is in fact coupled to the field. No Elsasser number (ambipolar, Ohmic or Hall)
is computed anywhere in the integration. The non-ideal terms enter only
through this geometric proxy and through the prescribed $\tau_B$.
The proxy itself is well supported. \citet{mori2025} find wind launching at
$\beta\sim1$ in global simulations carrying all three non-ideal terms with
radiation transport, and equation~(\ref{eq:z0mag}) reproduces
\citeauthor{baistone2013}'s measured base height to $2.3$~per cent
(Section~\ref{sec:bridge}). What is assumed is not the criterion but its
persistence: those calculations run for $\sim10^{2}$--$10^{4}$~yr, and we
apply their launching surface unchanged over $10^{7}$.
This has a sharper consequence than a missing diagnostic. Because
$\beta_{\mathrm{couple}}=1$ defines the magnetic base, wherever the base
is magnetically set one has $\beta_{\mathrm{base}}=1$ identically. The
plateau $\lambda\simeq6.8$ of Section~\ref{sec:leverarm} follows from that
definition, not from the wind solution. The cold-launch lever arm is
therefore a restatement of the coupling criterion, not an independent prediction
of it. A coupling height set instead by $\Lambda_{\mathrm A}\gtrsim1$ would
move the plateau, though, as shown there, not the contrast between the
branches, which is carried by the magnetothermal annulus where
$\beta_{\mathrm{base}}$ leaves unity.
The requirement can be checked externally instead of
internally. \citet{sarafidou2025} measure $\Lambda_{\mathrm A}\simeq1$--$8$ over
$1$--$15$~AU in a disc of this type, marginal inside $\sim2$~AU, and
\citet{mori2025} likewise launch from gas that is only marginally coupled,
$\mathrm{Am}\gtrsim1$. Both checks are borrowed, and both are taken at one epoch
of a disc that is not evolving: the coupling of a draining disc is precisely
what a non-ideal calculation would have to follow, and is the natural next
constraint on this closure.
The machinery for it exists. \citet{delage2022} compute Ohmic and ambipolar
diffusivities self-consistently from an ionisation network, with dust chemistry
and irradiation heating, in a $1{+}1$D vertically resolved disc.
\citet{steiner2025} carry the same class of calculation (thermal ionisation
together with cosmic rays, stellar X-rays and radioactive decay, feeding Ohmic
and ambipolar diffusivities) into the transport of a large-scale
field, evolving the induction equation for the poloidal flux instead of
prescribing it. Both calculations are steady-state. The first is MRI-driven,
as opposed to wind-driven, and the second explicitly omits the toroidal field
above the disc surface so as not to launch a wind, deferring outflows to a later
paper in that series. The combination this model would need, a coupling
height derived from a computed Elsasser number, carried through a secular
wind-driven evolution, therefore does not appear to have been done. It is a
concrete next step rather than an aspiration.

\subsubsection{Absolute normalisation of the accretion rate}

The fiducial disc mass ($0.02~M_\odot$) and stress ($\alpha_0=10^{-3}$) produce an
early accretion rate ($\sim 6\times10^{-9}~M_\odot~\mathrm{yr}^{-1}$) below typical
Class II observations.
However, since $\dot{M}_{\mathrm{acc}}$ scales linearly with $\alpha_0 M_{\mathrm{disc}}$,
our core findings are dimensionless and invariant to this vertical offset
(Appendix~\ref{app:highmdot}).
The scaling is bounded by the accretion timescale
($t_{\mathrm{acc}}=M_{\mathrm{disc}}/\dot{M}_\ast$), however: rates
$\sim 10^{-6}~M_\odot~\mathrm{yr}^{-1}$ would drain the disc faster than flux-loss
mechanisms can operate.

Similarly, the initial accretion rate scales linearly with the mid-plane
normalisation $T_{\mathrm{mid},0}$, not with the base temperature, which
Appendix~\ref{app:windbase-robustness} shows the launch to be insensitive to.
Cooling the disc delays dispersal and widens the separation between the branches
by roughly three quarters over $280\rightarrow200$~K, so a colder disc sharpens
the two-branch divide instead of blurring it.
However, these are upper bounds, as a realistic mid-plane temperature would evolve
with the disc's optical depth and would lead to a flatter profile,
accelerating the expansion of the screening radius and the clearing of the disc.

\subsubsection{The initial condition and the embedded, gravitoturbulent phase}

\textbf{We start from an already-viscous Class~II disc; its initial surface density is
imposed, not evolved from the earlier embedded phase. \citet{redkin2026} find
gravitational torques dominate that phase, with magnetic winds and viscosity taking over
only later, so our fiducial $\Sigma_0(R)$ and $M_{\mathrm{disc}}$ should be read as an
assumed hand-over state rather than a prediction of it.}

\subsubsection{External photoevaporation and the imposed thermal structure}

Every calculation presented here describes an isolated disc.
The FUV field of the surrounding cluster enters neither as a mass-loss sink beyond
a truncation radius nor as a heating term at the outer wind base.
That is not the situation of most discs, which form in regions where the ambient
field is elevated by orders of magnitude and where external irradiation is known to
truncate discs and shorten their lives \citep{haworth2018,winterhaworth2022,haworth2023}.
Whether a thermal wind then sweeps the residual disc is itself debated
\citep{owen2013,haworth2016}, so we treat it as a possible, not a guaranteed, terminal
channel.
The extension is quasi-linear in $\Sigma$ and therefore tractable within the same
Green's-function framework.
We defer it, with its numerical integration and its validation against the
\textsc{fried} grid, to a companion paper (Picogna et al., in prep.).

Two things make it substantial.
The \textsc{fried} rates are computed for unmagnetised, quasi-spherical outflows, so they
cannot simply be added to a magnetothermal wind that is already removing the same gas.
They have to be apportioned against it by the launch switch, exactly as the internal sink
is. And beyond $R_{\mathrm{c}}$ an external field would set the base temperature
while nothing in this treatment sets the base height there.
Equations~(\ref{eq:z0})--(\ref{eq:z0mag}) locate $z_0$ from the stellar X-ray penetration
column through a cold mid-plane structure, so photodissociation-region temperatures would
be applied at an altitude derived for a shielded, colder layer. Locating the physical base
under external irradiation needs an FUV penetration model that this vertically integrated
treatment does not carry, and that is a prerequisite as opposed to a refinement.

The second is the mid-plane temperature. We impose
$T_{\mathrm{mid}}=T_0(R/\mathrm{AU})^{-1/2}$ with $T_0=280$~K, held fixed in time. Only
the wind-base temperature evolves. The consequences of that
normalisation, and of freezing it, are bounded in the preceding subsection. What
those bounds do not supply is an external check, and the two benchmarks available separate
on exactly that point. Against \citet{sellek2024}, who likewise prescribe a temperature,
the agreement is excellent (Appendix~\ref{app:sellek2024-validation}: $0.6$ and $4.7$~per
cent in disc mass and accretion rate over $9$~Myr). Against a calculation that instead
solves the mid-plane temperature from irradiation, viscous heating and realistic
opacities \citep{coleman2021,haworth2023}, our disc drains viscously about twice as fast.
In comparisons we have run but do not present here, the accretion channel removes
$39\,M_{\mathrm J}$ over $4$~Myr against that calculation's $19$, and halving $T_0$ closes
almost all of that particular gap.

It does not, however, rescue the comparison as a whole, and the way it fails is the
reason we do not claim the external sink is validated. Repeating that benchmark at
$T_0=140$~K across all three ambient fields raises every one of its twelve
time-to-mass-threshold ratios, by factors of $1.4$--$1.7$, and more than doubles the
largest departure from the reference instead of reducing it
($\max|t_{\mathrm{model}}/t_{\mathrm{ref}}-1|$ rises from $0.44$ to $1.00$).
It also widens the spread among the three $G_0$ branches at every threshold.
A residual that varies systematically with $G_0$ cannot be produced by a mid-plane
temperature, which knows nothing about the ambient field.
The discrepancy is therefore not a single thermal offset, and no uniform choice of
$T_0$ removes it.
Either the sink implementation carries a field-dependent error of its own (the
mass-proportional partition, the treatment of cells falling outside the tabulated 
\textsc{fried} domain), or the reference's thermal structure is itself a function of
$G_0$, which this model has no way to represent.
Both readings point the same way: the external sink cannot be pronounced correct on the
strength of this benchmark, and could not be even with the mid-plane temperature
repaired.

These two limitations compound instead of adding. Any outer-disc sink competes for the same
gas as accretion. A model that accretes too quickly therefore leaves such a sink less
material and less time to act on, and would understate its importance. Adding external
photoevaporation to the present thermal structure would therefore not settle how much the
environment matters. The two have to be done together. That is the single clearest
direction in which this model should next be extended.

\subsection{Magnetically sustained photoevaporative gaps}

A further implication concerns the wind inside a forming photoevaporative gap.
Because the flux closure adopted here, $B_z\propto\Sigma^q$ with $q<1/2$, lets the
field decline more slowly than the gas, a gap carved by internal photoevaporation
becomes progressively more magnetised as it drains. $\beta\propto
(\Sigma/\Sigma_0)^{1-2q}$ falls with depletion, and $z_{0,\rm mag}$ drops below
$z_{0,\rm th}$, so the base reverts to a magnetothermal launch within the gap. The
residual gas can therefore keep driving a wind where a purely thermal treatment, in
which the outflow loses its driving as the gap empties, instead finds it choking
inside the gap \citep{weber2026}.

Whether this operates while the gap is still forming (and so could sustain
dispersal where a thermal-only wind stalls) turns out to depend on the branch,
because two competing terms set $\beta$. Depletion lowers it as
$(\Sigma/\Sigma_0)^{1-2q}$, while flux loss raises it as $\exp(t/\tau_B)$. When the
flux is retained the first term acts alone, and the magnetothermal region expands
into the draining disc at moderate depletion. Measured on the fiducial
flux-retained run, it grows monotonically from $47$ to $83$~AU over $5$~Myr, while
the gas at $3$~AU still holds $5$ per cent of its initial surface density. On the
flux-loss branch the same mechanism is defeated by the flux loss that defines the
branch. The magnetothermal region instead recedes to $\sim3$~AU by $3.5$~Myr, and
recovers only to $\sim6$~AU by $5$~Myr, in gas already depleted to
$\Sigma/\Sigma_0\sim10^{-13}$, that is, once the gap is empty, as opposed to while
it forms. A magnetically sustained gap is therefore a prediction for discs that
keep their flux, not a route by which a demagnetising disc rescues its own
dispersal.

Both statements are properties of the closure as much as of the model. The present
vertically-integrated treatment does not resolve the wind's vertical structure, so
whether a gap so magnetised actually launches is not settled here. A
two-dimensional non-ideal MHD calculation with photoevaporative heating
\citep[as in][]{sarafidou2024,sarafidou2025} would test it directly.

\section{Conclusions}
\label{sec:conclusions}

We presented a two-phase model that couples a magnetothermal wind, launched from an
irradiated base, to the secular evolution of a wind-dominated disc.
The central result is structural.
The standard flux-scaling closure $B_z\propto\Sigma^q$ (equation~\ref{eq:betaclosure})
is self-limiting: for $q\le1/2$, $\beta/\beta_0=(\Sigma/\Sigma_0)^{1-2q}$
(equation~\ref{eq:qomega}) cannot rise as the disc drains.
Gas depletion alone therefore cannot demagnetise a disc.
Photoevaporation-dominated disc dispersal is not automatic; it requires an
independent flux-transport dichotomy.
Efficient flux loss ($\mathcal{R}_m\lesssim1$) opens a photoevaporative gap.
Flux retention ($\mathcal{R}_m\gg1$) instead sustains accretion and defers dispersal.
Integrating numerically the two branches we find:

\begin{enumerate}
\item For a given initial magnetisation
(Fig.~\ref{fig:integrator}; Section~\ref{sec:bifurcation}), the flux-retained branch's
accretion rate collapses at $t\approx6.25$~Myr; the flux-lost branch's collapses at
$t\approx3.51$~Myr.
The split is a consequence of the magnetisation front $R_\beta$ evolution, which expands
from $47$ to $140$~AU on the flux-retained branch while it recedes to $3.6$~AU on
the flux-lost one.

\item $\mathcal{R}_m$ sets not only when the disc disperses, but in what state
(Section~\ref{sec:bifurcation}).
On the flux-retained branch, as the gas disperses it stops accreting.
On the flux-lost branch the two decouple -- accretion stops at $t\approx3.51$~Myr,
while the disc remains detectable for several Myr as a non-accreting transition disc.
This is a testable population-level robust prediction of the $\mathcal{R}_m$
separation.
However, the presence of an external FUV field might reduce the difference between the
two branches (see Section~\ref{sec:caveats}).

\item Setting the wind base from stellar irradiation does not break the
cold-launch assumption (equations~\ref{eq:epsb} and~\ref{eq:adwmatch};
Appendix~\ref{app:windbase-robustness}).
The base stays well below escape ($\varepsilon_{\mathrm b}\le0.16$ pre-collapse).
The torque is evaluated at the mid-plane sound speed (equation~\ref{eq:adwmatch}),
so the base temperature does not affect the accretion rate directly.
It reaches the launch only through the mass loading $\xi(\beta_{\mathrm{base}})$,
whose effect is important only in the magnetothermal band, which is limited in
time and space.
This holds for an equipartition-set coupling height specifically ($\beta_{\mathrm{base}}=1$).
An ambipolar Elsasser-number criterion would not pin $\beta_{\mathrm{base}}$, and
supplying one is the natural next step (Section~\ref{sec:caveats}).

\item The derived lever arm $\lambda=f(\beta_{\mathrm{base}})$ sits on a single
plateau ($\lambda=6.79$, $\xi=0.086$) wherever the base is magnetically set,
on both branches, for most of their lives (Fig.~\ref{fig:leverarm};
Section~\ref{sec:leverarm}).
The branches separate in the spread of $\lambda$, not in its typical value.
On the flux-retained branch the mass-weighted $\langle\lambda\rangle$ falls from
$6.7$ to $4.3$ by $5$~Myr as the magnetothermal shoulder widens and deepens.
On the flux-lost branch it stays within $2$~per cent of the plateau for the first
$2$~Myr and departs only in the final $0.7$~Myr.
Holding $\lambda$ fixed at the plateau value leaves the sustained accretion rate
unchanged on both branches, but delays the terminal collapse -- $5.8$~per cent on the
flux-retained branch, $1.6$~per cent on the flux-lost one.
Measured against \citet{kim2026}'s directly inferred $\lambda\simeq2.3$ for HOPS~358,
our launching-region value is a factor of three larger.
Since $\xi=1/[2(\lambda-1)]$, this is the sharpest quantitative test the launch
prescription currently faces.

\item Both branches clear inside-out, but by different mechanisms 
(Fig.~\ref{fig:integrator}, top panels; Section~\ref{sec:outerclear}).
Every depletion contour is anchored at the inner boundary and sweeps outward,
because the sinks are stronger in the inner disc.
On the flux-lost branch, internal photoevaporation alone disperses the disc down to
$5$~per cent of the initial mass by $7.75$~Myr, with no external field required.
The process self-accelerates when the cavity wall is directly irradiated by the
central star, increasing by an order of magnitude the local mass-loss rate 
(Appendix~\ref{app:cavitype}), and expanding the wall $3.2$ times faster than the
unperturbed profile would predict.
On the flux-retained branch the outer disc is instead cleared almost entirely by
the advance of the expanding magnetothermal wind (internal photoevaporation accounts
for only $7$~per cent of the removal by $5$~Myr): the same front that strips the
outer disc is the one whose collapse ends accretion, so inner and outer disc are
exhausted within a fraction of a Myr of each other, compared with the prolonged
transition phase the flux-lost branch shows (point~2).
Clearing inside-out fixes where the front is anchored, not how far behind the
outer disc lags.
That lag is set separately by whichever process reaches the outer disc, and it is
short here only because the wind itself is that process.

\item We do not recreate an accreting cavity (Section~\ref{sec:bifurcation}).
\citet{martel2022} and \citet{sarafidou2025} associate the flux-retained regime with a
long-lived, wind-fed cavity.
Here, the cavity that does open forms at the innermost grid cell and never separates from the star.
The wind has no viscous outer reservoir left to feed it once $R_\beta$ has screened the outer disc.
\citet{armitage2013}'s imposed-boundary experiment differs in exactly this respect: its outer
disc stays viscous, and does sustain a fed cavity.
How a real disc behaves depends on the residual turbulent stress beyond $R_\beta$, a quantity
neither calculation constrains.

\end{enumerate}

Every result above holds at a single point in parameter space: one $(q,\beta_0,\tau_B)$,
one stellar mass and X-ray luminosity, one disc mass and
$\alpha_{\mathrm{SS}}/\alpha_{\mathrm{DW}}$. Several of these fixed parameters set the
answer's normalisation instead of being passive choices: $M_{\mathrm d}$ and $\alpha_0$
set the absolute accretion rate, $T_{\mathrm{mid},0}$ additionally sets a timescale, and
$\beta_{\mathrm{crit}}$ moves the initial front by a factor $4.6$ per decade.
What is guaranteed to survive a systematic sweep is the structural identity
$\xi=1/[2(\lambda-1)]$, which forces the secular characteristics to recover the
steady-wind profile $\Sigma\propto R^{\xi-1}$ (equation~\ref{eq:sigprofile}).

Three developments would move beyond this parameterised dichotomy. The first is solving
the poloidal induction equation (equation~\ref{eq:fluxtransport}) directly, to replace the
imposed flux-decay timescale $\tau_B$ with an $\mathcal{R}_m$ set by disc microphysics.
The second is external photoevaporation together with a solved mid-plane temperature;
neither is present here, and the two interact, since a disc that accretes too fast leaves
an environmental sink less mass to remove.
The third is a population synthesis testing the $\mathcal{R}_m$ divide against the observed rarity and longevity ($\gtrsim10^6$~yr) of
mm-bright, high-accretion transition discs \citep{owen2016}. Such a synthesis would also
confront \citet{armitage2013}'s own demographic prediction -- that the spread in disc
lifetimes traces a spread in initial net flux.

\section*{Acknowledgements}

This research was supported by the Excellence Cluster ORIGINS which is funded by the
Deutsche Forschungsgemeinschaft (DFG, German Research Foundation) under Germany's
Excellence Strategy - EXC-2094 - 390783311.

For the integrator and postprocessing scripts we adopted \textsc{numpy} \citep{harris2020},
\textsc{scipy} \citep{virtanen2020} and \textsc{matplotlib} \citep{hunter2007}.

This work was carried out with the assistance of Claude Code (Sonnet 5).
It was used to help draft and edit the text of this paper, to write and document the
secular integrator and the figure-generation code.
All physical assumptions, model choices, numerical prescriptions and
interpretations are the authors' own; every quantity quoted in the text has been
checked and the authors accept full responsibility for the content of this paper.

\section*{Data availability}

The secular integrator is available from the corresponding author upon reasonable request.
The integrator output shown in Figure~\ref{fig:integrator}, together with the
transition-disc radiation-hydrodynamic runs underlying
Figure~\ref{fig:appendix-cavitykernel} (Appendix~\ref{app:cavitype}), is
archived at Zenodo \url{https://zenodo.org/records/22767933}.

\bibliographystyle{mnras}
\bibliography{discern}

\appendix

\section{Wind-base temperature sensitivity}
\label{app:windbase-robustness}

The wind-base temperature, prescribed by our self-consistent $\Xi$--$T$ prescription
(Section~\ref{sec:launch}), assumes a molecular mean weight of $\mu=2.34$, and produces
an effectively cold wind base.

\begin{figure*}
    \centering
    \includegraphics[width=\textwidth]{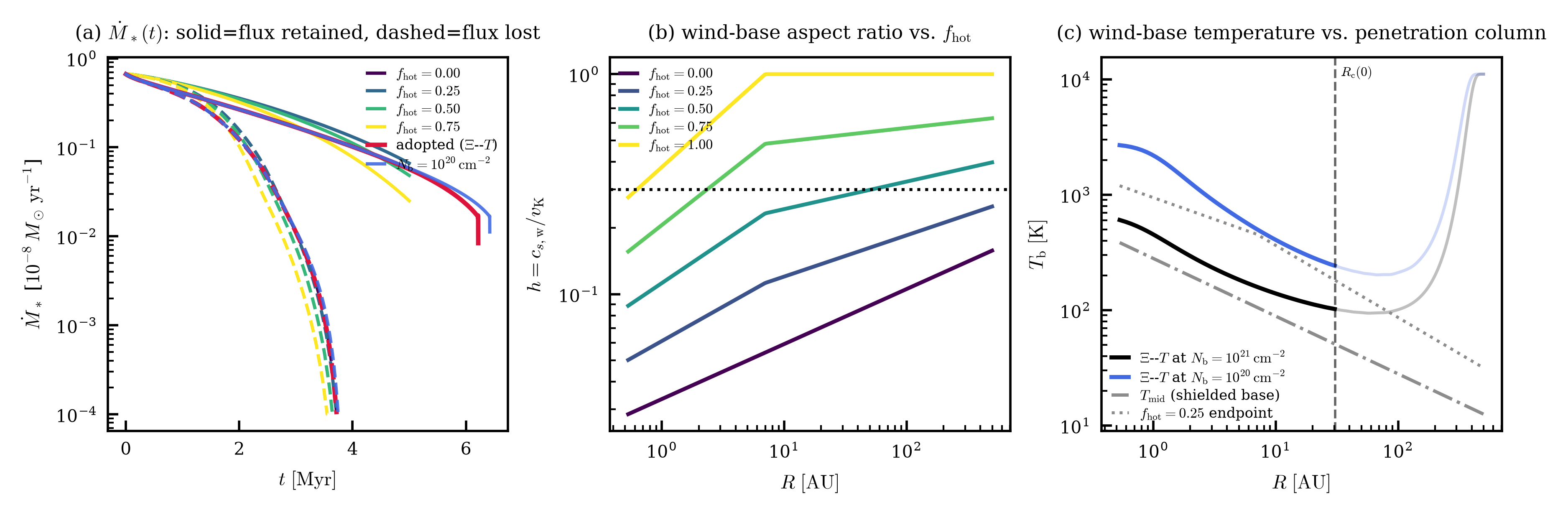}
    \caption{Wind-base temperature sensitivity. \emph{(a):} stellar accretion rate
    $\dot M_\ast(t)$ for both branches (solid: flux retained, dashed: flux lost),
    swept over $f_{\rm hot}\in\{0,0.25,0.5,0.75\}$ (colour), blending the wind-base
    sound speed between the cold mid-plane and capped escape-temperature
    endpoints. Each run stops at the first crossing of
    $\dot M_\ast<10^{-12}\,\Msun\,\mathrm{yr^{-1}}$ or at $t=5$~Myr. The adopted
    $\Xi$--$T$ prescription (red) lies essentially on the cold
    $f_{\rm hot}=0$ curve. Blue repeats the adopted prescription with the X-ray
    penetration column moved one decade, to
    $N_{\mathrm b}=10^{20}\,\mathrm{cm^{-2}}$ (run to $7$~Myr, since that is
    where the branches separate). Despite the much hotter base shown in
    \emph{(c)}, the accretion histories are nearly indistinguishable.
    \emph{(b):} the resulting wind-base aspect ratio
    $h(R)=c_{s,\rm w}/v_{\rm K}$, extended to the diagnostic-only
    $f_{\rm hot}=1$ endpoint; the dotted line marks $h=0.3$.
    \emph{(c):} the wind-base temperature given by the \citet{sellek2024}
    $\Xi$--$T$ relation at the penetration column
    $N_{\mathrm b}=10^{21}\,\mathrm{cm^{-2}}$ that defines the base (black),
    against the cold mid-plane and the
    $f_{\rm hot}=0.25$ endpoint (grey). Blue repeats the $\Xi$--$T$ relation at
    $N_{\mathrm b}=10^{20}\,\mathrm{cm^{-2}}$: less penetration puts the front
    higher, where the density is lower and $\Xi$ larger, and the base runs
    $2$--$5$ times hotter, the input whose consequences panel \emph{(a)}
    measures. The $\Xi$--$T$ curves are faint beyond
    $R_{\rm c}(0)$, where it is not applied: its rise there is the outer taper's
    vanishing density inflating $\Xi=L_{\rm X}/(n r^2)$, not a predicted
    $10^4$~K base. The jagged tails in (a) and the thresholds in (b) and (c) are
    discussed in the text.}
    \label{fig:appendix-windbase}
\end{figure*}

As shown in Figure~\ref{fig:appendix-windbase}a, the fiducial run mirrors a strictly
cold mid-plane launch. The peak accretion rates agree to better than one part in
$10^{3}$ ($6.644$ against $6.645\times10^{-9}\,\Msun\,\mathrm{yr^{-1}}$), and the
dispersal epoch to $0.03$~per cent on the flux-loss branch ($3.719$ against
$3.720$~Myr).
On the flux-retained branch the cold run reaches the $5$~Myr sweep horizon
without collapsing.
Using the molecular $\mu$ lowers the base sound speed squared ($c_{\mathrm{s,w}}^2$)
by a factor of $3.6$ relative to a fully ionised gas, which would otherwise produce a
hot-base acceleration.

The relation itself is the fit of \citet[][their eq.~(14)]{sellek2024},
\begin{equation}
\log_{10} T_{\mathrm b}=d+\frac{a-d}
{\left[1+\left(\log_{10}\Xi/c\right)^{b}\right]^{m}} ,
\label{eq:xitfit}
\end{equation}
for which we adopt their Table~2 $T_{\rm gas}>T_{\rm dust}$ parameter set,
$(a,b,c,d,m)=(1.1960,-2.8975,-0.1658,4.0450,1.0374\times10^{4})$, instead of their
all-cells set, since it is the irradiation-heated gas that sets the wind base.

This cold-base conclusion relies on three safeguards:
\begin{enumerate}
    \item Screening: The ionisation parameter $\Xi$ is screened beyond the cut-off
     radius $R_{\mathrm{c}}(t)$, since $\Xi$ otherwise assumes an unshadowed line of sight
     to the star that \citet{picogna2026} argue breaks down there.
    \item Outer-disc reversion: Beyond $R_{\mathrm{c}}(t)$, the base reverts to
     the cold mid-plane temperature.
    \item Physical intercept: The heated base temperature is applied only where the
     magnetic base physically reaches the irradiation front
     ($z_0=z_{\mathrm{0,th}}$, equation~\ref{eq:Nrad}).
     Applying the hot temperature uniformly across the inner disc triggers an
     unphysical, rapid global draining.
\end{enumerate}

The isolated runs are not given a corresponding ambient interstellar field, even though
every disc sits in one.
Applying that field's heating without a matching mass-loss sink would be inconsistent.
The outer wind base in Section~\ref{sec:results} should therefore be read as a lower bound.

We test this outcome against the assumed penetration column, lowering it
from $N_{\mathrm{b}}=10^{21}$ to $10^{20}\;\mathrm{cm}^{-2}$.
A lower column shifts the irradiation front higher up, where lower density drives $\Xi$ up:
the median $\Xi$ over the disc rises by a factor $7.2$ and the base temperature by factors of
$2$--$5$ (Figure~\ref{fig:appendix-windbase}c).
However, this elevated front is intercepted by the magnetic base over a much narrower radial
extent: at $t=0$ from $R>67$~AU, compared with $R>40$~AU.
The two effects cancel almost perfectly out.
The accretion rates over $0.5$--$3$~Myr remain unchanged to three significant digits on both
branches as shown in Figure~\ref{fig:appendix-windbase}a.
The dispersal epochs do not match exactly.
On the flux-loss branch the apparent $+1.5$ per cent shift is the criterion sensitivity of
Section~\ref{sec:bifurcation} as opposed to a physical delay.
On the flux-retained branch, however, the hotter base delays collapse, from $5.45$ to
$6.80$~Myr.
Radial confinement of the heated region outweighs its higher temperature, so a warmer base
over a narrower annulus is not a route back to hot-base acceleration.
The cold-base outcome does not rest on the choice $N_{\mathrm{b}}=10^{21}\,\mathrm{cm^{-2}}$.

Numerically, coupling the local surface density to the base temperature via the
\citet{sellek2024} $\Xi$--$T$ fit requires two regularisations to maintain a well-posed secular
integration.
First, we evaluate $\Xi$ at a smoothed $\log_{10}\Sigma$ (Gaussian, $\sigma\sim5$ grid cells).
The very large exponent $m$ makes equation~(\ref{eq:xitfit}) extremely steep near the
transition, so grid-scale noise in $\Sigma$ would otherwise be amplified into a spurious sharp
front.
Second, we apply the recombination-time relaxation (Section~\ref{sec:bridge}) to prevent a
positive-feedback loop (hotter base $\to$ faster mass loss $\to$ lower $\Sigma$ $\to$ hotter
base) that would otherwise collapse the explicit timestep.

Finally, to bracket the base temperature independent of the $\Xi$--$T$ relation, we use a
simpler log-space blend between the cold mid-plane temperature $T_{\rm mid}$ and
the capped escape temperature $T_{\rm esc}$ (equation~\ref{eq:tesc}) through a single
parameter $f_{\mathrm{hot}}\in[0,1]$, applied to the wind-torque sound speed and to the
magnetic-coupling height $z_{0,\rm mag}$, but not to $z_{0,\rm th}$'s attenuation
column, which describes gas the same photons have not yet heated. $f_{\rm hot}=0$ recovers
a strictly cold-base launch exactly.
The sweep is nearly flat. Peak accretion is unchanged across its whole range, $6.645$ to
$6.652\times10^{-9}\,\Msun\,\mathrm{yr^{-1}}$ from $f_{\mathrm{hot}}=0$ to $0.75$.
On the flux-loss branch, the dispersal epoch moves from
$3.720$ to $3.552$~Myr, $4.5$ per cent, over a sweep that takes the base from the cold
mid-plane to three quarters of the way to the escape temperature.

The sweep is flat because the torque is evaluated consistently at the mid-plane sound speed,
and the base temperature acts only through the mass loading $\xi(\beta_{\mathrm{base}})$.
The residual $4.5$ per cent is that surviving channel, integrated over the disc's life.
These winds are effectively cold, and an hot base is not a route to rapid dispersal in this
model.
The flux-retained arm bounds rather than measures: all four of its runs reach the $5$~Myr
horizon without collapsing, so they establish only that no $f_{\rm hot}\le0.75$ brings
that branch's collapse inside a window its fiducial ($6.3$~Myr under the same criterion)
already sits outside.
We integrate $f_{\rm hot}\in\{0,0.25,0.5,0.75\}$, stopping each run at the first crossing of
$\Mdot_\ast<10^{-12}\,\Msun\,\mathrm{yr^{-1}}$.
This parameter space is in any case bounded by the thin-disc assumption of the
\citet{tabone2022} secular operator.
As shown in Figure~\ref{fig:appendix-windbase}b, values of $f_{\mathrm{hot}} \ge 0.5$ drive
the aspect ratio $h=c_{\mathrm{s,w}}/v_{\mathrm{K}}$ above $0.3$ across most of the disc, so
the operator is valid only for $f_{\mathrm{hot}} \lesssim 0.25$.
That bound costs little since the response to $f_{\rm hot}$ is weak and it would not
change the conclusion even if they could be trusted.
Our self-consistent $\Xi$--$T$ prescription sits at the cold end of the valid regime regardless.

\section{A first test of the poloidal flux-transport closure}
\label{app:flux-transport-test}

In Section~\ref{sec:flux}, we bypassed a direct solution of the induction equation
(equation~\ref{eq:fluxtransport}) in favour of the phenomenological $\tau_B$ proxy
(equation~\ref{eq:betaclosuredecay}).
Here, we report an exploratory test of directly solving for flux transport using the
vertically-averaged ``simple reference model'' of \citet{guilet2014} (their
eqs.~17--18), with field-line advection and diffusion rates
$v_{\rm adv}=-1.5\nu/R$ and $v_{\rm diff}=\eta_{\rm eff}/H$.

Their model defines the local magnetic Reynolds number by comparing field-line
advection to diffusion:
\begin{equation}
\mathcal{R}_m(R) = \frac{|v_{\rm adv}|}{v_{\rm diff}} = \frac{3}{2}\,P_{\rm B}\,\frac{H(R)}{R},
\label{eq:rm-local}
\end{equation}
where $P_{\rm B}\equiv\nu/\eta_{\rm eff}$ is the turbulent magnetic Prandtl number and
$\eta_{\rm eff}=\nu/P_{\rm B}$ is the diffusivity entering
equation~(\ref{eq:fluxtransport}).
Equation~(\ref{eq:rm-local}) is not the same ratio as the large-scale
$\mathcal{R}_m=|v_\psi|R/\eta_{\rm eff}$ of Section~\ref{sec:flux}.
The two differ by a factor $H/R$: the radial redistribution of
$B_z$ that sets $\tau_B\sim R_{\rm disc}^2/\eta_{\rm eff}$ occurs over $R$, whereas the
competition between advection and diffusion that decides whether flux can be dragged
inward at all is governed by how fast flux diffuses vertically out of a thin
disc, over $H$ \citep{lubow1994,guilet2012}.
For $P_{\rm B}\sim1$ the large-scale form would give $\mathcal{R}_m\sim1.5$, marginally
advection-dominated. Equation~(\ref{eq:rm-local}), the form appropriate to the
thin-disc problem, gives much smaller values: for this disc's aspect ratio
($H/R\simeq0.028$ at the inner edge to $0.157$ at the outer), $\mathcal{R}_m\sim0.04$
--$0.24$ everywhere on the grid.
It is the latter that the test in this appendix is measured against, using the
order-unity threshold $\mathcal{R}_m^{\rm crit}\sim1$ implied by the $D$ parameter of
\citet{lubow1994} (their $\mathcal{R}_m=(2/3)/D$) and the turbulent magnetic Prandtl
number $P_{\rm B}\sim1$ that shearing-box MHD turbulence simulations and theoretical
expectations both favour \citep{guilet2012}.

\begin{figure}
\centering
\revisedfig{width=\columnwidth}{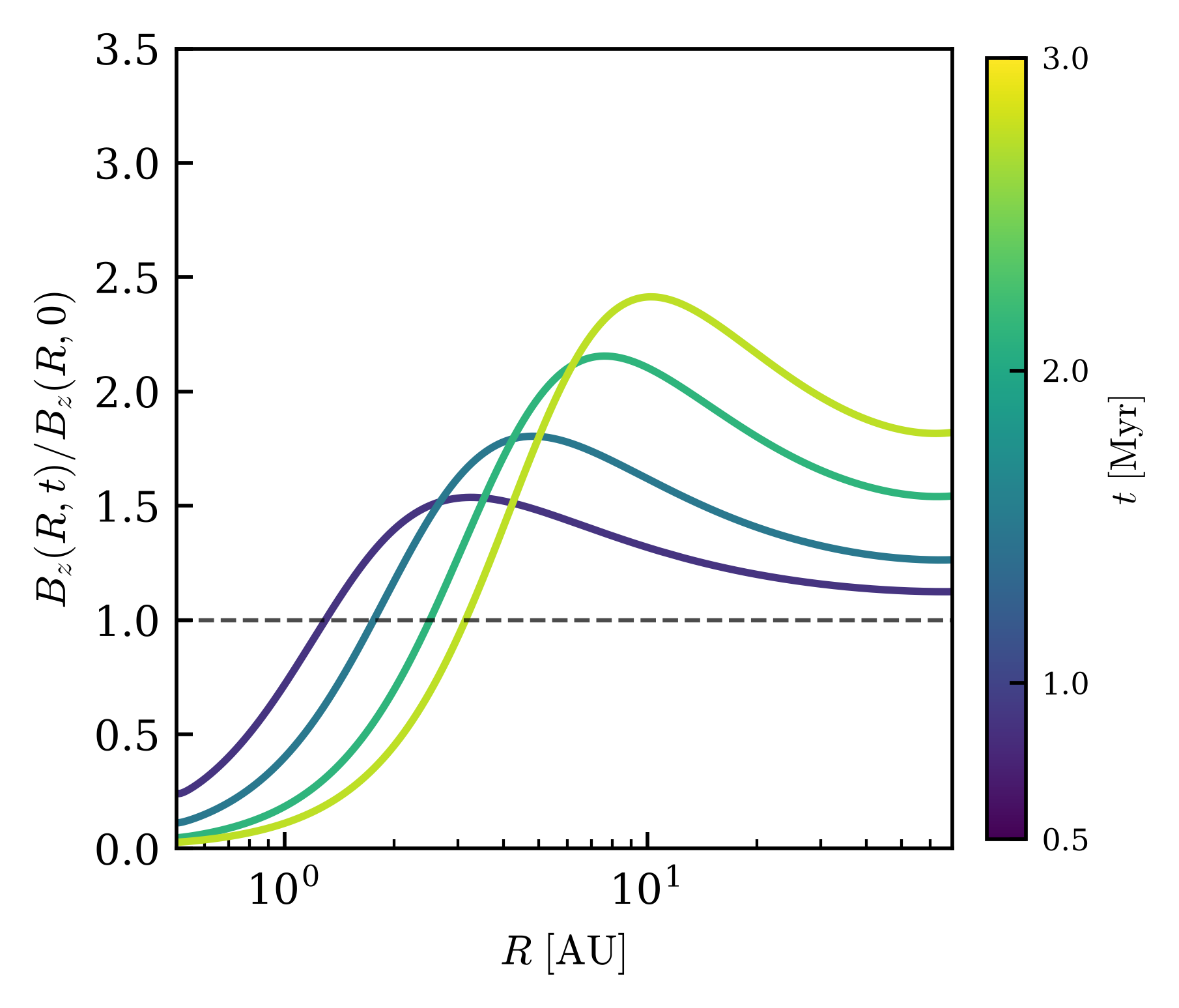}
\caption{\textbf{$B_z(R,t)/B_z(R,0)$ from directly integrating
equation~(\ref{eq:fluxtransport}) on this disc's actual $t=0$ field profile
(Pm$_{\rm B}=1$), at $t=0.5$, $1$, $2$ and $3$~Myr (colour). The dashed
line marks no change. The inner disc demagnetises (ratio $<1$) while the
field strengthens broadly outside it, peaking around $R\sim5$--$10$~AU. The
plot is cropped at $R=70$~AU, each curve's own local minimum: beyond that
radius $B_z(R,0)$ has fallen several decades below its inner value, and the
finite-difference $B_z=(1/R)\,\mathrm{d}\Psi/\mathrm{d}R$ becomes dominated
by floating-point cancellation rather than physics from $R\sim190$~AU on.}}
\label{fig:appendix-flux-transport}
\end{figure}

Integrating equation~(\ref{eq:fluxtransport}) directly on this disc's actual,
sharply radius-peaked initial field profile $B_{z,0}(R)$ does not reproduce the
outside-in demagnetisation assumed by equation~(\ref{eq:betaclosuredecay}).
The flux redistributes outward, demagnetising the inner disc while strengthening
the field over the surrounding disc\textbf{, peaking around $R\sim5$--$10$~AU
(Figure~\ref{fig:appendix-flux-transport})}.
This is not a numerical artefact (total flux is conserved to a few--$20$ per cent
over the test window), and it is not a contradiction of the literature cited in
Section~\ref{sec:flux} favouring inward flux concentration. It follows directly from
\citet{guilet2012}'s own analysis of equation~(\ref{eq:rm-local}), since
$\mathcal{R}_m\ll1$ places the whole grid in the diffusion-dominated regime.
\citet{guilet2012} note explicitly that significant field bending, and hence inward
concentration, instead requires an ``unexpectedly large'' Prandtl number,
$P_{\rm B}\sim R/H\sim6$--$36$ here, well above the shearing-box-turbulence value. They
point instead to genuinely non-axisymmetric mechanisms \citep[e.g.][]{spruit2005} as
the more likely route to the concentration seen in global simulations.

Two ingredients absent from this reference model bear directly on that conclusion, and
both act in the same direction:
\begin{enumerate}
    \item the wind torque: the reference model advects field lines solely via
     viscous inflow ($v_{\mathrm{adv}}=-1.5\nu/R$), whereas \citet{leung2019} find that
     it is a wind outflow that drives flux inward, with Hall drift significantly
     enhancing that inward transport in the positive-polarity case
     $\eta_H(\mathbf{B}_z\cdot\mathbf{\Omega})>0$, while Ohmic and ambipolar
     diffusivities drive transport outward. Coupled only to an outwardly inclined field,
     and with no wind, the Hall effect drives no flux transport at all.
     The advective side of equation~(\ref{eq:rm-local}) is therefore the one term of the
     budget the reference model most clearly understates, and the values quoted above
     should be read as a lower bound on the true advection-to-diffusion ratio, not as an
     estimate of it.
    \item the Hall effect: the model carries a single scalar diffusivity,
     $\eta_{\rm eff}=\eta_{\rm turb}+\eta_{\rm AD}$, so its transport is blind to field
     polarity, whereas Hall drift is not a diffusivity at all: it advects the field
     with the electron fluid, and its sense is set by the sign of
     $\mathbf{B}_z\cdot\mathbf{\Omega}$ \citep{bai2014,baistone2017,leung2019}.
\end{enumerate}

Adding Hall drift is not a viable route to recovering inward concentration in the
quasi-steady state.
\citet{baistone2017} find that for an aligned field the Hall
effect rapidly drags flux inward at the mid-plane (the global manifestation of
the Hall-shear instability), but that the resulting radially elongated field promotes
rapid outward transport by ambipolar diffusion. The instability therefore saturates,
and the quasi-steady rate is outward, comparable to the Hall-free case. For the
anti-aligned polarity the Hall effect transports flux outward throughout, about twice as
fast as the aligned case.
On that evidence the outward redistribution found here is qualitatively consistent with
the polarity-resolved simulations, not in tension with them, and the aligned polarity is
the one that would favour inward transport, through the wind term, not through Hall
drift acting alone.

The calculation closest to the configuration modelled in this paper is
\citet{sarafidou2024}, which is polarity-resolved and carries internal X-ray
heating. It is a 2D-axisymmetric non-ideal MHD calculation with Ohmic, Hall and
ambipolar terms plus an X-ray temperature parametrisation, run at both orientations
($M_\ast=0.7\,\Msun$, $|\beta_P|=10^5$ at the mid-plane, and
$L_X=2\times10^{30}\,\mathrm{erg\,s^{-1}}$, the same X-ray luminosity adopted here).
They confirm the polarity asymmetry in the sense described above: the two orientations
differ by about two decades in laminar accretion rate ($\simeq4\times10^{-9}$ versus
$\simeq2\times10^{-11}\,\Msun\,\mathrm{yr^{-1}}$), and the wind-launching radius moves
from $\simeq1$~au (aligned) to $\simeq2.5$~au (anti-aligned).
Their result that matters most here, however, is that the sign of the radial
displacement is not fixed by the polarity alone. Within the same aligned run they
recover the inward ``pinch'' only for $R<5$~au, and find an outward ``repulsion'' of
flux at $5<R<8$~au. They trace this to the sign of $J_\phi B_z$, and leave open whether
the convex curvature is physical or an artefact of an initial condition with vanishing $B_\phi$.

Either way, transport whose sign changes with radius within one calculation, at
fixed polarity cannot be captured by a scalar $\mathcal{R}_m$, however that scalar is
chosen. That limitation is sharper than the absence of a Hall term by itself, and
applies equally to the closure of Section~\ref{sec:flux} and to the reference model
tested here.

The net position is therefore weaker than a negative result. This test omits both terms
that drive flux inward, so its outward redistribution is what their absence predicts.
It does not adjudicate the phenomenological closure, and the sign of
$\mathbf{B}_z\cdot\mathbf{\Omega}$ (unconstrained by observation, and fixed once per
disc) remains a parameter of $\mathcal{R}_m$ that this work does not resolve.
A test that could adjudicate the closure needs the wind torque in the flux budget, a
polarity-resolved Hall term, and the vertically-resolved diffusivity structure
\citet{guilet2012} show is needed for realistic concentration even in an axisymmetric
treatment. Until then, the uniform-coefficient, wind-free, Hall-free reference model
tried here does not support replacing the phenomenological closure, which we therefore
retain.

\section{The numerical scheme, and its agreement with the closed form}
\label{app:scheme}

The numerical integrations of Section~\ref{sec:results} solve equation~(\ref{eq:tabone})
directly, with the conservative finite-volume discretisation set out below.
They do not evaluate the Green's function of equation~(\ref{eq:greens}).
That closed form rests on a background, quasi-static
treatment of the sink terms, and the point of integrating numerically is to avoid
inheriting it. The two are therefore independent descriptions of the same equation, and
this appendix checks the first against the second in the limit where the analytic
solution is exact.

\subsection{Discretisation and timestep control}

The evolved variable is the cell mass $A_j\Sigma_j$. The one term left explicit is the
internal photoevaporative source, which is local, not stiff, and enters only through
the right-hand side,
\begin{equation}
(A\Sigma)^{\rm exp}_j=A_j\Sigma_j-\Delta t\,A_j\Sigdot_{\rm PE,\,j}.
\label{eq:fvexp}
\end{equation}
Writing $a_j$ for the advective coefficient at face $j$ and $V$ for the discrete
viscous operator, the implicit update is
\begin{equation}
\left[A_j(1+\Delta t\,k_j)-\Delta t\,a_{j}\right]\Sigma^{\rm new}_j
 +\Delta t\,a_{j+1}\,\Sigma^{\rm new}_{j+1}
 -\Delta t\,(V\Sigma^{\rm new})_j
 =(A\Sigma)^{\rm exp}_j ,
\label{eq:fvimp}
\end{equation}
with $k_j$ the rate collecting the linear wind sink.
The face value in the advective term is reconstructed by first-order donor-cell
upwinding. Because the wind velocity is inward everywhere, the donor for face $j$ is
always cell $j$, so advection couples each cell only to its outer neighbour. The
viscous operator couples both neighbours symmetrically, with non-positive
off-diagonal weights. The system is therefore tridiagonal with a positive diagonal
and non-positive off-diagonals (an M-matrix), and is solved directly in $O(N)$.
That structure is what makes the solve positivity-preserving at any $\Delta t$.

The timestep is then set by accuracy alone.
For a local sink $\mathrm d\Sigma/\mathrm dt=-k\Sigma$, backward Euler returns
$1/(1+k\Delta t)$ against the exact $\mathrm e^{-k\Delta t}$, a local error
$\sim(k\Delta t)^2/2$, so requiring one per cent per step gives $k\Delta t\le0.14$.
Three further dimensionless coefficients bound the advective rate $|a_j|/A_j$, the
viscous diagonal rate, and the explicit photoevaporative removal
$\Sigdot_{\rm PE}\Delta t/\Sigma$, at $0.03$, $0.5$ and $0.1$ respectively.
The first two are set by convergence against the independently implemented
operator-split scheme, the third so that no cell loses more than a tenth of its
surface density in one step.
All four are evaluated only on cells still holding one per cent of their initial
surface density, together with their immediate neighbours, for the reason given in
Section~\ref{sec:numsetup}.

\subsection{Agreement with the closed form}

Section~\ref{sec:char} solves the same equation in closed form in the wind-dominated
limit. To enforce that limit we strip the transport down to pure wind-driven advection.
The diffusive term is removed ($\aSS=0$, i.e.\ $\psi\to\infty$), internal
photoevaporation is switched off, $\aDW$ is held constant (which the depletion closure
gives for $q=1/2$), $\xi$ is held constant, and the launch cutoff $S(\beta)$ is disabled
(otherwise $\aDW$ acquires a radial dependence and the characteristics curve). The
wind base is taken cold, so that $\cs^2$ is the pure power law the derivation assumes.
Each of these follows from the closed form itself, not from a preference about
the physics. The characteristics are then straight, $R(t)=R_i-At$ with
$A=\tfrac32\aDW\cs^2/\vK$ constant, and the homogeneous solution along them is
\begin{equation}
\Sigma(R,t)=\Sigma_0(R+At)\left(\frac{R}{R+At}\right)^{\xi-1}.
\label{eq:charsol}
\end{equation}
Equation~(\ref{eq:charsol}) is only the homogeneous kernel, so agreeing with it would
test the advection operator and nothing else. The quantity of interest is the full
equation~(\ref{eq:greens}), whose second term convolves the photoevaporative source
along the same characteristics. Along $R(t')=R_i-At'$ the integrating factor closes
analytically, $\mathcal I(t')=(R(t')/R_i)^{1-\xi}$, so equation~(\ref{eq:greens}) can be
evaluated by quadrature against the same $\Sigdot_{\rm PE}$ and the same co-moving
cut-off $R_{\rm c}(t)$ the integration uses. We therefore run the comparison twice: once
with the source off, against equation~(\ref{eq:charsol}), and once with internal
photoevaporation active, against equation~(\ref{eq:greens}). The second run uses the
additive combination in place of the fiducial suppressed one, because
$\beta_{\rm crit}\to\infty$ sends $S\to1$ and would switch the suppressed source off
entirely, leaving nothing to test. That is the one respect in which the check departs
from the production configuration.

\begin{figure*}
\centering
\includegraphics[width=\textwidth]{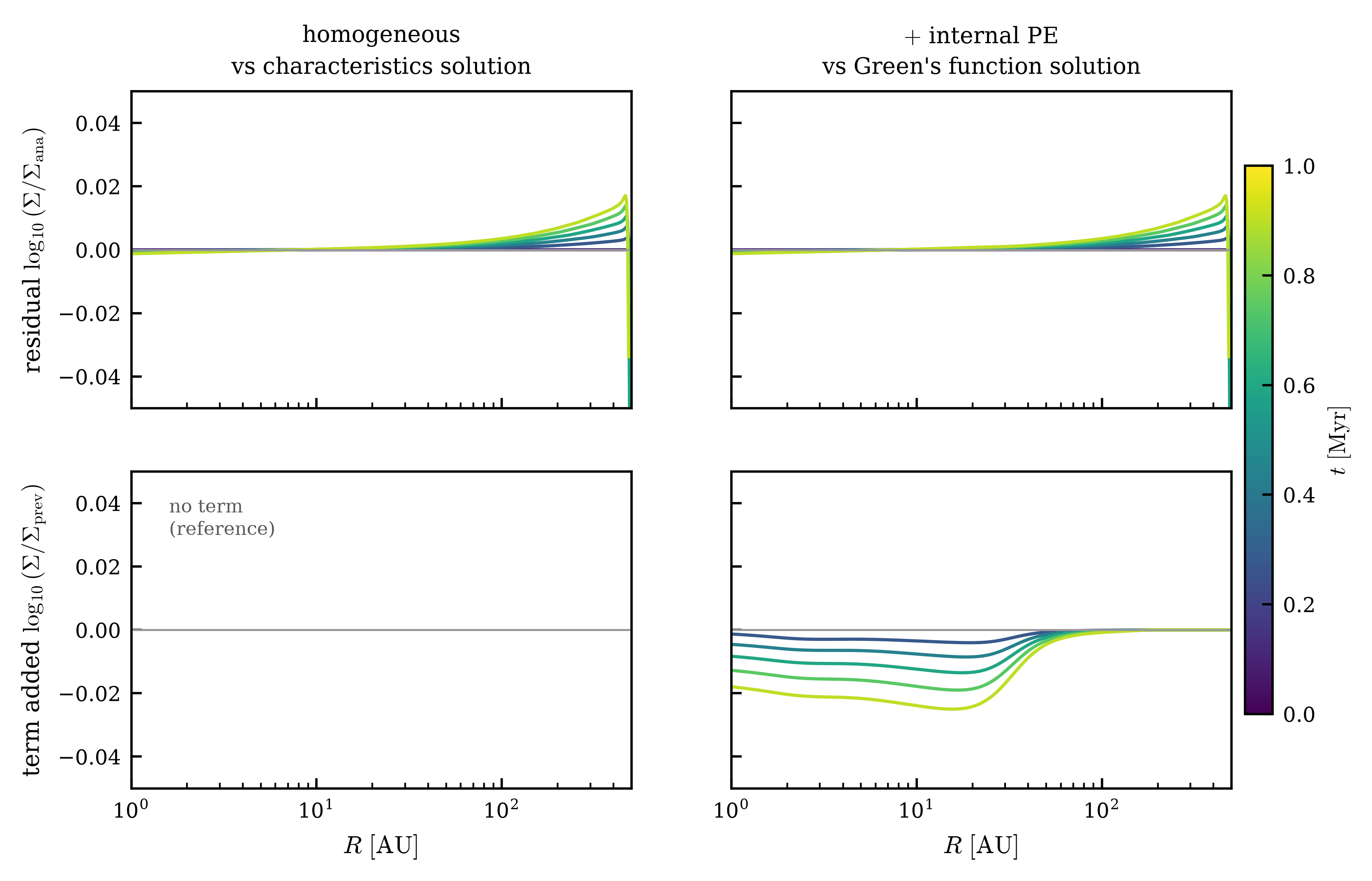}
\caption{The transport solver against both closed forms, in the limit
where they are exact, plotted against radius $R$ and coloured by time from
$t=0$ to $1$~Myr (bar at right). \emph{Columns:} the homogeneous
characteristics solution (equation~\ref{eq:charsol}) and the Green's function
with internal photoevaporation (equation~\ref{eq:greens}). \emph{Top:} the
residual $\log_{10}(\Sigma/\Sigma_{\rm analytic})$ against each. \emph{Bottom:}
the size of the term that column adds, $\log_{10}(\Sigma/\Sigma_{\rm prev})$,
on a symmetric-log scale. Configuration and interpretation in
Appendix~\ref{app:scheme}.}
\label{fig:appendix-scheme}
\end{figure*}

Figure~\ref{fig:appendix-scheme} shows the result for $A=10.49$~AU~Myr$^{-1}$ and
$\xi=0.25$. Over $459$ cells spanning $1$--$500$~AU the numerical solution reproduces the
closed form to a median $|\Delta\log_{10}\Sigma|$ below $10^{-4}$~dex at $t=0$, rising only
to $0.0010$~dex, under a quarter of a per cent, after $1$~Myr, with a maximum of
$0.034$~dex. The
two runs agree with their respective analytic solutions equally well.
The residual is not distributed uniformly, and neither is the term being tested. Inside
$\approx30$~AU, where the source acts, the photoevaporative correction grows to $5.6$~per
cent of $\Sigma$ by $1$~Myr ($0.025$~dex, dashed in the right-hand panel) while the
residual there is $0.0012$~dex. The convolution term is therefore recovered to about
five per cent of its own magnitude. Outside $\approx50$~AU the source has vanished and the residual is
instead set by advection, accumulating outward because the characteristics import
material from ever larger $R_i=R+At$, where the initial profile is falling exponentially.
That outer residual is a property of any finite-mass disc, not of the scheme, and
it bounds the comparison, not the integration.

\section{A high-accretion run from a wind-consistent initial condition}
\label{app:highmdot}

Two objections can be raised at once against the fiducial integrations of
Section~\ref{sec:results}. The first is that their absolute accretion rate
($\sim6\times10^{-9}\,\Msun\,\mathrm{yr^{-1}}$) sits below the bulk of young Class~II
accretors (Section~\ref{sec:caveats}). The second is that they start from the viscous
\citet{lyndenbell1974} profile $\Sigma_0\propto R^{-1}\exp(-R/R_{\rm c})$, which is the
$\psi\to0$ initial condition, in a model whose transport is wind-dominated.
To confirm the results are not artefacts of either choice, we answer both with a single
pair of runs: both branches restarted from the wind self-similar profile of
\citet{tabone2022}, $\Sigma_0\propto(R/R_{\mathrm{c}})^{\xi-1}\exp(-R/R_{\mathrm{c}})$,
normalised to the high end of the observed Class~II range.

Only five aspects of the setup differ from the fiducial run:
\begin{enumerate}
    \item Ejection index: unlike the fiducial's $\Sigma_0\propto R^{-1}$ profile,
     which does not depend on $\xi$, the wind-consistent profile here does.
     It is then initially set to the same plateau value as the fiducial
     run: $\xi=0.0864$ ($\lambda=6.79$).
    \item Disc mass: raised to $0.1\,\Msun$, which gravitational stability caps
     ($Q_{\rm min}=3.96$ at $0.1\,\Msun$; $Q_{\rm min}=2.02$, marginal, at
     $0.15\,\Msun$).
    \item Transport stress: $\aDW^0=5\times10^{-3}$, calibrated by short
     integrations on the actual profile instead of scaled analytically. With the disc
     mass above this gives $\Mdot_\ast(0)=1.25\times10^{-7}\,\Msun\,\mathrm{yr^{-1}}$.
    \item Preserved magnetisation: $\beta_0(1\,\mathrm{AU})=30$ is held, so
     the launching state is preserved: $R_\beta(0)=47.2$~AU, against the fiducial's
     $47.16$~AU.
    \item Time scaling: $\tau_B$ and $t_{\rm end}$ are rescaled by the ratio of
     accretion timescales, $f=t_{\rm acc}/t_{\rm acc}^{\rm fid}=0.954/3.241=0.294$,
     giving $\tau_B=0.15$~Myr and the same per-branch horizons the fiducial uses,
     $t_{\rm end}=3.0$ and $1.5$~Myr for the flux-retained and flux-loss branches.
     That rescaling is not trivial. The two-branch divide is governed by
     $\tau_B/t_{\rm acc}$, not by $\tau_B$ alone (it asks whether flux is lost within
     a disc lifetime), so holding $\tau_B=0.5$~Myr while $t_{\rm acc}$ shortens by
     $3.4$ would change the regime rather than the timescale. With the rescaling,
     $\tau_B/t_{\rm acc}$ and $t_{\rm end}/t_{\rm acc}$ match the fiducial's to $1.9$
     per cent.
\end{enumerate}

\subsection*{Maximize the initial accretion rate}

The accretion rate cannot be raised freely, and the obstruction is not numerical.
For the irradiated profile $\cs^2\propto R^{-1/2}$ the wind-driven drift
$v_{\rm DW}=\tfrac32\aDW\cs^2/\vK$ is radius-independent, so for
$\Sigma\propto R^{-1}$ the accretion rate is too, and the accretion timescale is
\begin{equation}
t_{\rm acc}\equiv\frac{M_{\rm disc}}{\Mdot_\ast}
\;\approx\;2.9~\mathrm{Myr}\left(\frac{10^{-3}}{\aDW^0}\right)
\left(\frac{R_{\rm c}}{30~\mathrm{AU}}\right).
\label{eq:tacc}
\end{equation}
in which $M_{\rm disc}$ cancels: raising the disc mass raises the reservoir and
the rate in exact proportion, so it cannot repair the mismatch.
Fixing $\Mdot_\ast$ and $R_{\rm c}$ therefore fixes
$t_{\rm acc}=M_{\rm disc}/\Mdot_\ast$ (equation~\ref{eq:tacc}), and the only remaining
freedom, $M_{\rm disc}$, is capped near $0.1\,\Msun$ by gravitational stability.
Demanding $\Mdot_\ast\sim10^{-6}\,\Msun\,\mathrm{yr^{-1}}$ therefore forces
$t_{\rm acc}\lesssim0.1$~Myr. The disc empties before $\tau_B$ can act at all, so the
model's central two-branch divide cannot be evaluated at that rate.
A $5$~Myr run at $10^{-6}\,\Msun\,\mathrm{yr^{-1}}$ would need
$M_{\rm disc}\approx5\,\Msun$.

\subsection*{The two-branch divide survives}

\begin{figure*}
\centering
\includegraphics[width=0.82\textwidth]{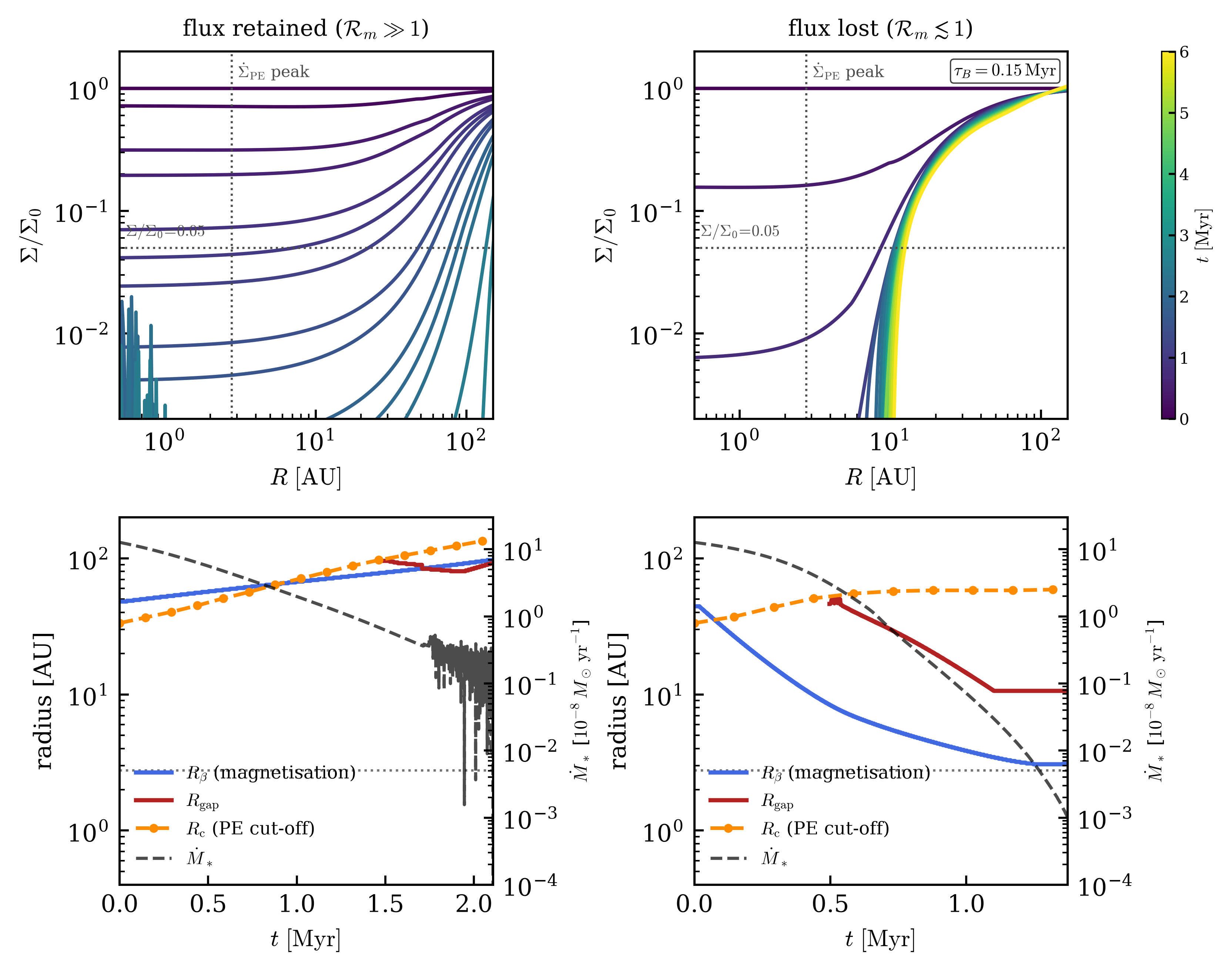}
\caption{As Fig.~\ref{fig:integrator}, but for both branches restarted from the
wind-consistent \citet{tabone2022} initial profile, normalised to the high end
of the observed Class~II accretion range ($17$ times the fiducial
$\Mdot_\ast(0)$), with $\beta_0(1\,\mathrm{AU})$ held so that $R_\beta(0)$ is
unchanged, and $\tau_B$ and $t_{\rm end}$ rescaled so that the dimensionless
control ratios $\tau_B/t_{\rm acc}$ and $t_{\rm end}/t_{\rm acc}$ match
Fig.~\ref{fig:integrator}'s. Holding those ratios fixed does not make the
outcome dimensionless, and the appendix text does not claim it does. What this
figure tests is that the two-branch divide occurs at all at this accretion rate.
Panels and colour scale follow Figure~\ref{fig:integrator}: surface density at a
regular cadence carried to the end of each run, coloured by the shared bar above,
and curves thinned to at most $16$ per panel where the stored cadence is finer.
Neither branch reaches the $1\,\mathrm{g\,cm^{-2}}$ dispersal threshold within
the integrated horizon: at this disc mass the photoevaporative depletion time
is a factor $\sim5$ longer than in the fiducial, for the reason given in the
appendix text. Parameters, and the post-collapse caveats, are given there.}
\label{fig:appendix-highmdot}
\end{figure*}

As shown in Figure~\ref{fig:appendix-highmdot}, the two-branch divide survives intact at $17$
times the fiducial accretion rate and from a different initial profile. The
magnetisation front expands on flux retention ($R_\beta:47.2\to87.7$~AU before collapse)
and recedes on flux loss ($47.2\to3.0$~AU). The photoevaporative gap opens on both
($1.47$~Myr at $97.7$~AU, and $0.50$~Myr at $52.5$~AU). Finally, the flux-retained branch
outlives the flux-loss one, collapsing at $1.92$~Myr against $1.37$~Myr, each comfortably
inside its own horizon. Every qualitative element of Section~\ref{sec:bifurcation}
therefore reappears at an accretion rate an order of magnitude higher and from a
wind-consistent instead of a viscous initial profile.

The sizes of these intervals are a different matter, and we do not claim they transfer.
Expressed in units of each run's own $t_{\rm acc}$, the flux-retained collapse
does reproduce the fiducial's ($2.05$ against $2.01$). The flux-loss collapse, however,
comes $34$ per cent later, the gap opens $62$ and $93$ per cent later on the two branches,
and the branch separation is $36$ per cent smaller ($0.58$ against $0.91\,t_{\rm
acc}$). The wind-consistent profile is shallower in the inner disc than the viscous one,
and drains correspondingly more slowly. Holding $\tau_B/t_{\rm acc}$ and
$t_{\rm end}/t_{\rm acc}$ fixed does not by itself make the outcome dimensionless: the
two-branch divide also depends on where the launching region sits relative to the mass, which
the two initial profiles place differently.
This appendix therefore establishes that the two-branch divide is not an artefact of the
low-accretion normalisation adopted in the main text, not that its magnitude is a universal
function of the dimensionless parameters.

One dimensional scale in particular refuses to be absorbed, and it is the reason this
run never disperses. The internal photoevaporation rate is set by the X-ray luminosity,
not by the disc. $\dot\Sigma_{\rm PE}(R)$ is the same absolute profile in both
configurations, delivering $5.1$ and $5.6\times10^{-10}\,\Msun\,\mathrm{yr^{-1}}$ after
screening, a $10$ per cent difference against a factor $5$ in disc mass. The local
depletion time $\Sigma/\dot\Sigma_{\rm PE}$ over $8$--$40$~AU is consequently
$26$~Myr in the fiducial and $121$~Myr here. Once flux loss has collapsed $R_\beta$ to a
few AU, photoevaporation is the only sink left, and at that rate the disc is static on
the horizon integrated. It retains $33$ per cent of its initial mass at $5.3$~Myr, with
$\max\Sigma$ still at $29\,\mathrm{g\,cm^{-2}}$, against the $1\,\mathrm{g\,cm^{-2}}$
dispersal threshold. The cavity profile of Appendix~\ref{app:cavitype} never engages
either, since it amplifies the loss at an advancing wall and no wall advances. $R_{\rm
cav}$ wanders non-monotonically between $3$ and $10$~AU for the whole run. The
implication is a sharper form of the disc-lifetime problem discussed in
Section~\ref{sec:caveats}: at fixed $L_{\rm X}$, internal photoevaporation cannot
disperse a massive, high-accretion disc on an observable timescale, because its rate is
a stellar property while the mass to be removed is a disc property.

\section{Independent literature validation}
\label{app:literature-validation}

We now test the numerical integration against two independent, externally published
results, each isolating one core component: the transport solver \citep{tabone2022} and
the internal photoevaporative sink \citep{sellek2024}. For each test we adopt the
published configuration and switch off all unrelated physics: the magnetothermal wind
coupling and the screening among it.

\subsection{The transport solver: Tabone et al. (2022)}
\label{app:tabone2022-validation}

We benchmark against the exact closed-form self-similar solutions for disc mass
$M_{\mathrm{D}}(t)$ derived by \citet{tabone2022}, which isolates the advective and
diffusive operators. We test two families: a constant-$\aDW$ ``hybrid'' family
parameterised by the transport ratio $\psi=\alpha_{\mathrm{DW}}/\alpha_{\mathrm{SS}}$
(their Figure 6), and an evolving $\aDW(t)\propto\Sigma_{\mathrm{c}}(t)^{-\omega_B}$
family parameterised by the field-decay index $\omega_B$ of Section~\ref{sec:validity}
(their Figure 7).

Because the wind-drift velocity ($\varv_{\rm DW}=\tfrac{3}{2}\aDW\epsilon c_{\mathrm{s}}$,
their eq.~15) and the viscous flux in our integrator are algebraically identical to
their formulation, we compute the comparison timescale $t_{\mathrm{acc,0}}$ and the
exact ejection index $\xi(\psi,\lambda)$ directly from the integrator physics, without
fitted parameters. That index is their exact eq.~(14), not the
$\psi\to\infty$ limit $\xi=1/[2(\lambda-1)]$ used elsewhere in this paper, so the
benchmark does not assume the identity the rest of the model relies on.

\begin{figure*}
    \centering
    \includegraphics[width=\textwidth]{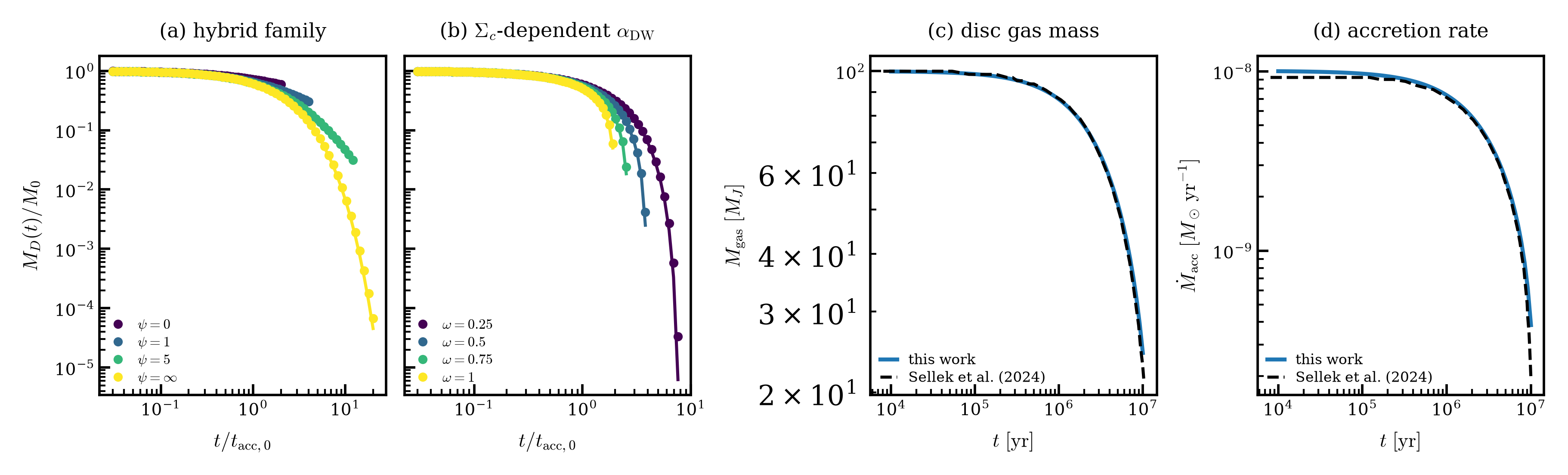}
    \caption{Validation against \citet{tabone2022} \emph{(a, b)} and
    \citet{sellek2024} \emph{(c, d)}. \emph{(a, b)}: disc mass evolution
    $M_D(t)/M_0$ vs.\ $t/t_{\rm acc,0}$, this integrator's numerical solution
    (points) against \citeauthor{tabone2022}'s exact closed-form self-similar
    solutions (lines): \emph{(a)} their Fig.~6 hybrid family (constant
    $\aDW$) at $\psi=\alpha_{\rm DW}/\alpha_{\rm SS}=0$ (pure viscous), $1$,
    $5$ and $\infty$ (pure wind); \emph{(b)} their Fig.~7 $\Sigma_c$-dependent
    $\aDW(t)\propto\Sigma_c(t)^{-\omega_B}$ family at
    $\omega_B=0.25,0.5,0.75,1.0$, each terminating near their predicted finite
    dispersal time $t_{\rm disp}=2t_{\rm acc,0}/\omega_B$. \emph{(c, d)}: disc
    gas mass and stellar accretion rate for the pure-viscous, unscreened
    configuration of \citeauthor{sellek2024}'s own fig.~15 (``this work''
    curve), this work (blue) against their published curve (black dashed,
    digitised). Agreement is at the few-per-cent level (median
    $0.6\%$/$4.7\%$) over the first $9$~Myr, diverging only in the final
    $\sim1$~Myr as the disc approaches near-complete depletion (see text).}
    \label{fig:appendix-tabone2022}
    \label{fig:appendix-sellek2024}
\end{figure*}

As shown in Figure~\ref{fig:appendix-tabone2022}, the numerical integration reproduces
the analytic $M_{\mathrm{D}}(t)$ across all eight parameter regimes, to a median of
$0.1$--$1$ per cent in each. The larger excursions are confined to the late-time tail
where both curves approach zero mass. The comparison window for finite-$\psi$ runs is
truncated once viscous spreading carries the disc beyond the integrator's $\beta_0(R)$
magnetic launch switch (Section~\ref{sec:flux}):
a physical constraint present in our numerical setup, set by this paper's own
fiducial $\beta_0(R)$ instead of by any free parameter of this comparison, and absent
in the idealised analytic model.

\subsection{The internal photoevaporation sink: Sellek et al. (2024)}
\label{app:sellek2024-validation}

To isolate the internal X-ray-driven mass-loss sink, we replicate the plot from
Fig.~15 of \citet{sellek2024}.
The configuration is a purely viscous disc
($\alpha_{\mathrm{SS}}=10^{-3}$, $\alpha_{\mathrm{DW}}=0$) with a $1~M_\odot$ star,
initial mass $100~M_{\mathrm{Jup}}$, cut-off radius $R_{\mathrm{c}}=100$~AU, and
$L_{\mathrm{X}}=2\times10^{30}~\mathrm{erg}~\mathrm{s}^{-1}$, integrated to $t=10$~Myr.
We disable the co-moving screening $R_{\mathrm{c}}(t)$ of \citet{picogna2026} adopted
elsewhere in this paper, leaving internal photoevaporation continuous and unscreened as
\citeauthor{sellek2024} originally modelled it.

Figure~\ref{fig:appendix-sellek2024} compares our numerical output to the digitised
reference curve. The two agree closely over the first $9$~Myr, with median offsets of
$0.6\%$ in gas mass and $4.7\%$ in accretion rate. They diverge only in the final
$\sim 1$~Myr as the disc depletes, where small timing offsets become large fractional
errors near zero mass. This validates the
internal sink and the viscous transport term in isolation: the magnetothermal wind, the
$\Xi$--$T$ wind-base prescription and the $R_{\rm c}(t)$ screening used elsewhere in
this paper are all inactive here.

\section{Grid-resolution convergence}
\label{app:resolution-check}

\begin{figure*}
    \centering
    \includegraphics[width=\textwidth]{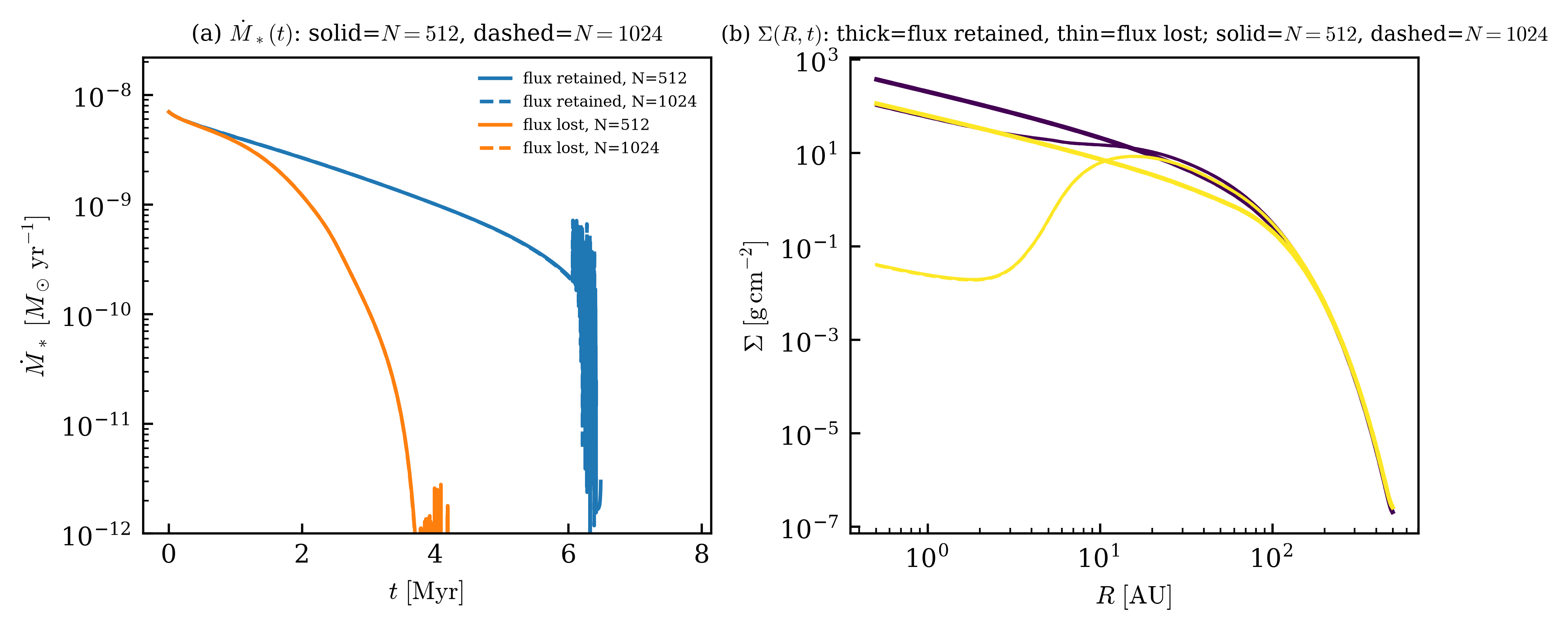}
    \caption{Grid-resolution convergence, $N=512$ (fiducial) vs.\ $N=1024$,
    log spacing, for both named branches of Figure~\ref{fig:integrator}.
    \emph{Left:} $\dot M_\ast(t)$ (solid: $N=512$; dashed: $N=1024$; blue:
    flux retained; orange: flux lost), with the ordinate floored at
    $10^{-12}\,\Msun\,\mathrm{yr^{-1}}$. Below the collapse criterion both
    resolutions oscillate over many decades on gas carrying negligible mass, and
    that range is excluded from interpretation in any case
    (Appendix~\ref{app:trust-near-dispersal}). The structure retained just above
    the floor near $6.5$~Myr is not numerical scatter but the threshold
    re-crossing described in Section~\ref{sec:bifurcation}, which is why the
    collapse epoch is quoted with a criterion attached. \emph{Right:} $\Sigma(R,t)$ at $t=2.0$ and
    $3.5$~Myr (thick: flux retained; thin: flux lost; solid: $N=256$; dashed:
    $N=512$). Both epochs are pre-collapse on both branches, and the two
    resolutions record snapshots on a common $0.5$~Myr grid, so each pair
    compares identical times.}
    \label{fig:appendix-resolution-check}
\end{figure*}

Figure~\ref{fig:appendix-resolution-check} evaluates convergence by doubling the
fiducial grid resolution from $N=512$ to $N=1024$ (logarithmic spacing).
Both resolutions are run to the same per-branch horizon and record snapshots on the
same $0.5$~Myr grid, so every comparison below is between identical epochs.
Across both evolutionary branches, the stellar accretion rate $\dot{M}_\ast(t)$
overlaps almost perfectly up to terminal collapse, whose epoch shifts by $-0.6\%$
(flux-retained) and $-0.01\%$ (flux-loss). These are differential shifts
measured under the fixed criterion of Section~\ref{sec:bifurcation}, not a claim of
that precision on the absolute epoch, which that criterion leaves far less
well determined.

Prior to collapse, the surface density profiles $\Sigma(R,t)$ agree across the full
radial range to a median of $0.6\%$ ($2.0$~Myr) and $1.3\%$ ($3.5$~Myr) on the
flux-retained branch, and $0.9\%$ and $3.8\%$ on the flux-loss one.
The magnetisation front $R_\beta(t)$ is best compared in grid cells, one cell being
worth $1.4$~per cent in $R$ at $N=512$ and $0.7$ at $N=1024$. Measured over the whole
pre-collapse trajectory rather than at sampled epochs, the two resolutions agree to a
median of $0.25$ cells on both branches, and lie within one cell for $89$~per cent
(flux-retained) and $100$~per cent (flux-loss) of that trajectory.
Because the quantities quoted in the main text are converged at the per-cent level,
$N=512$ is adopted as the fiducial resolution.

\subsection{What can be trusted near dispersal}
\label{app:trust-near-dispersal}

Grid convergence degrades only during the terminal collapse phase. Rather than record
that as a generic numerical artefact, we set out where it degrades, what the disc is
doing there, and why further grid refinement is the wrong tool for it.

The flux-loss branch traverses its dispersal phase fully converged.
Its median $\Sigma$ discrepancy stays below
$0.8\%$ at every epoch sampled, and $R_\beta$ agrees to a median of $0.25$ grid
cells and stays within a single cell for the full simulation.
The flux-retained branch is converged in the same sense over most of its life
(also a median of $0.25$ cells, and within one cell $89$ per cent of the time),
but degrades over the approach to collapse. The disagreement exceeds two cells for
$3.0$ per cent of the pre-collapse trajectory, in brief excursions of $3$--$109$~kyr
between $t_{\rm coll}-0.48$ and $t_{\rm coll}-0.30$~Myr. These peak at $2.8$ cells
($3.7$ per cent in radius) while the front is sweeping outward from $110$ to
$140$~AU. Local $\Sigma$ errors reach $16\%$ over the final $0.05$~Myr, along the
steepest edges of the collapsing density profile.
These values are quoted for $N=512\rightarrow1024$, but where the same comparing
$N=256$ to $512$ grid cells. So $R_\beta$ converges at first order in radius.

When the flux-retained branch loses convergence, it retains $0.70\%$ of its
initial mass at $t_{\rm coll}-0.15$~Myr and $0.58\%$ at $t_{\rm coll}-0.05$, and
$80\%$ of the remaining active grid cells (carrying $47\%$ of what mass is left)
have dropped below $\Sigma/\Sigma_0=10^{-2}$.
That threshold is a hard boundary: it is the lower limit of the
\citet{sellek2024} $\Xi$--$T$ calibration, below which the model reverts to the
cold-base fallback of Section~\ref{sec:bridge}. The $\beta$--$\Sigma$ closure
(\ref{eq:betaclosure}) is extrapolated by a comparable margin.\footnote{The
flux-loss branch escapes this: $36\%$ of its cells are also sub-floor at
collapse, but they carry only $0.1\%$ of the remaining gas mass and so are
dynamically irrelevant. What matters is not how many cells are extrapolated but how
much mass they hold.}

The residual disagreement is largely a phase offset. Comparing the two
resolutions at equal time to collapse, as opposed to at equal time, removes most
of it. On the flux-retained branch the median $\Sigma$ discrepancy falls from
$9.2$ to $2.3$ per cent at $t_{\rm coll}-0.15$~Myr, and from $16.4$ to $0.7$ per
cent at $t_{\rm coll}-0.05$. The two grids therefore evolve through nearly the same
sequence of states and reach them at slightly different times, instead of
diverging onto different solutions.

Grid refinement is not the remedy. On the flux-loss branch the
dispersal epoch is definition-limited, as opposed to grid-limited, by a wide margin.
Doubling the resolution moves it by $0.5$~kyr, whereas moving the $\dot{M}_\ast$
collapse criterion from $10^{-10}$ to $10^{-13}~M_\odot~\mathrm{yr}^{-1}$ moves
it by $963$~kyr, a factor of $1800$. On the flux-retained branch refinement moves
the epoch by $40$~kyr and would buy something.
What it would not buy is the thing that limits the answer.
Inside the terminal window the physical closures are being extrapolated decades
outside their calibrated density range, so a numerically converged result there is
a more precise solution to the wrong equations.

We therefore make no quantitative claims about the terminal $0.3$~Myr of the
flux-retained branch, nor about its post-collapse plateau.
What the paper does quote (the dispersal epochs to two significant digits with
their criterion range attached, the branch separation, the sustained accretion rates,
and $\Sigma(R,t)$ and the front trajectories outside that window) is extracted
exclusively from the converged windows, and is converged there at the per-cent level or
better. It is also differential, in a way that makes the residual sensitivity
common to both branches. The two-branch divide the paper rests on is therefore unaffected
even where the absolute epochs are not well determined.

\section{The mass-loss profile at a cavity wall}
\label{app:cavitype}

The photo-evaporative profile applied above is the one computed for a full disc,
held fixed for all time.
That is the right choice while the disc is primordial, and it brakes down
once a cavity has opened.
The cavity wall is directly illuminated, and the mass-loss rate per unit area there
is enhanced over the primordial value at the same radius.
\citet{picogna2019}-family radiation-hydrodynamic calculations of
transition discs (Picogna et al., in prep.) give that enhancement as a function
of the wall radius $R_{\rm cav}$. Expressed as a local, dimensionless ratio to
the primordial profile at the same radius, it is
\begin{equation}
E(u)=1+5.0\,(u-1)^{-0.23}\exp\!\left[-(u-1)/0.15\right],
\qquad u\equiv R/R_{\rm cav},
\label{eq:cavitykernel}
\end{equation}
applied for $R_{\rm cav}\le R<2R_{\rm cav}$, with $\Sigdot_{\rm PE}=0$ inside the
wall and the primordial profile recovered exactly beyond .
The enhancement is sharply confined, which is the behaviour
\citet[][their Figure A.1]{monsch2021} report for the same family of calculations,
showing the profile returning to the primordial one just outside the wall. It peaks at
$11$--$14\times$ the primordial rate at the wall itself, nearly independently of
$R_{\rm cav}$, while raising the integrated rate by only $2$--$27$ per
cent over the range of cavity radii sampled.

\begin{figure}
\centering
\includegraphics[width=\columnwidth]{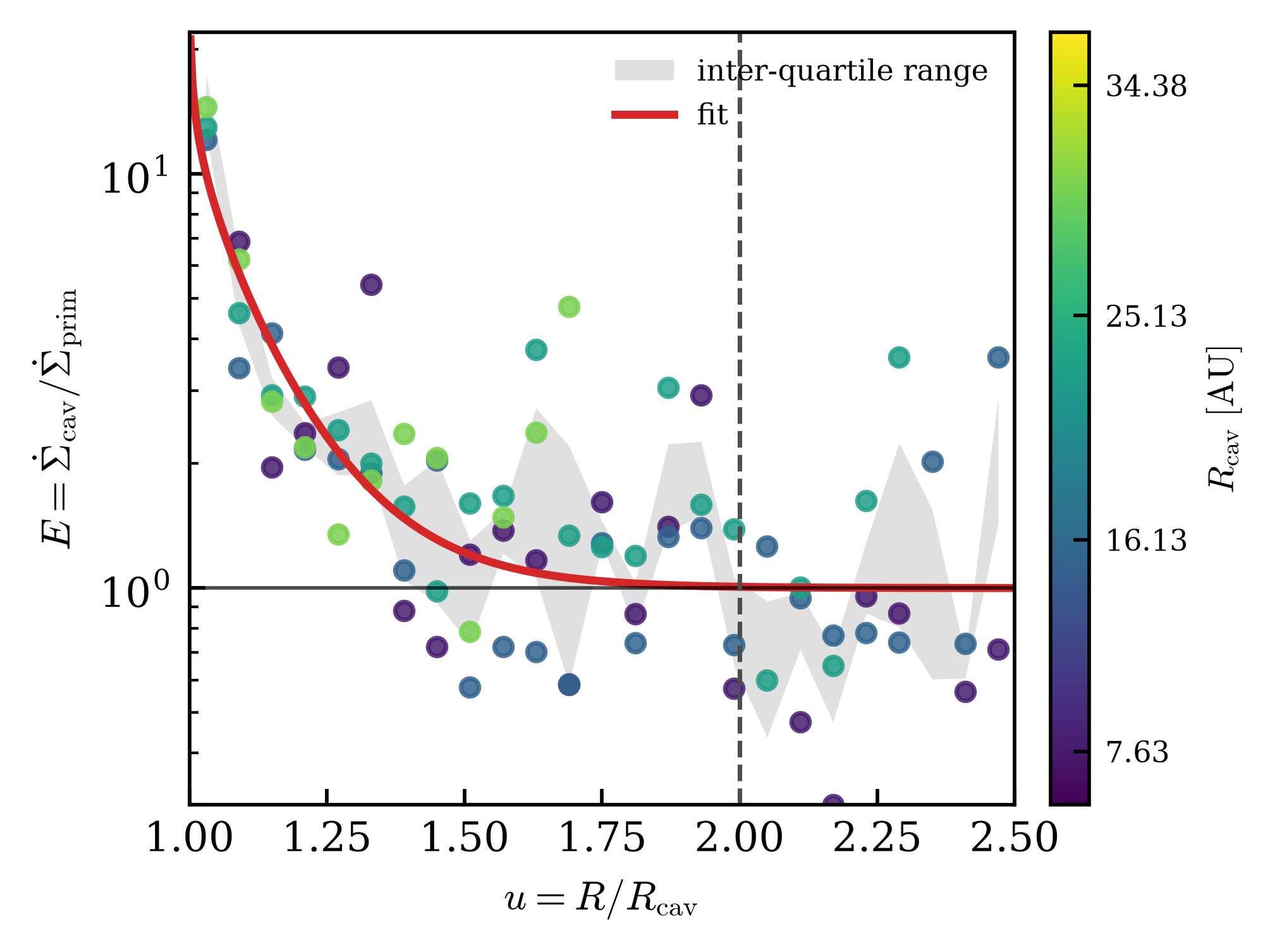}
\caption{Equation~(\ref{eq:cavitykernel}) (red) against the four transition-disc
runs it is fitted to (coloured points, $R_{\rm cav}=7.6$--$34.4$~AU), each
measured as the ratio of the cavity run's mass-loss rate to the primordial
run's at the same radius, binned in $u=R/R_{\rm cav}$. The shaded band is the
inter-quartile range across runs at each $u$; the dashed line marks $u=2$,
beyond which the profile reverts to the primordial one by construction.}
\label{fig:appendix-cavitykernel}
\end{figure}

$E$ is a ratio measured against the \citet{picogna2019} primordial profile but
applied here to the \citet{sellek2024} one. The two peak within $0.5$~AU of each
other ($3.2$ against $2.8$~AU), which is what makes the transfer defensible, but
their integrated rates differ by a factor $9.4$.
Figure~\ref{fig:appendix-cavitykernel} shows the fit of equation~(\ref{eq:cavitykernel})
overlayed on the mass-los rates derived from the hydrodynamic numerical simulations
(Picogna et al., in prep.), which we publish alongside the rest of this paper's archived
data (Data availability).

On both branches a genuine density cavity forms well before the accretion
collapse (at $t=2.05$~Myr on the flux-loss branch and $3.74$~Myr on the
flux-retained one, see Section~\ref{sec:bifurcation}), and outside the terminal window
Appendix~\ref{app:trust-near-dispersal} excludes.
Its effect on the clearing is significant.
Repeating the flux-loss integration with equation~(\ref{eq:cavitykernel}) active
and nothing else changed, the cavity wall advances at
$5.3\pm0.3$~AU~Myr$^{-1}$ against $1.7\pm0.1$ with the primordial profile, over
the common window $3.8$--$5.5$~Myr in which both runs have a cavity. At matched
epochs the ratio of the two cavity radii grows from $1.4$ at $t=4.1$~Myr to
$1.9$ at $5.5$~Myr, so the difference is still widening when the comparison
ends. The retained mass, by contrast, differs by only $2.6$ percentage points at
$5.5$~Myr ($23.5$ against $26.1$ per cent). The cavity-edge enhancement
therefore changes where the gas is removed far more than how much
is removed. This is what equation~(\ref{eq:cavitykernel}) predicts, since it
raises the integrated rate by at most a few tens of per cent while amplifying
the local rate at the wall by an order of magnitude.

\bsp
\label{lastpage}
\end{document}